\pdfoutput=1
\documentclass[PAPER, atlasdraft=false, UKenglish, cernpreprint, texmf, orcidlogo]{atlasdoc}
 
\usepackage[subfigure]{atlaspackage}
\usepackage{atlasbiblatex}
 
\usepackage[process,hepprocess,jetetmiss,snippets]{atlasphysics}

\graphicspath{{logos/}{figures/}}
 
\usepackage{ANA-JETM-2023-01-PAPER-defs}
\usepackage{MC_snippets-defs}

\AtlasTitle{Improving topological cluster reconstruction using calorimeter cell timing in ATLAS}

\AtlasAbstract{
Clusters of topologically connected calorimeter cells around cells with large absolute signal-to-noise ratio (\textit{topo-clusters}) are the basis for calorimeter signal reconstruction in the ATLAS experiment.
Topological cell clustering has proven performant 
in LHC Runs~1 and~2.
It is, however, susceptible to \textit{out-of-time pile-up} of signals from soft collisions outside the 25~ns proton-bunch-crossing window associated with the event's hard collision.
To reduce this effect, a calorimeter-cell timing criterion was added to the signal-to-noise ratio requirement in the clustering algorithm.
Multiple versions of this criterion were tested by reconstructing hadronic signals in simulated events and Run~2 ATLAS data.
The preferred version is found to reduce the out-of-time pile-up jet multiplicity by ${\sim}50\%$ for jet $p_\text{T}\sim20$~GeV and by ${\sim}80\%$ for jet $p_\text{T} \gtrsim 50$~GeV, while not disrupting the reconstruction of hadronic signals of interest, and improving the jet energy resolution by up to 5\% for $20 < p_\text{T} < 30$~GeV.
Pile-up is also suppressed for other physics objects based on topo-clusters (electrons, photons, $\tau$-leptons), reducing the overall event size on disk by about $6\%$ in early Run~3 pile-up conditions.
Offline reconstruction for Run~3 includes the timing requirement.
}
 
\author{The ATLAS Collaboration}
 
\AtlasRefCode{JETM-2023-01}
 
\PreprintIdNumber{CERN-EP-2023-207}

\AtlasJournal{Eur.\ Phys.\ J.\ C}
\AtlasJournalRef{Eur.\ Phys.\ J.\ C 84 (2024) 455}
\AtlasDOI{10.1140/epjc/s10052-024-12657-1}

\hypersetup{pdftitle={ATLAS document},pdfauthor={The ATLAS Collaboration}}

\begin{document}
 
\maketitle
 
\tableofcontents

% The next lines are included from the .//IntroductionAndCut_pap.tex input file
\newcommand{\AtlasCoordFootnote}{
ATLAS uses a right-handed coordinate system with its origin at the nominal interaction point (IP)
in the centre of the detector and the \(z\)-axis along the beam pipe.
The \(x\)-axis points from the IP to the centre of the LHC ring,
and the \(y\)-axis points upwards.
Cylindrical coordinates \((r,\phi)\) are used in the transverse plane,
\(\phi\) being the azimuthal angle around the \(z\)-axis.
The pseudorapidity is defined in terms of the polar angle \(\theta\) as \(\eta = -\ln \tan(\theta/2)\).
Angular distance is measured in units of \(\Delta R \equiv \sqrt{(\Delta\eta)^{2} + (\Delta\phi)^{2}}\).
The rapidity \(y=\frac{1}{2}\ln{\left( \frac{E+p_z}{E-p_z} \right) }\) can also be used for kinematic distributions instead of $\eta$, as it accounts for the particle's mass.}

\section{Introduction}
 
The ATLAS experiment~\cite{PERF-2007-01}, one of the four major experiments at the LHC~\cite{Evans:2008zzb}, relies on the precise measurement of particle showers as the starting point for physics object reconstruction. Calorimeter showers are reconstructed by clustering groups of topologically connected calorimeter cells, the first step in this process involving cells with high absolute energy relative to calorimeter noise~\cite{PERF-2014-07}. The produced clusters are referred to as \textit{topo-clusters} and provide the input to the reconstruction of a number of physics objects, including jets, electrons, $\tau$-leptons and missing transverse momentum (\met).
The topological cell clustering algorithm is well established and performed very well during the LHC Run~1 and Run~2 data-taking periods, from 2010 to 2012 and 2015 to 2018 respectively. However, further improvement is possible, especially in view of the increased luminosity foreseen in LHC Run~3 and beyond.
 
At the LHC, bunches of about $10^{11}$ protons collide every 25~ns, giving rise to multiple $pp$ interactions in each \textit{bunch-crossing}.\footnote{The average number of \pp collisions per bunch-crossing was 25 in 2016, 38 in 2017 and 36 in 2018 data-taking~\cite{DAPR-2021-01}.} Signals from particles produced in additional soft \pp collisions pile up on top of those from the \pp \textit{hard-scattering} that triggered the ATLAS data-recording process. If the additional particles are produced in the same bunch-crossing that produced the recorded event's \pp hard-scattering, they are referred to as \textit{in-time \pileup}, whereas if they are produced in the previous or next bunch-crossings they are called \textit{out-of-time \pileup}.
Calorimeter signals are sensitive to both in-time and out-of-time \pileup, which can result in the formation of additional topo-clusters, as well as spurious contributions to those originating from the hard scattering.
Multiple \pileup suppression techniques have been studied in ATLAS, such as the grooming algorithms used for the reconstruction of boosted objects~\cite{Krohn:2009th, Larkoski:2014wba, Dasgupta:2013ihk} or the Constituent Subtraction~\cite{Berta:2014eza} and Soft Killer~\cite{Cacciari:2014gra} methods used prior to jet reconstruction~\cite{JETM-2018-06}. These methods, however, are all applied after  topo-cluster reconstruction, whereas in this \mypaper\ a low-level mitigation technique is explored with the goal of reducing \pileup contributions at cell level while building the topo-clusters.
 
The ATLAS calorimeters provide a measurement of the signal time, in addition to the deposited energy. In this \mypaper, the topo-clustering algorithm is modified by applying a calorimeter time-measurement selection criterion (the \textit{time cut} in the following) to cells with large signal-to-noise ratio in order to reject contributions from out-of-time \pileup.
Multiple versions of the time cut were compared and tested on both Monte Carlo (MC) simulated events and a sample of ATLAS data events. An Upper Limit, switching off the time cut for very high energy signals, was explored and it was adopted as a safety measure to avoid rejecting potential new-physics processes which might produce out-of-time signals.
The performance of the new algorithm has been tested in the context of jet reconstruction, focusing on the suppression of \pileup-initiated jets as well as its effects on jet energy calibration and resolution.
 
The rest of this \mypaper \ is organised as follows. Section~\ref{sec:atlas} describes the detector. Section~\ref{sec:samples} lists the data and MC samples used for this analysis. Section~\ref{sec:topoclusteringalgo} describes the current topo-cluster reconstruction algorithm and Section~\ref{sec:thecut} introduces the time cut. Section~\ref{sec:mcperf} discusses the time cut's performance on topo-cluster and jet kinematics as evaluated using MC samples. Section~\ref{sec:dataperf} discusses the time cut's performance on ATLAS data taken during Run~2 of the LHC, including checks for possible time-cut inefficiency and the cut's impact on the ATLAS event size. Conclusions are presented in Section~\ref{sec:conclusion}.

\section{ATLAS detector}
\label{sec:atlas}
 
The ATLAS detector~\cite{PERF-2007-01} at the LHC covers nearly the entire solid angle around the collision point.\footnote{\AtlasCoordFootnote}
It consists of an inner tracking detector surrounded by a thin superconducting solenoid, electromagnetic and hadron calorimeters,
and a muon spectrometer incorporating three large superconducting air-core toroidal magnets.
 
The inner-detector system is immersed in a $2~\textrm{T}$ axial magnetic field
and provides charged-particle tracking in the range \(|\eta| < 2.5\).
The high-granularity silicon pixel detector covers the vertex region and typically provides four measurements per track,
the first hit normally being in the insertable B-layer~\cite{ATLAS-TDR-19,PIX-2018-001} installed before Run~2.
It is followed by the silicon microstrip tracker, which usually provides eight measurements per track.
These silicon detectors are complemented by the transition radiation tracker (TRT),
which enables radially extended track reconstruction up to \(|\eta| = 2.0\).
The TRT also provides electron identification information
based on the fraction of hits (typically 30 in total) above a higher energy-deposit threshold corresponding to transition radiation.
 
The calorimeter system covers the pseudorapidity range \(|\eta| < 4.9\).
Within the region \(|\eta|< 3.2\), electromagnetic (EM) calorimetry is provided by barrel and
endcap high-granularity lead/liquid-argon (LAr) calorimeters,
with an additional thin LAr presampler covering \(|\eta| < 1.8\)
to correct for energy loss in material upstream of the calorimeters.
Hadron calorimetry is provided by the steel/scintillator-tile calorimeter (\textit{Tile calorimeter} in the following),
segmented into three barrel structures within \(|\eta| < 1.7\), and two copper/LAr hadronic endcap calorimeters.
The solid angle coverage is completed with forward copper/LAr and tungsten/LAr calorimeter modules
optimised for electromagnetic and hadronic energy measurements respectively.
Electromagnetic and hadronic calorimeters based on LAr technology are collectively referred to as the \textit{LAr calorimeter} in the following.
Additional details about the timing measurements in the ATLAS calorimeters are provided in Section~\ref{sec:calotiming}.
 
The muon spectrometer comprises separate trigger and
high-precision tracking chambers measuring the deflection of muons in a magnetic field generated by the superconducting air-core toroidal magnets.
The field integral of the toroids ranges between \num{2.0} and $6.0~\textrm{T}\cdot\textrm{m}$
across most of the detector.
Three layers of precision chambers, each consisting of layers of monitored drift tubes, cover the region \(|\eta| < 2.7\),
complemented by cathode-strip chambers in the forward region, where the background is highest.
The muon trigger system covers the range \(|\eta| < 2.4\) with resistive-plate chambers in the barrel, and thin-gap chambers in the endcap regions.
 
Interesting events are selected by the first-level (L1) trigger system implemented in custom hardware,
followed by selections made by algorithms implemented in software in the high-level trigger (HLT)~\cite{TRIG-2016-01}.
The first-level trigger accepts events from the $40~\textrm{MHz}$ bunch-crossings at a rate below $100~\textrm{kHz}$,
which the high-level trigger reduces in order to record events to disk at about $1~\textrm{kHz}$. Triggers with acceptance rates that are too large are prescaled, i.e.\ only a fraction of the events satisitying the trigger are written to disk. The prescale factor is then applied to estimate the original rate.
An extensive software suite~\cite{ATL-SOFT-PUB-2021-001} is used in data simulation, in the reconstruction
and analysis of real and simulated data, in detector operations, and in the trigger and data acquisition
systems of the experiment.
 
\subsection{Calorimeter timing measurement}
\label{sec:calotiming}
\begin{figure}
\centering
\includegraphics[width=0.4\linewidth]{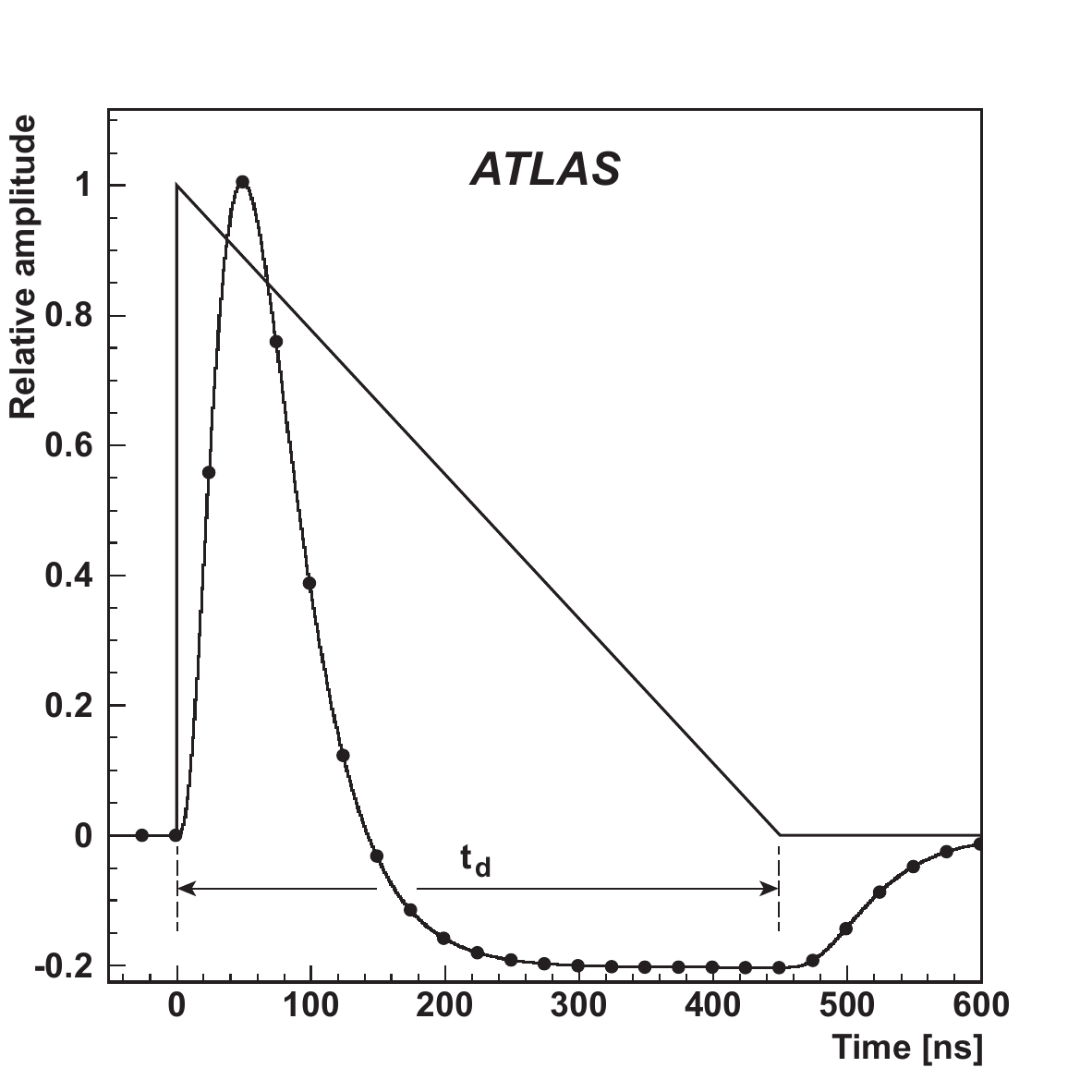} 
\caption{Example of the LAr (central EM barrel) calorimeter pulse shape from Ref.~\cite{PERF-2014-07}. The unipolar triangular pulse is the one generated by fast ionising particles in the liquid argon. Its characteristic time is the drift time $t_\textrm{d} \simeq 450~\textrm{ns}$ in the example shown. The shaped pulse is superimposed, with a characteristic duration of $t_{\textrm{signal}} \simeq 600~\textrm{ns}$. The full circles on the shaped pulse indicate the nominal bunch-crossings at 25~ns intervals.
}
\label{subfig:larPS}
\end{figure}

In addition to the energy deposited in each cell, both the LAr and Tile calorimeters can measure the time of arrival of the particle depositing the energy.
LAr signals are read out using Front End Boards (FEB). They shape the signal and sample it at $40~\textrm{MHz}$, four samples are then digitised if the event passes the L1 trigger. The shape of the LAr signal is optimised to minimise the noise contribution~\cite{NJBuchanan_2008}. An example of the typical LAr calorimeter pulse shape is shown in Figure~\ref{subfig:larPS}.
The long negative tail of the shaped LAr pulse implies that out-of-time \pileup provides a negative energy contribution on average, while the contribution from in-time \pileup is positive on average. Because of this, the average \pileup energy per event is zero.
While this feature is very useful in correcting for the average \pileup energy deposition, suppression methods like the one presented in this \mypaper \ are still very useful in suppressing \pileup contributions to individual physics objects.
 
The signal amplitude $A$ (proportional to its energy) and timing ($t$) are both reconstructed using an optimal filtering algorithm~\cite{Cleland:1994upr} applied to the digitised samples $S_i$:
\begin{equation}
A = \sum_{i=1}^{n_{\text{samples}}} a_i S_i, \qquad
t = \frac{1}{A} \sum_{i=1}^{n_{\text{samples}}} b_i S_i,
\label{eq:optFilt}
\end{equation}
where the optimal filtering coefficients $a_i$ and $b_i$ are computed from the predicted pulse shape and measured noise.
The cell time is only measured if the detected energy is above a certain configurable threshold. Typically, threshold values equal to three times the cell noise ($3\sigma_{\textrm{noise}}$) are used. If the reconstructed energy is below threshold, then the cell time is not computed and $t=0$ is stored.
 
The LAr time measurement is synchronised with the LHC clock and fine-tuned at the FEB level~\cite{LARG-2009-01}. Time alignment corrections are recalculated using beam-splash and early collision data~\cite{LAr_public_plots}. The measured time is also monitored as part of the data quality assessment~\cite{LARG-2013-01} in order to ensure its stability.
The LAr time resolution is typically parameterised as the sum in quadrature of a constant term and a ${\sim}1/E$ noise term:
\begin{equation*}
\sigma_t = \frac{p_1}{E} \oplus p_0.
\end{equation*}
In Run~2 and after applying calibration constants obtained from data, the constant term $p_0$ was found to reach ${\sim}200 \text{ ps}$, while the noise term $p_1$ was $\mathcal{O}(1~\text{GeV ns})$~\cite{LAr_public_plots}.
 
The Tile calorimeter signal is also shaped, amplified and sampled at $40~\textrm{MHz}$ by the front-end electronics.
Unlike the LAr calorimeter, seven rather than four samples are digitised and the cell energy and time are reconstructed via an optimal filtering method as described by Eq.~(\ref{eq:optFilt}).
If the signal amplitude is below 5 ADC counts,\footnote{In a few Tile cells affected by higher radiation (\textit{gap scintillators}~\cite{TCAL-2010-01}), the minimum amplitude for time reconstruction is 15 ADC counts.} then the timing is not measured and a default value of $t=0$ is stored.
Since most Tile calorimeter cells are read out by two independent channels, the average of the two times is taken as the measured cell time~\cite{TCAL-2010-01}.
The Tile calorimeter uses its Laser calibration system~\cite{TILE-LASER-RUN2} to monitor the timing of all channels during physics runs. Laser events are recorded
during empty bunch-crossings in
physics data-taking and the stability of the timing is monitored as part of the data quality assessment, allowing corrections for possible misconfiguration to be made in the main data processing.
Channels that exhibit a large number of timing jumps are flagged as \textit{bad timing} and are not used for further physics object reconstruction~\cite{TILE-LASER-RUN2}. Cells flagged as Tile \textit{bad timing} are not considered when implementing the time cut.
The Tile timing resolution is described with a functional form similar to the one commonly used for the calorimeter energy resolution:
\begin{equation*}
\sigma_t = p_0  \oplus \frac{p_1}{E} \oplus \frac{p_2}{\sqrt{E}}.
\end{equation*}
In Run~2, the constant term was found to be ${\sim}300~\text{ps}$, while $p_1$ and $p_2$ are $\mathcal{O}(1~\text{GeV ns})$ and $\mathcal{O}(1~\sqrt{\GeV} \text{ ns})$, respectively, with variations depending on the gain~\cite{Tile_public_plots}.
 
\section{Data and Monte Carlo samples}
\label{sec:samples}
 
The time cut's performance has been studied on simulated multi-jet event samples as well as \pp collision data from the ATLAS detector.
The data sample used corresponds to one run, taken by ATLAS in 2017 during Run~2 of the LHC at a centre-of-mass energy $\sqrt{s}=13$~\TeV, with an integrated luminosity of $\mathcal{L} = 76 \pm 2~\si{\ipb}$. Because the goal of this study is only to compare jet kinematic distributions, a larger data sample is not needed.
In the selected run the LHC provided collisions from bunch trains with a nominal bunch-crossing spacing of 25~ns. The recorded average mean number of interactions per bunch-crossing $\langle \mu \rangle = 38.8$ is consistent with the average $\langle \mu \rangle$ recorded during the 2017 data-taking period.
Data quality requirements are applied to ensure that all detector components were fully operational~\cite{DAPR-2018-01}.
Data were collected using multiple single-jet triggers (see Section~\ref{sec:dataperf}). Events selected by each trigger are scaled by the corresponding trigger prescale factor. In a few \textit{Luminosity Blocks}\footnote{A Luminosity Block (LB) is a period of time during which the instantaneous luminosity, detector and trigger configuration, and data quality conditions are considered constant. In general, one LB corresponds to a time period of 60~s, although LB duration is flexible and actions that might alter the run configuration or detector conditions trigger the start of a new LB before 60~s have elapsed. LB start and end timestamps are assigned in real time during data-taking by the ATLAS Central Trigger Processor~\cite{DAPR-2018-01}.} at the beginning of the run, the prescale factors for the chosen single-jet triggers were found to be very large, resulting in an apparent loss of statistical precision due to recording few events with very large trigger weights. To avoid this issue, the Luminosity Blocks in question were excluded from the analysis. The loss of integrated luminosity due to this additional rejection amounts to $6.5\%$. The integrated luminosity after data quality requirements are applied is $\mathcal{L} = 71 \pm 2~\si{\ipb}$. The uncertainty in the 2017 integrated luminosity is assumed for this run; it amounts to $2.4 \%$~\cite{DAPR-2021-01}, obtained using the LUCID-2 detector~\cite{LUCID2} for the primary luminosity measurements.
 
Additional data samples were used to check for inefficiencies caused by time miscalibration and to test the ATLAS event size for reductions due to the time cut. For time-cut inefficiency checks, one Luminosity Block of data collected in 2018 was used, with an average number of interactions per bunch-crossing of $\langle \mu \rangle = 42.5$.
The data used for the event size test also consists of one Luminosity Block, taken in 2022. The average \pileup for this run is $\langle \mu \rangle = 41.3$.
 
Multi-jet production was modelled using \PYTHIA[8.230]~\cite{Sjostrand:2014zea} with leading-order matrix elements
for dijet production which were matched to the parton shower.
The renormalisation and factorisation scales were set to the geometric
mean of the squared transverse masses of the two outgoing particles in the matrix element,
\(\sqrt{(\pTX[2][1] + m_1^2) (\pTX[2][2] + m_2^2)}\).
The \NNPDF[2.3lo] set of parton distribution functions (PDFs)~\cite{Ball:2012cx} was used in
the matrix element generation, the parton shower, and the simulation of the
multi-parton interactions. The A14~\cite{ATL-PHYS-PUB-2014-021}
set of tuned parameters was used.
The modelling of fragmentation and
hadronisation was based on the Lund string
model~\cite{Andersson:1983ia,Sjostrand:1984ic}.
The effect of both in-time and out-of-time \pileup was modelled by overlaying the simulated hard-scattering event with
inelastic proton--proton (\(pp\)) events generated with \PYTHIA[8.186]~\cite{Sjostrand:2007gs}
using the \NNPDF[2.3lo] PDF set~\cite{Ball:2012cx} and the
A3 set of tuned parameters~\cite{ATL-PHYS-PUB-2016-017}.
 
The MC events have been reweighted to account for the difference between the simulated number of interactions per bunch-crossing ($\mu$) profile and the one from the single run used for this \mypaper. The correction factor is defined in bins of $\mu$, as the ratio of the data to MC sample $\mu$ spectra, normalised to unity and prior to any event selection. This reweighting is not applied when performing MC-only studies, to allow a wider $\mu$ range to be covered.
 
An additional MC sample was used to assess the ATLAS event size reduction due to the time cut.  This MC sample describes $t\overline{t}$ production in the fully hadronic decay channel. The production of \ttbar events was modelled using the
\POWHEGBOX[v2]~\cite{Frixione:2007nw,Nason:2004rx,Frixione:2007vw,Alioli:2010xd}
generator at NLO with the \NNPDF[3.0nlo]~\cite{Ball:2014uwa} PDF set
and the \hdamp parameter
set
to 1.5\,\mtop~\cite{ATL-PHYS-PUB-2016-020}.  The events were interfaced
to \PYTHIA[8.307]~\cite{Sjostrand:2014zea} to model the parton shower,
hadronisation, and underlying event, with parameters set according
to the A14 tune~\cite{ATL-PHYS-PUB-2014-021} and using the \NNPDF[2.3lo]
set of PDFs~\cite{Ball:2012cx}. The decays of bottom and charm hadrons
were performed by \EVTGEN[2.1.1]~\cite{Lange:2001uf}.

\section{The topo-cluster reconstruction algorithm}
\label{sec:topoclusteringalgo}
The cell time cut is introduced into the existing ATLAS topo-cluster reconstruction algorithm, a detailed description is provided in Ref.~\cite{PERF-2014-07}. In short, ATLAS topo-clustering is based on the cell signal significance $\cellsig$, defined as the ratio of the cell signal energy to the average expected noise (Eq.~(\ref{eq:cellsig})). Both are measured at the electromagnetic (EM) scale, defined by calibrating the calorimeter energy measurement to the energy deposited by an electron or photon, without any correction for the non-compensating response of the calorimeter.
\begin{equation}
\cellsig= \frac{E_{\textrm{cell}}^{\textrm{EM}}}{\sigma_{\textrm{noise,cell}}^{\textrm{EM}}}.
\label{eq:cellsig}
\end{equation}
Topo-clusters are formed by a growing-volume algorithm, configured by three threshold parameters $\{S, \ N, \ P\}$:
\begin{eqnarray}
& |\cellsig|>S & (\textrm{primary seed threshold}) \label{eq:seedcells},\\
& |\cellsig|>N & (\textrm{threshold for growth control}) \label{eq:neighcell},\\
& |\cellsig|>P & (\textrm{principal cell filter}) \label{eq:allcell};
\end{eqnarray}
set to $\{S=4, \ N=2, \ P=0\}$ in Runs 1, 2 and 3.
The algorithm is seeded by cells (\textit{seed} cells in the following) whose signal significance exceeds a threshold $S$ (Eq.~(\ref{eq:seedcells})) and which have not been marked as having either read-out or general signal extraction problems in the actual run conditions. Seed cells are then sorted from highest to lowest energy significance and topo-clusters are grown by adding all neighbouring cells that satisfy Eq.~(\ref{eq:allcell}).
 
Two cells are considered to be neighbours if they are directly adjacent in a given sampling layer, or, if in adjacent layers, if they at least partially overlap in the plane. If a neighbouring cell has a signal significance above the threshold $N$ in Eq.~(\ref{eq:neighcell}), then the procedure is iterated over its neighbours, while if it satisfies Eq.~(\ref{eq:allcell}) but not Eq.~(\ref{eq:neighcell}), its neighbours are not included in the growing topo-cluster.
This algorithm iterates over further neighbours until no more neighbouring cells satisfying Eq.~(\ref{eq:neighcell}) are found.
Since the $S$, $N$ and $P$ thresholds are applied to the absolute significance, cells of either positive or negative energy can seed a cluster or be included in one. As discussed in Section~\ref{sec:calotiming}, negative cell signals in the ATLAS calorimeters are often the result of out-of-time \pileup, since the residual signal trace produced by out-of-time particles is scaled by the negative tail of the LAr calorimeter signal-shaping function~\cite{NJBuchanan_2008}. Only clusters with an overall positive energy are used as inputs for jet reconstruction.
The algorithm described so far, however, has no limitations to how large a cluster can grow and it could potentially lead to very large, unphysical clusters. For this reason, clusters are then split around local energy maxima, as detailed in Ref.~\cite{PERF-2014-07}.

\section{The cell time cut}
\label{sec:thecut}
\begin{figure}
\centering
\subfigure[]{\label{fig:cellplotLArCentral2D}
\includegraphics[width=0.56\linewidth]{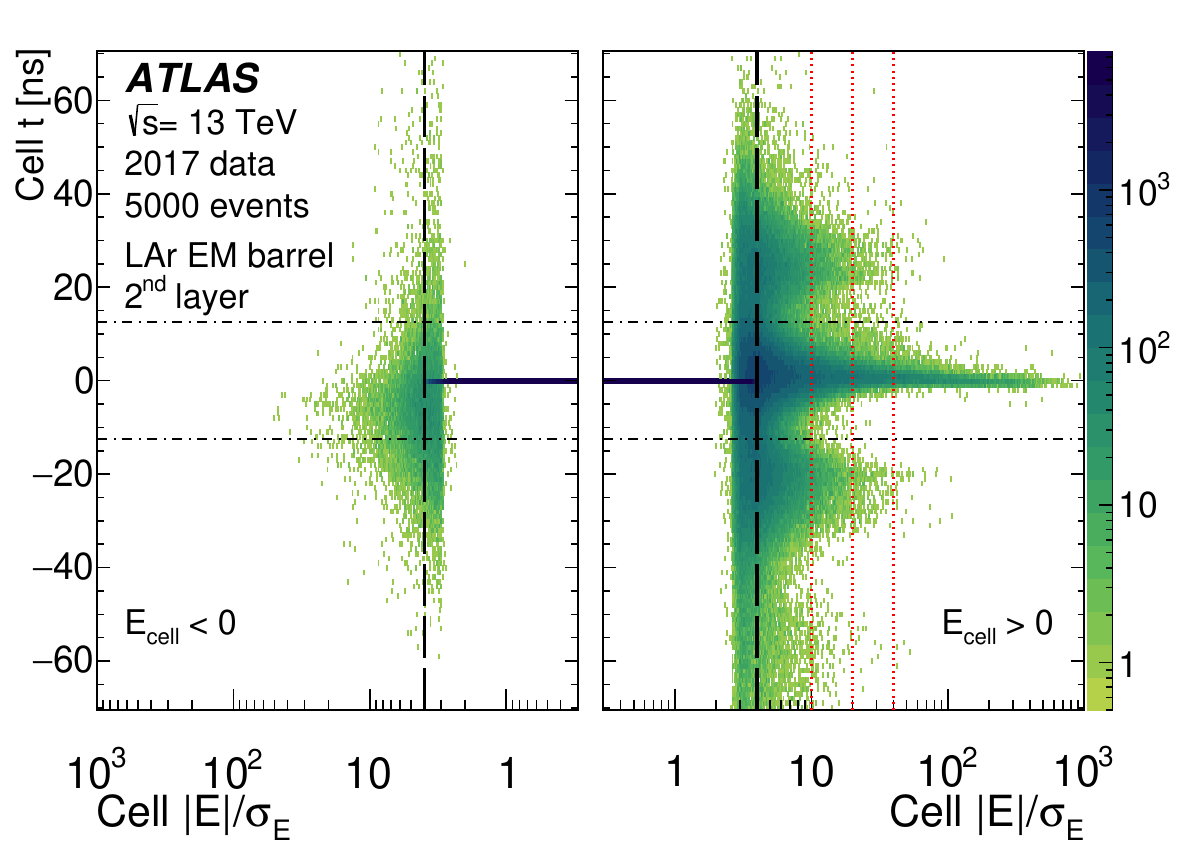}}
\raisebox{0.02\height}{
\subfigure[]{\label{fig:cellplotLArCentral1D}
\includegraphics[width=0.41\linewidth]{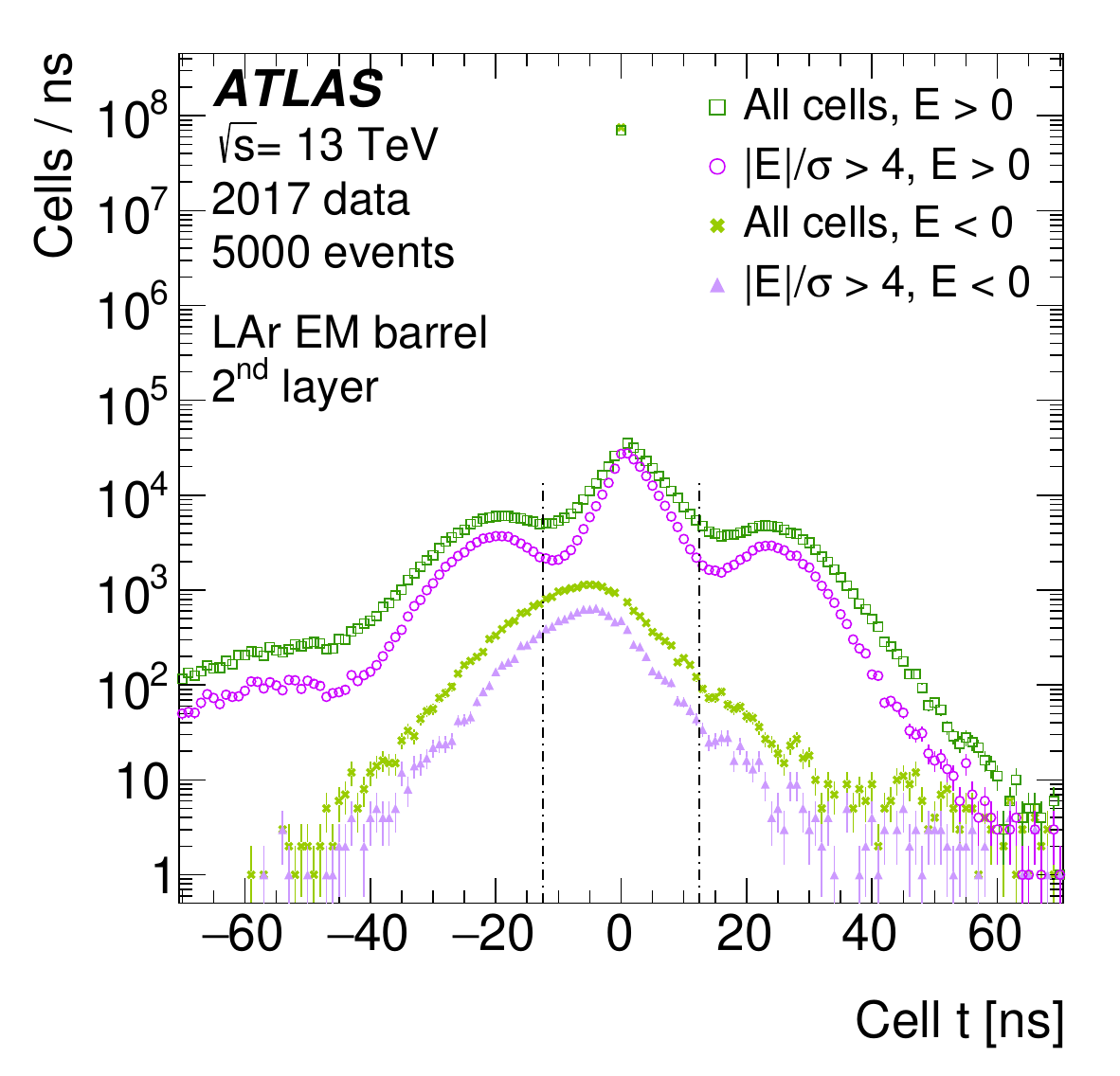}}}
\caption{Time and energy significance for calorimeter cells in the second layer of the LAr EM barrel calorimeter. \subref{fig:cellplotLArCentral2D} shows the two-dimensional cell time vs significance spectrum. The cell time is computed if the cell energy exceeds a given threshold (typically $3\cdot \sigma_{\textrm{noise,cell}}^{\textrm{EM}}$), below which a default value of $t=0$ is stored. The dashed vertical line represents the seed candidate requirement $S=4$ (Eq.~(\ref{eq:seedcells})), while the horizontal lines represent the cell time rejection limits of $\pm 12.5~\textrm{ns}$. The three dotted vertical lines represent the three possible values considered for the Upper Limit (see Section~\ref{sec:UL}). \subref{fig:cellplotLArCentral1D} shows a comparison between the inclusive cell time spectrum and the one for seed candidates ($|E|/\sigma_E>4$). The vertical lines represent the cell time rejection limits of $\pm 12.5~\textrm{ns}$.}
\label{fig:cellplotLArCentral}
\end{figure}

As discussed in Section~\ref{sec:calotiming}, both the ATLAS LAr and Tile calorimeters provide a time measurement for signals of sufficiently large energy. Figure~\ref{fig:cellplotLArCentral} shows the distribution of the energy significance and time of calorimeter signals in real data. Calorimeter cells belonging to the LAr EM barrel region are picked as an example, but similar distributions occur in other calorimeter regions.
At high energy significance, two secondary peaks at $\pm 25$ ns are clearly visible, associated with out-of-time \pileup from the previous and next bunch-crossings. The asymmetry visible between positive and negative cell time is due to the LAr calorimeter being affected by a larger number of bunch-crossings before the current one than after it.
 
At lower energy significance the time resolution is poorer, and the secondary peaks cannot in general be distinguished. When limiting the study to candidate \textit{seed} cells, i.e.\ those with energy significance greater than $S=4$,  the three peaks are separated well enough to justify a selection on the absolute cell time, requiring it to be within $\pm 12.5$~ns of the midpoint between the secondary peaks.
Multiple approaches to implementing a cell time cut in topo-cluster building were explored. The most relevant are detailed below.
 
\subsubsection{The \textit{Seed} time cut}
The requirement for seeding a topo-cluster is modified. Seed cells must satisfy Eq.~(\ref{eq:tcSeed}), i.e.\ have both an absolute cell signal significance larger than $S=4$ and a cell time within $\pm 12.5$~ns:
\begin{equation}
\left\{ \begin{array}{l}
|\cellsig|>S \quad (\textrm{primary seed threshold, default }S = 4)
\\
|t_{\textrm{cell}}| < 12.5 \textrm{ ns}.
\end{array} \right.
\label{eq:tcSeed}
\end{equation}
The cluster growth and splitting stages of the topo-clustering algorithm are left untouched.
Since this selection is exclusively applied to topo-cluster seeds, it is referred to as the \textit{Seed cut} in the following.
 
\subsubsection{The \textit{Seed Extended} time cut}
In the Seed cut implementation, a cell exceeding the signal significance threshold in Eq.~(\ref{eq:tcSeed}), but failing the time cut, would be prevented from seeding a cluster. It could, however, be included in another cluster as a neighbouring cell if it falls in the vicinity of an in-time signal. Due to the high level of \pileup at the LHC, this phenomenon is not unlikely.
To prevent their inclusion, a tighter version of the cut was defined, in which candidate seed cells that satisfy Eq.~(\ref{eq:seedcells}), but not Eq.~(\ref{eq:tcSeed}), are vetoed from being included in growing topo-clusters. In the following, this version of the time cut is referred to as the \textit{Seed Extended} cut.
 
\subsubsection{The Upper Limit}
\label{sec:UL}
Although this \mypaper \ focuses on the reduction of out-of-time \pileup, delayed calorimeter signals with substantial energy can also arise from new physics, for example from the decays of slow-moving long-lived particles (LLPs) that are predicted by several theories for physics beyond the Standard Model~\cite{Alimena:2019zri, Lee:2018pag}.
Optimising the time cut for LLP signals would require taking into account the many signatures considered for LLP searches and is outside the scope of this \mypaper. A conservative approach was therefore adopted, where the time cut is not applied to signals with very large significance, thus leaving the decision of whether or not to reject such signals to a later stage of the reconstruction chain, where differentiation between standard and LLP-dedicated selections is possible.
 
The time cut was therefore modified by introducing an Upper Limit (UL). The time cut is turned off if the cell signal significance $\cellsig = E_{\textrm{cell}}^{\textrm{EM}}/\sigma_{\textrm{noise,cell}}^{\textrm{EM}}$ is positive and larger than a given value $X_{\textrm{{UL}}}$, i.e.\ if:
\begin{equation*}
\left\{ \begin{array}{l}
|\cellsig|>S \quad (\textrm{primary seed threshold, default }S = 4)
\\
|t_{\textrm{cell}}| < 12.5 \textrm{ ns}
\end{array} \right. \quad \textrm{OR} \quad \cellsig>X_{\textrm{UL}}.
\end{equation*}
A lower value of $X_{\textrm{UL}}$ implies a larger phase space left untouched by the time cut. Three values $X_{\textrm{UL}}=10$, 20 and 40 were tested in order to estimate how low the Upper Limit can be set without reintroducing too much \pileup into the event. While the Upper Limit could in principle be combined with either the Seed cut or the Seed Extended cut, in this \mypaper \ it is only studied in combination with the Seed Extended cut, as the latter appeared more promising than the Seed cut (Section~\ref{sec:mcperf}).
The Upper Limits $X_{\textrm{UL}}=10$, 20 and 40 are shown as dotted lines in Figure~\ref{fig:cellplotLArCentral}. As expected, $X_{\textrm{UL}}=40$ affects a very limited number of out-of-time cells and it is expected to have a negligible effect, while $X_{\textrm{UL}}=10$ is expected to have the largest impact.

% End of text imported from the .//IntroductionAndCut_pap.tex input file

\FloatBarrier

% The next lines are included from the .//MCperformance_pap.tex input file
\section{Performance of the time cut on MC simulation}
\label{sec:mcperf}
 
The performance of the time cut was evaluated using a \PYTHIA[8.230] dijet MC sample, as described in Section~\ref{sec:samples}. The event selection is kept as loose as possible in order to observe ample ranges of cluster energy and jet energy.
 
\subsection{Performance on topo-clusters}
\begin{figure}
\centering
\includegraphics[width=0.52\linewidth]{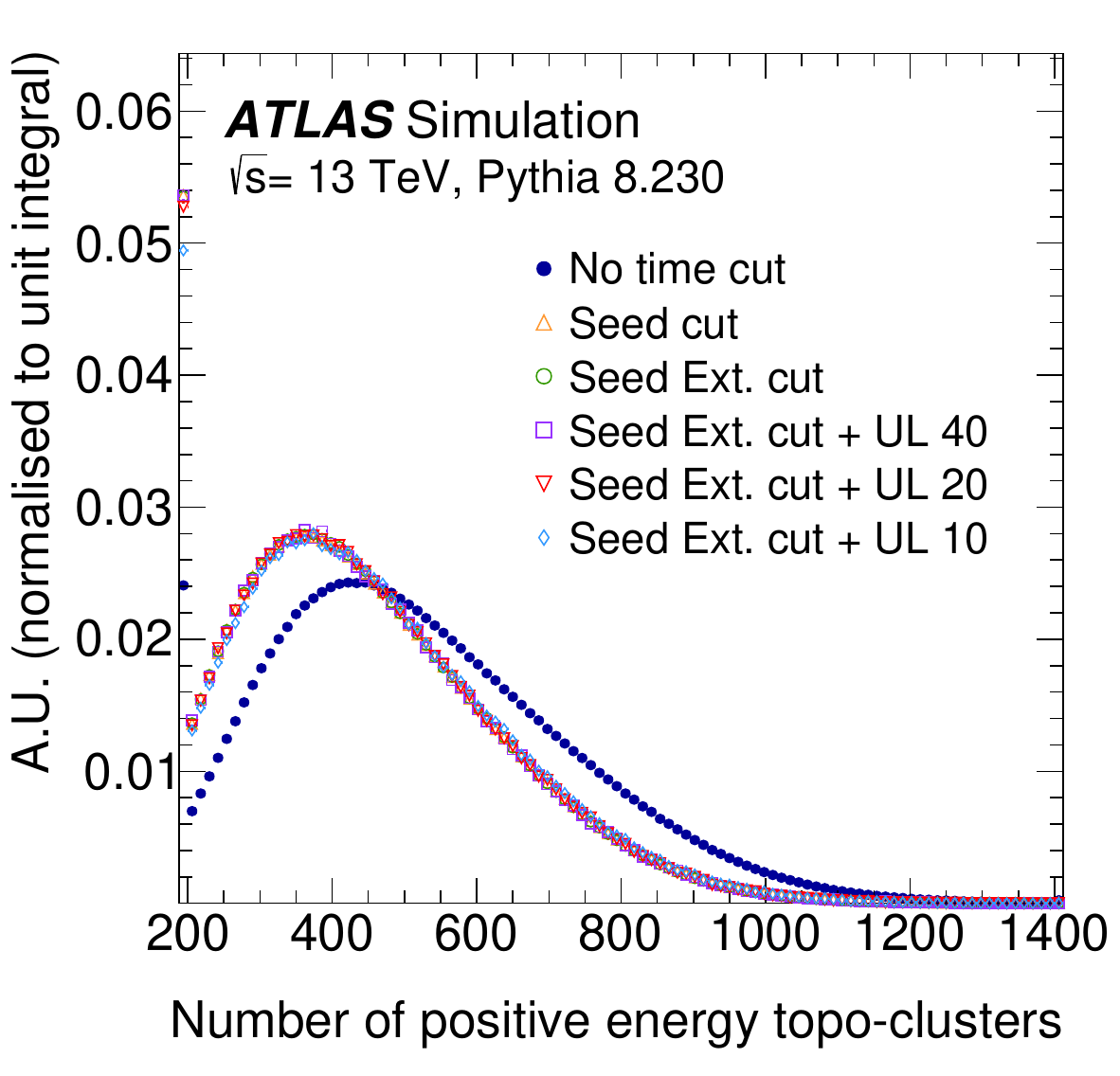}
\caption{Number of positive energy ($E_{\textrm{cl}}>0$) reconstructed topo-clusters per event: comparison between no time cut and multiple choices for the time cut. The first and last bins show the overflow and underflow.}
\label{JZx_clN}
\end{figure}

Since the time cut is applied in the topo-clustering algorithm, the overall number of clusters is expected to be reduced, because the applied cut leads to a smaller number of cluster seeds. It should be noted, however, that the topo-clusters built with the time cut applied are not necessarily a subsample of those built using only the cell energy significance. Excluding out-of-time signals may change the number of local maxima, prompting additional cluster splitting.
Figure~\ref{JZx_clN} shows the number of clusters reconstructed per event; only positive-energy topo-clusters (i.e.\ eligible inputs to the jet reconstruction) are shown.
The number of positive-energy clusters in each event is on average ${\sim}15\%$ smaller when the time cut is applied, with very little difference between the different choices for the cut. The Seed and Seed Extended cuts differ only in their treatment of out-of-time signals with significant energy when these signals are included as neighbouring cells, and hence the two algorithms can be expected to differ more strongly in their effect on topo-cluster properties than on the number of reconstructed topo-clusters. Since the UL only truncates the relatively sparsely populated high-energy tail of the topo-cluster spectrum, its effect then becomes subdominant when considering the inclusive number of clusters.
 
\begin{figure}
\centering
\subfigure[]{\label{subfig:JZx_clusT}
\includegraphics[width=0.49\linewidth]{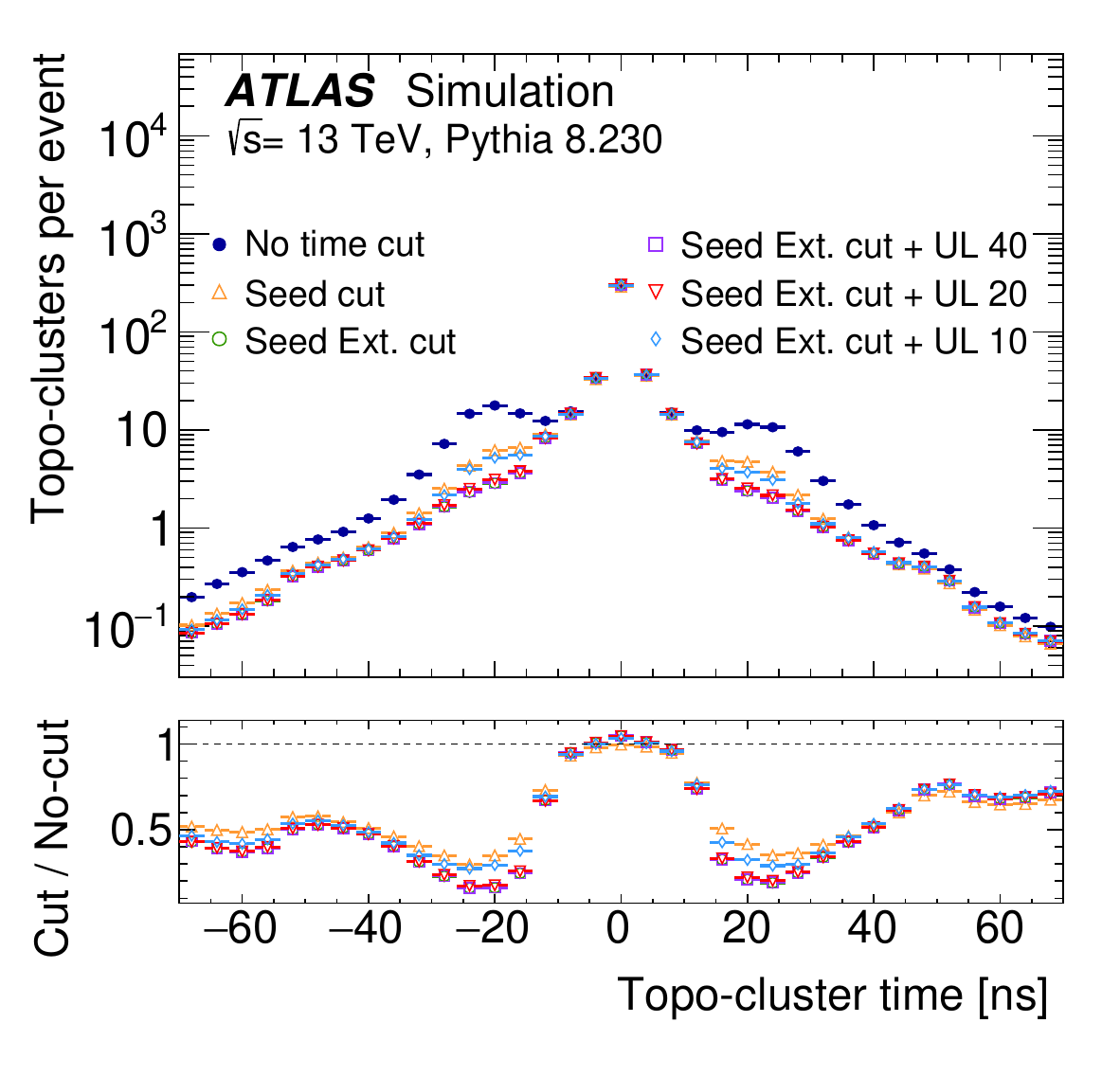}}
\subfigure[]{\label{subfig:JZx_clusRapidity}
\includegraphics[width=0.49\linewidth]{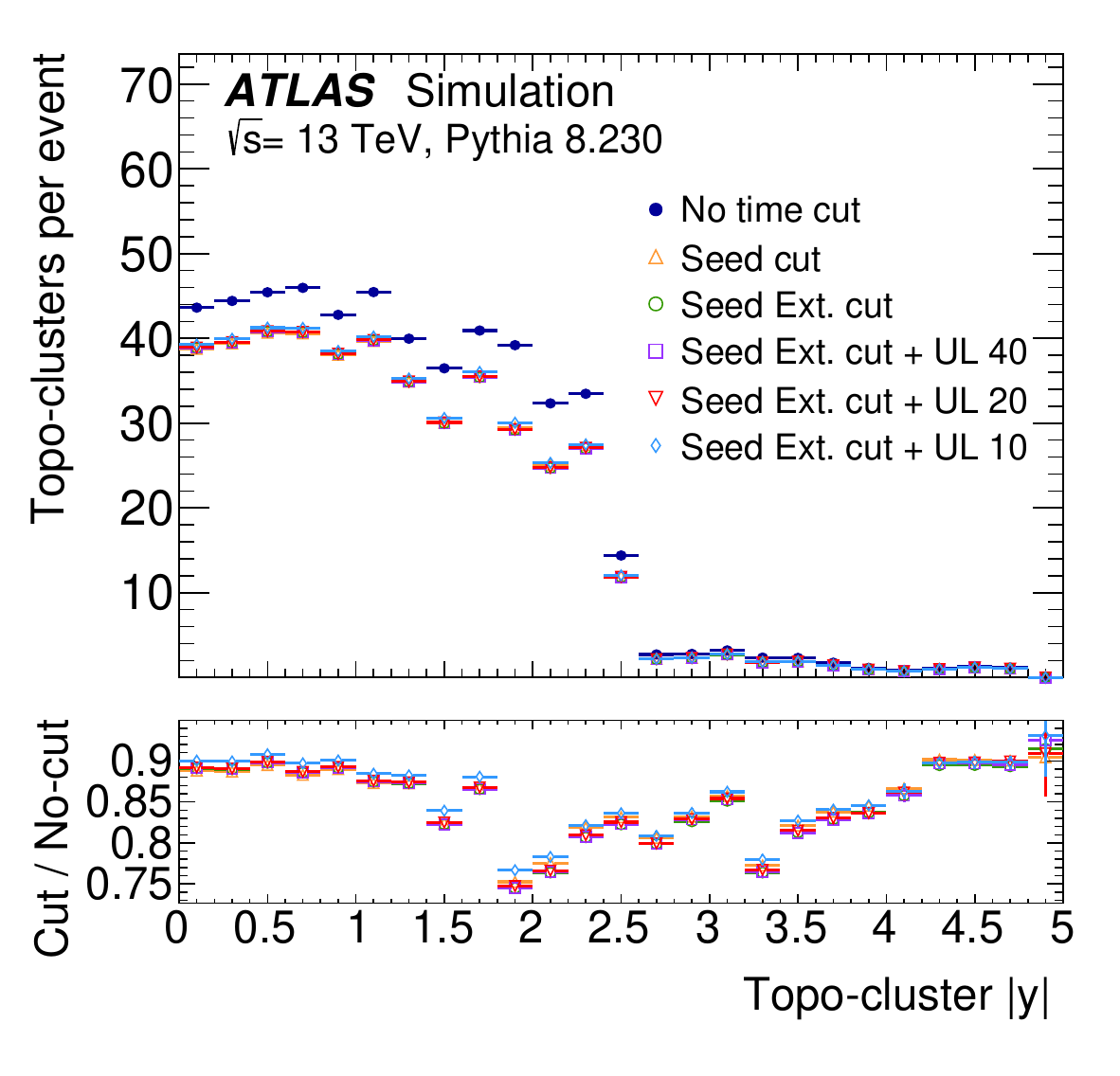}}
\subfigure[]{\label{subfig:JZx_clusE}
\includegraphics[width=0.49\linewidth]{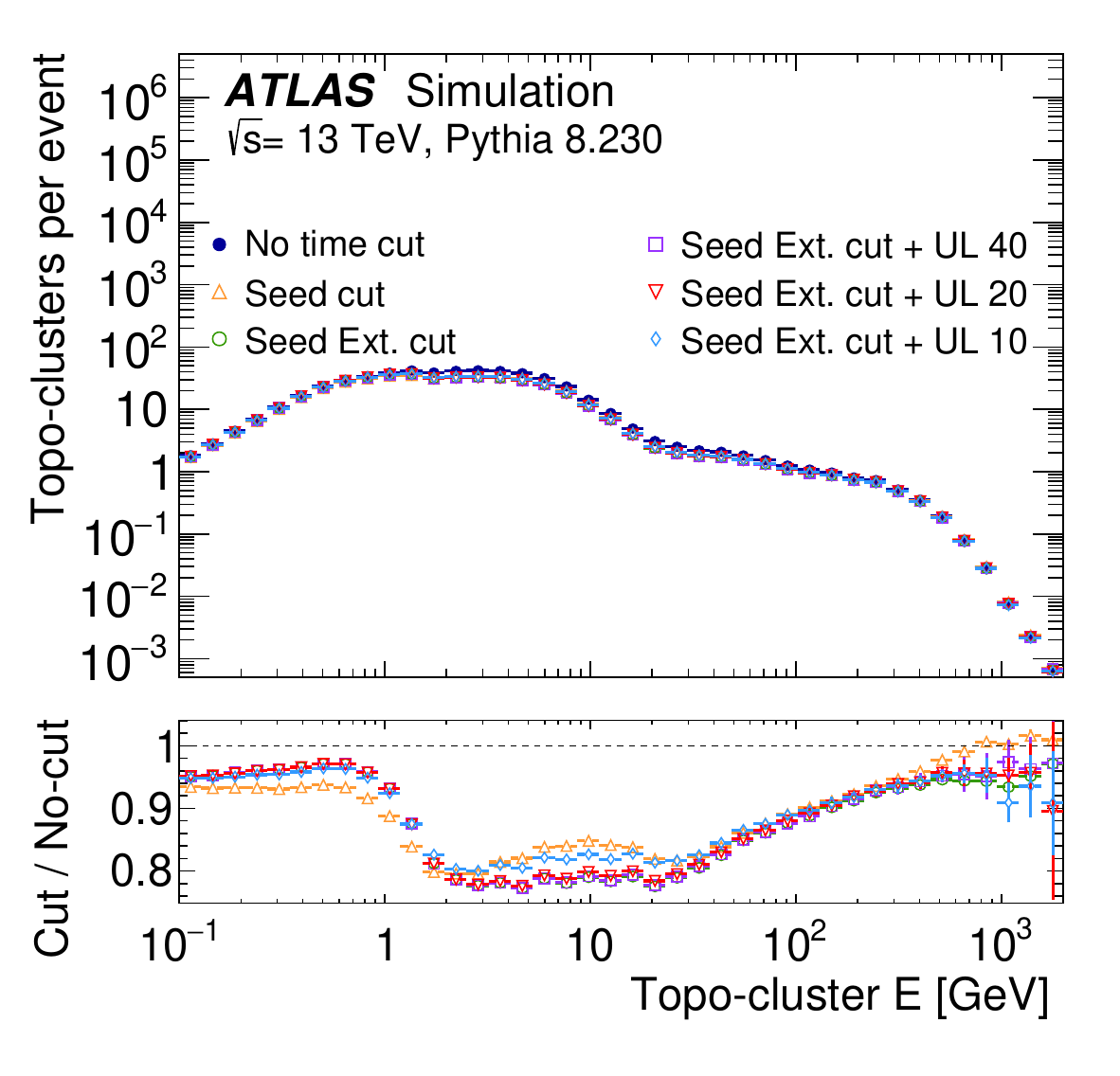}}
\subfigure[]{\label{subfig:JZx_clusPt}
\includegraphics[width=0.49\linewidth]{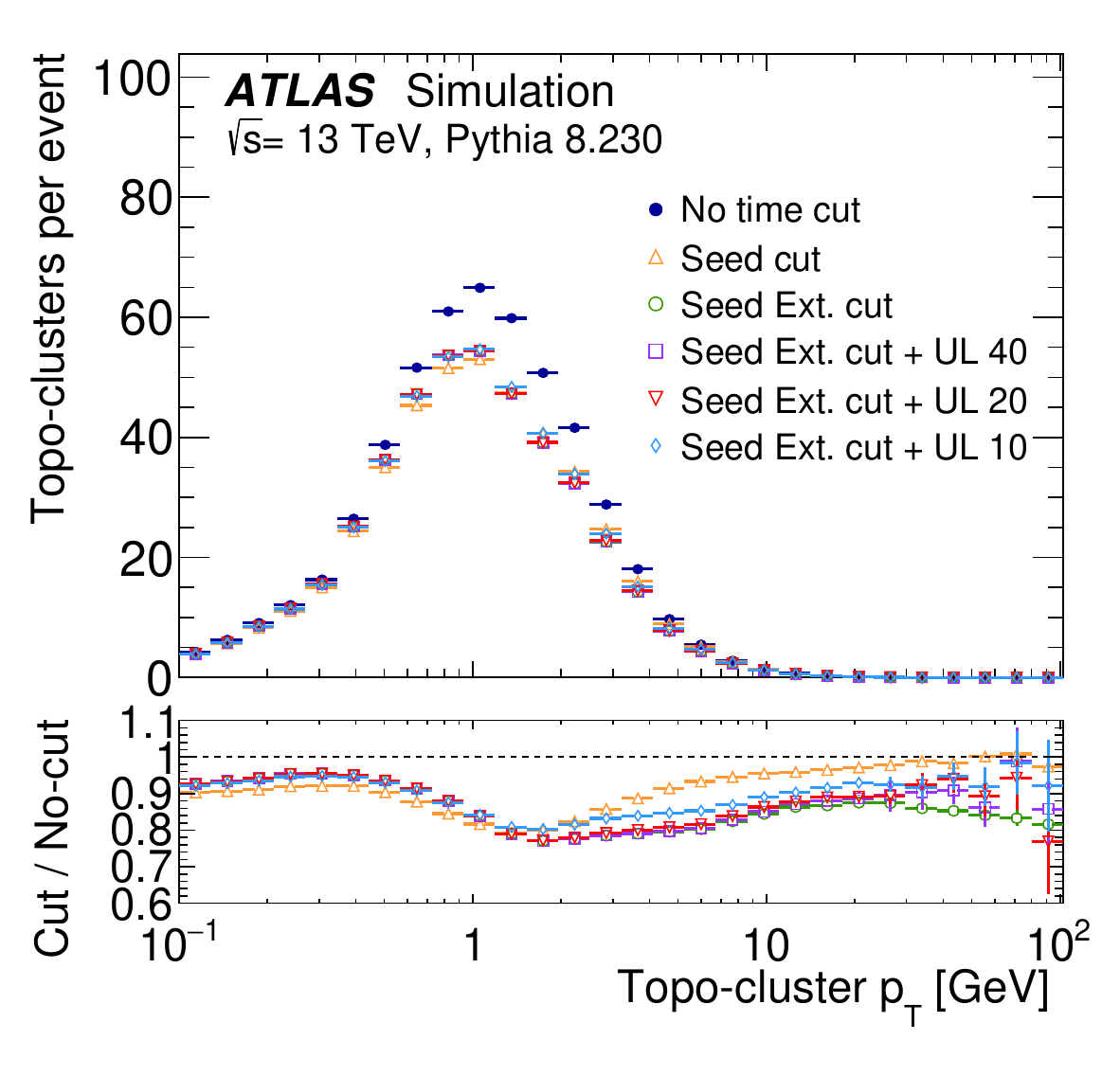}}
\caption{Kinematics of reconstructed topo-clusters, compared between no time cut, the Seed cut, the Seed Extended cut and the Seed Extended cut combined with multiple choices of Upper Limit: $X_{\textrm{UL}}=40$, $X_{\textrm{UL}}=20$ and $X_{\textrm{UL}}=10$.
Each kinematic moment of the topo-clusters is defined as the weighted average of the property of interest over the cells contributing to the cluster~\cite{PERF-2014-07}. Topo-clusters with energy $E_{\textrm{cl}}>0$ are shown. Plots are normalised to the total number of positive-energy topo-clusters per event. Uncertainties in the cut~/~no-cut ratios are computed by splitting the available sample into subsamples and recomputing the cut~/~no-cut ratio for each subsample. The standard deviation of the distribution of the ratio is used to estimate the ratio's uncertainty.}
\label{fig:JZx_clKine_UL}
\end{figure}

Removing out-of-time contributions can also result in a change in the kinematic properties of the reconstructed topo-clusters. This effect is expected to be more significant when applying the Seed Extended time cut as opposed to the Seed time cut.
Figure~\ref{fig:JZx_clKine_UL} shows the multiplicity of topo-clusters that have various kinematic properties, comparing the various time cut options.
Each kinematic moment of the topo-clusters is defined as the weighted average of the property of interest over the cells contributing to the cluster. A full description of the topo-clusters' properties is given in Ref.~\cite{PERF-2014-07}. An additional calibration using the local cell weighting (LCW) scheme is applied\footnote{The LCW scheme is not propagated to the jets, in order to mimic the approach most commonly followed by ATLAS physics analyses.} to take into account the non-compensating response of the calorimeter, out-of-cluster energy deposits and energy deposited in the dead material within the detector~\cite{PERF-2014-07}. Only the topo-clusters with overall positive energy in each event are shown. Bottom panels show the \enquote{cut~/~no-cut} ratio plot.
Uncertainties in the ratios must take into account that the distributions with and without the time cut are obtained from the same events, and hence are correlated in a non-trivial way. To properly account for the correlations, uncertainties are computed by splitting the available sample into subsamples of approximately 10\,000 events each, and the cut~/~no-cut ratio is recomputed for each subsample. The ratio's uncertainty is estimated from the standard deviation of the resulting distribution.
The estimated uncertainty is propagated to the inclusive ratio as a relative uncertainty. This method is used to compute the uncertainties in the cut~/~no-cut ratio plots throughout this \mypaper.
 
Figure~\ref{subfig:JZx_clusT} shows the cluster time. When no time cut is applied, two shoulders are clearly visible near $\pm25$~ns, consistent with being due to out-of-time \pileup from the previous and next bunch-crossings. Once the time cut is applied, the number of clusters around $\pm 25~\text{ns}$ is reduced by more than $80\%$ by the Seed Extended time cut, smoothing out the two shoulders, while the main peak at 0~ns appears unchanged. The Seed cut has a smaller impact, reducing the two peaks by ${\sim}70\%$, while the impact of the Seed Extended plus Upper Limit $X_{\textrm{UL}}=10$ cut lies between those of the Seed and Seed Extended cuts.
Overall, topo-clusters with absolute time greater than 12.5~ns are not removed entirely, since out-of-time contributions can still occur due to predominantly low-energy cells being included in topo-clusters as neighbouring cells.
 
The relative effect of the time cut on the number of clusters is also found to vary with jet rapidity (Figure~\ref{subfig:JZx_clusRapidity}), the largest effect being a ${\sim}25\%$ reduction in the cluster multiplicity for $2 \lesssim |y| \lesssim3.5$. The energy deposited by \pileup is expected to increase at larger absolute rapidity, but the time cut has only a ${\sim}10\%$ effect above $|y|\sim3.5$ because each forward calorimeter cell covers a larger pseudorapidity interval. This geometrical effect implies that larger numbers of both in-time particles and out-of-time particles deposit energy in the same cell, thus resulting in a less effective distinction between in-time and out-of-time signals.
The effect of the time cut is also seen to vary with the cluster energy (Figure~\ref{subfig:JZx_clusE}). The number of clusters with energy between about 1~\GeV\ and 100~\GeV\ is reduced by ${\sim}25\%$, with the reduction decreasing to ${\sim}10\%$ at both very low and very high energies.
Finally, the dependence on the cluster \pT (Figure~\ref{subfig:JZx_clusPt}) provides the best way to observe the effect of the Upper Limit. As expected, $X_{\textrm{UL}}=40$ has hardly any effect, while $X_{\textrm{UL}}=20$ produces a small difference and it starts to diverge from the Seed Extended cut for clusters with $p_{\textrm{T}}\gtrsim 15$~\GeV. However, $X_{\textrm{UL}}=10$ has a larger impact, which starts to appear for clusters with $p_{\textrm{T}}\gtrsim2$~\GeV. At $\pT=5~\GeV$ for instance, the Seed Extended cut reduces the number of clusters by ${\sim}20\%$, while the Seed Extended plus $X_{\textrm{UL}}=10$ cut reduces their number by only ${\sim}15\%$.
 
\begin{figure}
\centering
\subfigure[]{\label{subfig:JZx_clusIsol}
\includegraphics[width=0.49\linewidth]{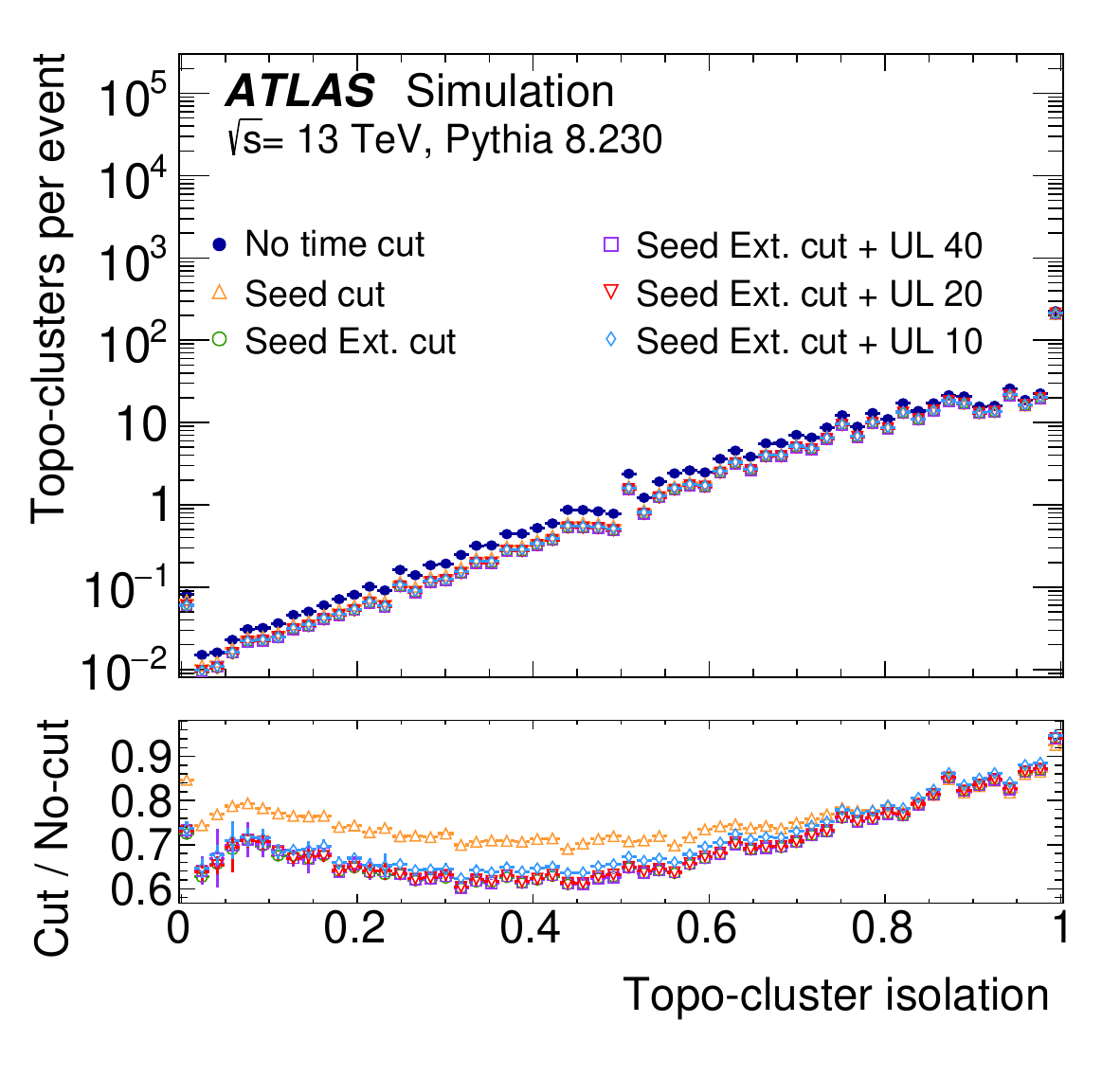}}
\subfigure[]{\label{subfig:JZx_clusClambda}
\includegraphics[width=0.49\linewidth]{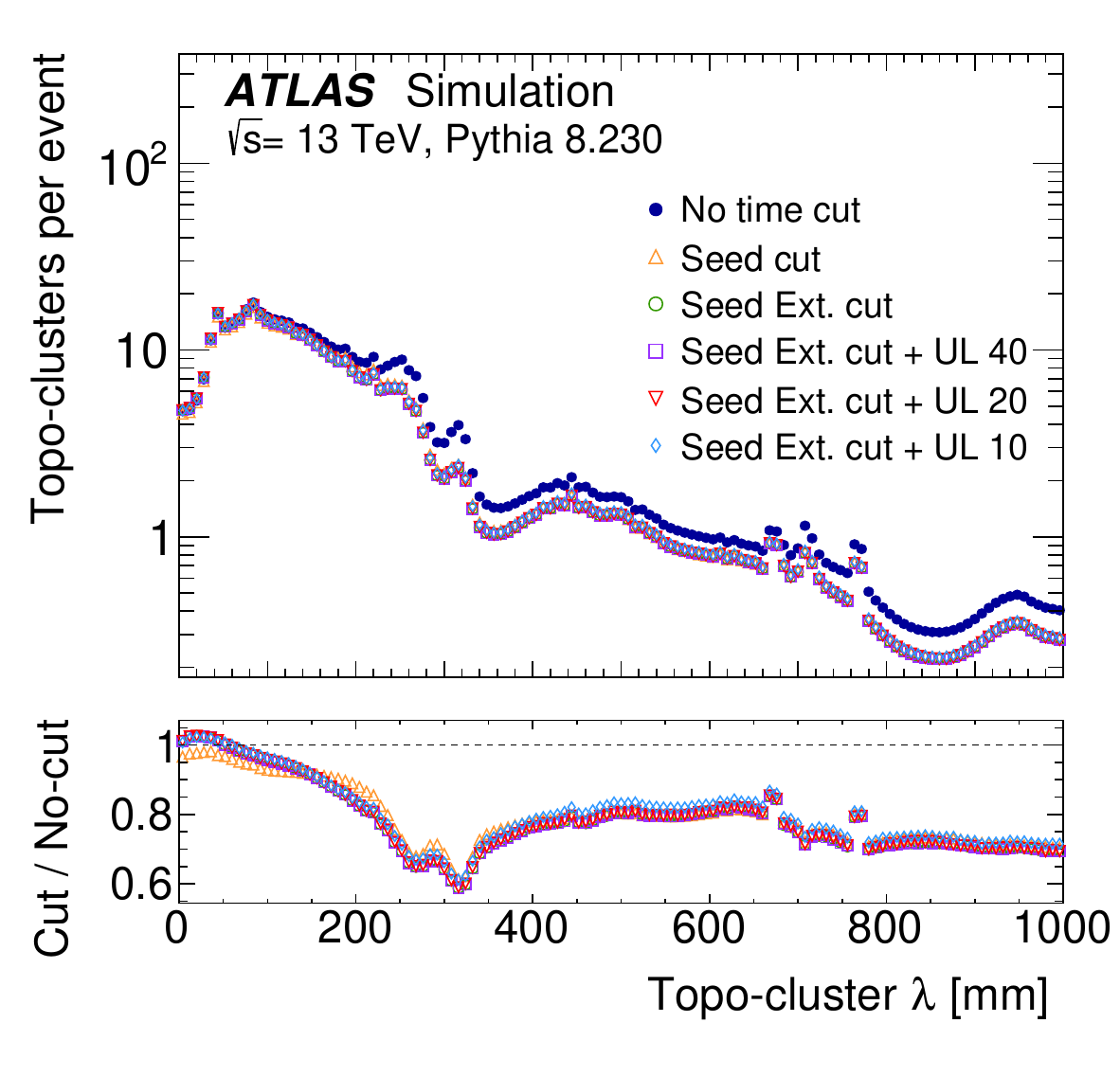}}
\caption{Topo-cluster \subref{subfig:JZx_clusIsol} isolation and \subref{subfig:JZx_clusClambda} distance $\lambda$ from the calorimeter front face. A comparison between no time cut, the Seed cut, the Seed Extended cut and the Seed Extended cut combined with $X_{\textrm{UL}}=40$, $X_{\textrm{UL}}=20$ and $X_{\textrm{UL}}=10$ is shown.
Each kinematic moment of the topo-clusters is defined as the weighted average of the property of interest over the cells contributing to the cluster~\cite{PERF-2014-07}. Topo-clusters with energy $E_{\textrm{cl}}>0$ are shown. Plots are normalised to the total number of positive-energy topo-clusters per event. Uncertainties in the cut~/~no-cut ratios are computed by splitting the available sample into subsamples and recomputing the cut~/~no-cut ratio for each subsample. The standard deviation of the distribution of the ratio is used to estimate the ratio's uncertainty.}
\label{fig:JZx_clKineMoments_UL}
\end{figure}

Figure~\ref{fig:JZx_clKineMoments_UL} shows two topo-cluster moments.
The topo-cluster isolation (Figure~\ref{subfig:JZx_clusIsol}) is computed from the number of non-clustered cells on the outer perimeter of the topo-cluster~\cite{PERF-2014-07}: an isolation value close to 1 indicates an isolated topo-cluster, while it tends to 0 when the topo-cluster is not isolated. Since the lower values of the isolation spectrum are affected the most by the time cut, it can be concluded that topo-clusters reconstructed with the time cut tend to be more isolated than those from the standard algorithm. This observation can be explained as a combination of two effects. Firstly, isolated topo-clusters are mostly produced by EM showers, and are less sensitive to \pileup because of their smaller size. Secondly, the \pileup reduction introduced by the timing cut is likely to produce more isolated topo-clusters. Specific studies are needed to further understand the interplay between these two effects.
The distance $\lambda$ of the cluster's centre of gravity from the calorimeter's front face~\cite{PERF-2014-07} is shown in Figure~\ref{subfig:JZx_clusClambda}. The region of $\lambda \lesssim 100$~mm is known to be dominated by EM clusters, while clusters located at $\lambda \gtrsim 400$~mm are predominantly hadronic~\cite{PERF-2014-07}. In Figure~\ref{subfig:JZx_clusClambda} very little difference is visible in the region dominated by EM clusters while a relatively uniform reduction of about 20\% is visible for predominantly hadronic clusters. The area between $\lambda \sim 100$~mm and $\lambda \sim 400$~mm corresponds to a mixture of EM and hadronic clusters. Here, peaks are visible in the spectrum of standard clusters, while they are almost entirely removed by the time cut. This effect hints that the timing cut may increase the separation between hadronic and EM showers in the presence of \pileup.

\subsection{Performance on jets}
Topo-clusters serve as inputs to the reconstruction of many physics objects in ATLAS. This \mypaper \ focuses on the effect of the time cut on jets reconstructed by applying the \antikt clustering algorithm with radius parameter $R=0.4$~\cite{Cacciari:2008gp}.
In ATLAS, jets are most often reconstructed from \textit{particle-flow} objects~\cite{PERF-2015-09, JETM-2018-05}, which are built by combining topo-clusters with matching inner-detector tracks. Despite it being the preferred method in ATLAS, particle flow adds a layer of complexity to the jet reconstruction: jets built directly from topo-clusters provide a more straightforward way of evaluating the time cut's performance and are hence used in this \mypaper.
Preliminary studies show that the effect of the time cut on $R=0.4$ particle-flow jets is about the same size as the effect on those built directly from topo-clusters. Since the ATLAS tracking is performed within the triggered bunch crossing, it is reasonable to expect that most out-of-time calorimeter signals would not match tracks, so the time cut can be expected to primarily affect objects classified as neutral by the particle-flow algorithm, which in turn consist only of topo-clusters, with no track-based correction.
 
\subsubsection{Jet calibration}
\label{sec:jetCalib}
Jets are calibrated to correct for both the \pileup contribution and the detector response. A complete description of jet calibration in ATLAS can be found in Ref.~\cite{JETM-2018-05}.
The calibration procedure consists of three stages. First, a two-step \pileup correction is made. Then a MC-based calibration factor is applied to correct for the detector response (MC jet energy scale: \textit{MC-JES}) and improve the jet resolution by reducing the calibration's dependence on additional jet properties (\textit{Global Sequential Calibration}). Finally, an \textit{in situ} correction is applied to account for discrepancies between data and MC events.
 
The first stage of the calibration procedure starts with an \textit{area-based} correction in which the average \pileup contribution is subtracted from the jet's transverse momentum:
\begin{equation*}
\pT^{\text{area}} = \pT - \rho \times A,
\end{equation*}
where \pT is the jet \pT, $A$ is the jet catchment area and $\rho$ is the median \pT density of the event. The value of $\rho$ is calculated using jets reconstructed by applying the \kt algorithm~\cite{PhysRevD.48.3160, CATANI1993187} to positive-energy topo-clusters with $|\eta|<2$.
A residual \pileup correction is then applied to account for residual dependence of the jet \pT on \pileup activity in the event. Two $\eta$-dependent correction terms are derived independently by fitting the dependence of the jet \pT on $\mu$ and the number of reconstructed primary vertices per event, $N_{\text{PV}}$.
In general, $\mu$ provides an estimate of the amount of out-of-time \pileup in the event, while $N_{\text{PV}}$ estimates the amount of in-time \pileup.
The residual correction is then applied as:
\begin{equation*}
\pT^{\text{residual}} = \pT^{\text{area}} - \alpha \times (N_{\text{PV}}-1) - \beta \times \mu,
\end{equation*}
where $\alpha$ and $\beta$ are the correction factors estimated from MC simulation.
 
\begin{figure}
\centering
\subfigure[]{\label{subfig:rho}
\includegraphics[width=0.49\linewidth]{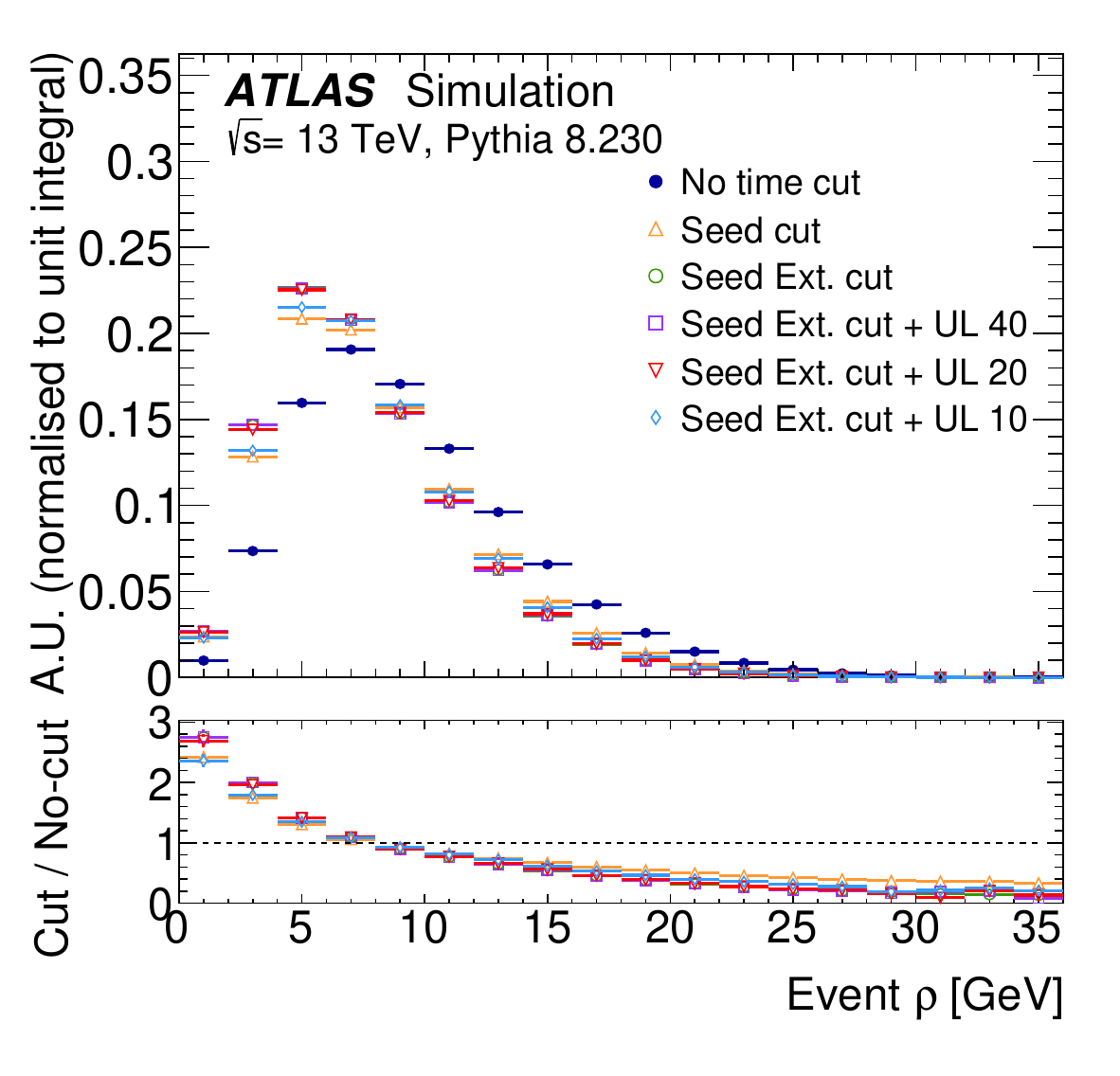}}
\subfigure[]{\label{subfig:rhovsmu}
\includegraphics[width=0.49\linewidth]{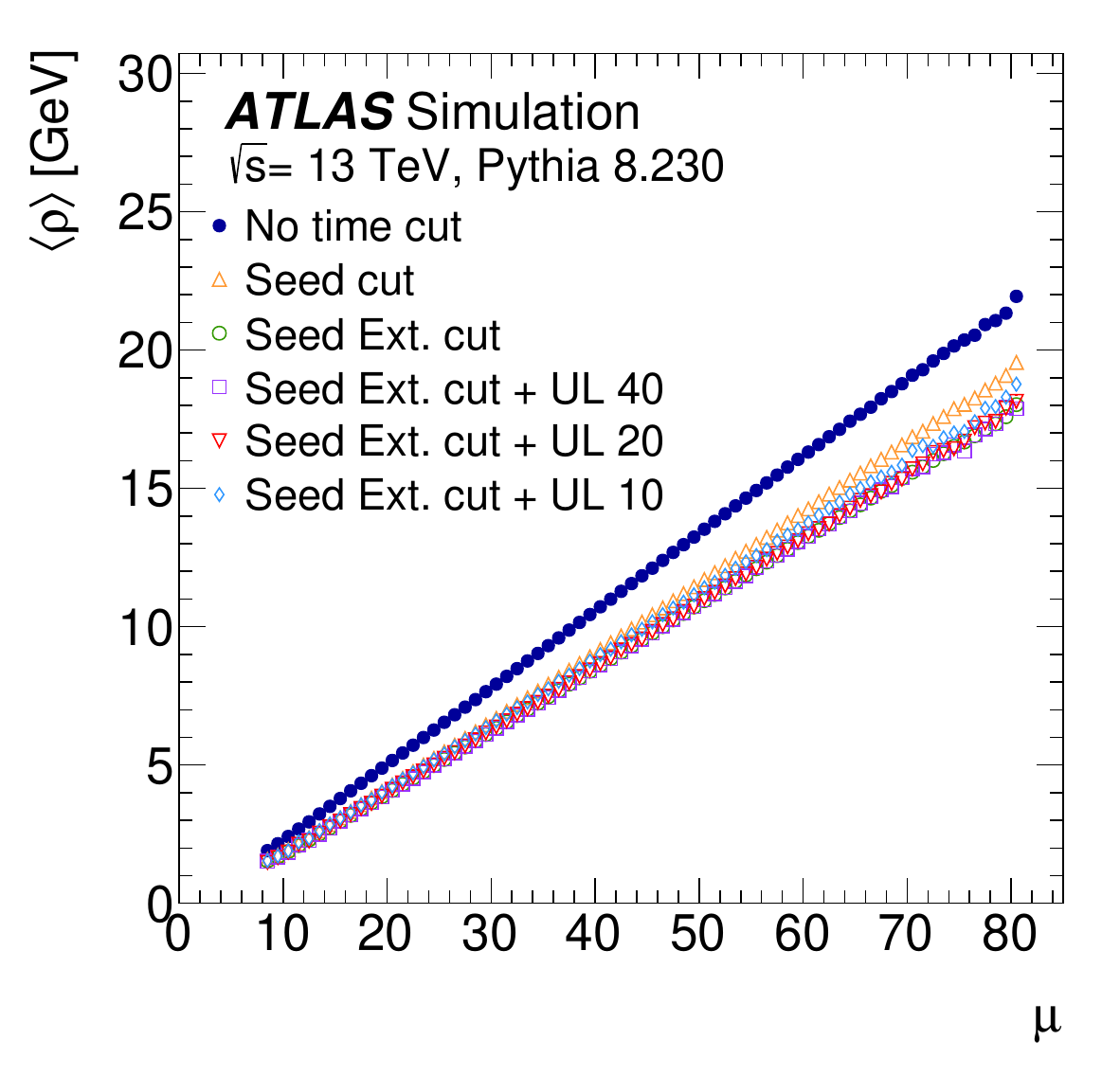}}
\caption{Topo-cluster median \pT density ($\rho$):  comparison between the no time cut and the considered time cut options. \subref{subfig:rho}: $\rho$ spectrum. \subref{subfig:rhovsmu}: average $\rho$ in bins of the number of interactions per bunch-crossing ($\mu$). Uncertainties in the cut~/~no-cut ratios are obtained by splitting the available sample into subsamples and recomputing the cut~/~no-cut ratio for each subsample. The standard deviation of the distribution of the ratio is used to estimate the ratio's uncertainty.}
\label{fig:JZx_evtRho}
\end{figure}

The \pileup contribution to jets is expected to be affected by the time cut. Figure~\ref{fig:JZx_evtRho} shows the spectrum of the median \pT density $\rho$, as well as the dependence of $\rho$ on $\mu$. When the time cut is applied, $\rho$ is found to be ${\sim}20\%$ smaller overall, and also to increase more slowly as a function of $\mu$. Since $\rho$ is built from topo-clusters, the improved \pileup rejection of the time cut leads to smaller $\rho$ values, decreasing the \pileup-dependent contribution to the jet energy.
To account for possible remaining data--MC differences in the \pileup contribution to jets, not covered by the jet-area correction, the residual correction is recalcuted after the application of the time cut. The standard procedure summarised above is repeated for both the Seed and Seed Extended cuts. Since the effect of the Upper Limit is smaller and limited to high-energy regions, the residual correction is not recalculated for any of the $X_{\textrm{UL}}$ values, and the residual correction calculated for the Seed Extended cut is used.
 
\subsubsection{Jet selection}
\label{sec:jetSelection}
Reconstructed jets are calibrated by applying the \pileup correction and the MC-JES calibration. They are then preselected, requiring them to lie within $|\eta|<4.5$ and satisfy $\pT>7$~\GeV. Jets are also required to be isolated, i.e.\ there must be no other preselected jet satisfying $\Delta R(j_1,j_2)<1.5\, R_{\text{jet}}$, where $j_{1}$ and $j_{2}$ are any two reconstructed jets and $R_{\text{jet}}$ is the jet radius parameter used in the \antikt algorithm (here $R_{\text{jet}}=0.4$). Since this study focuses on both signal and \pileup jets, \pileup suppression requirements such as the one using the Jet Vertex Tagger~\cite{ATLAS-CONF-2014-018} are not applied.
 
In order to distinguish jets produced by the hard-scattering from jets originating from \pileup, reconstructed jets are matched to \textit{truth-level jets} in MC events. Truth-level jets are reconstructed by applying the \antikt algorithm ($R=0.4$) to stable final-state particles, defined as having $c\tau > 10~\textrm{mm}$ in the generator's event record, excluding muons and neutrinos.
Three types of truth jets are distinguished. First, truth jets originating from the hard scattering (\textit{HS-truth jet}) are obtained by clustering stable particles from the simulated hard-scattering event, excluding particles from \pileup interactions. Next, two types of \pileup truth jets are obtained by clustering stable particles from the minimum-bias events which are overlaid on the hard scattering in order to simulate \pileup~\cite{SIMU-2020-01}. Jets are clustered for each minimum-bias event individually and then grouped into two \pileup jet collections: jets from in-time overlay events are included in the in-time \pileup jet collection (\textit{IT-truth jet} in the following), while those from out-of-time overlay events are included in the out-of-time \pileup jet collection (\textit{OOT-truth jet} in the following). The latter are built from 32 bunch-crossings before the current one and 6 after~\cite{SIMU-2020-01}.
 
Default cuts requiring $\pT>10~\GeV$ and $\pT>15~\GeV$ are applied to IT-truth jets and OOT-truth jets respectively. HS-truth jets are also preselected by requiring $\pT>7$~\GeV. All truth jets must satisfy $|\eta|<5$. It should be noted that while a \pT match can be expected between HS-truth jets and reconstructed jets after the calibration is applied, this is less the case for OOT-truth jets: because of negative energy contributions, one might expect the reconstructed energy of an out-of-time \pileup jet to be lower than that of the truth jet.
This is the reason for having a higher \pT selection for truth \pileup jets than for HS-truth jets.
Truth-level jets are also required to be isolated: there must be no other truth jet satisfying $\Delta R(j_1,j_2) < 2.5\, R_{\text{jet}}$. Here $j_{1}$ and $j_{2}$ are any two truth jets, all three truth-jet categories being considered together when computing the isolation.
Reconstructed jets are matched to truth jets via a geometrical matching requirement: a reconstructed jet $j_\text{r}$ is said to match a truth jet $j_\text{t}$ if their separation satisfies $\Delta R(j_\text{r},j_\text{t})<0.3$. Reconstructed jets are first checked for matches with a HS-truth jet. If no such match is found, then matches with IT-truth and OOT-truth jets are attempted.

\subsubsection{Jet kinematics}
\begin{figure}
\centering
\subfigure[]{\label{subfig:JZx_withAllUL_calijetHSmPt}
\includegraphics[width=0.49\linewidth]{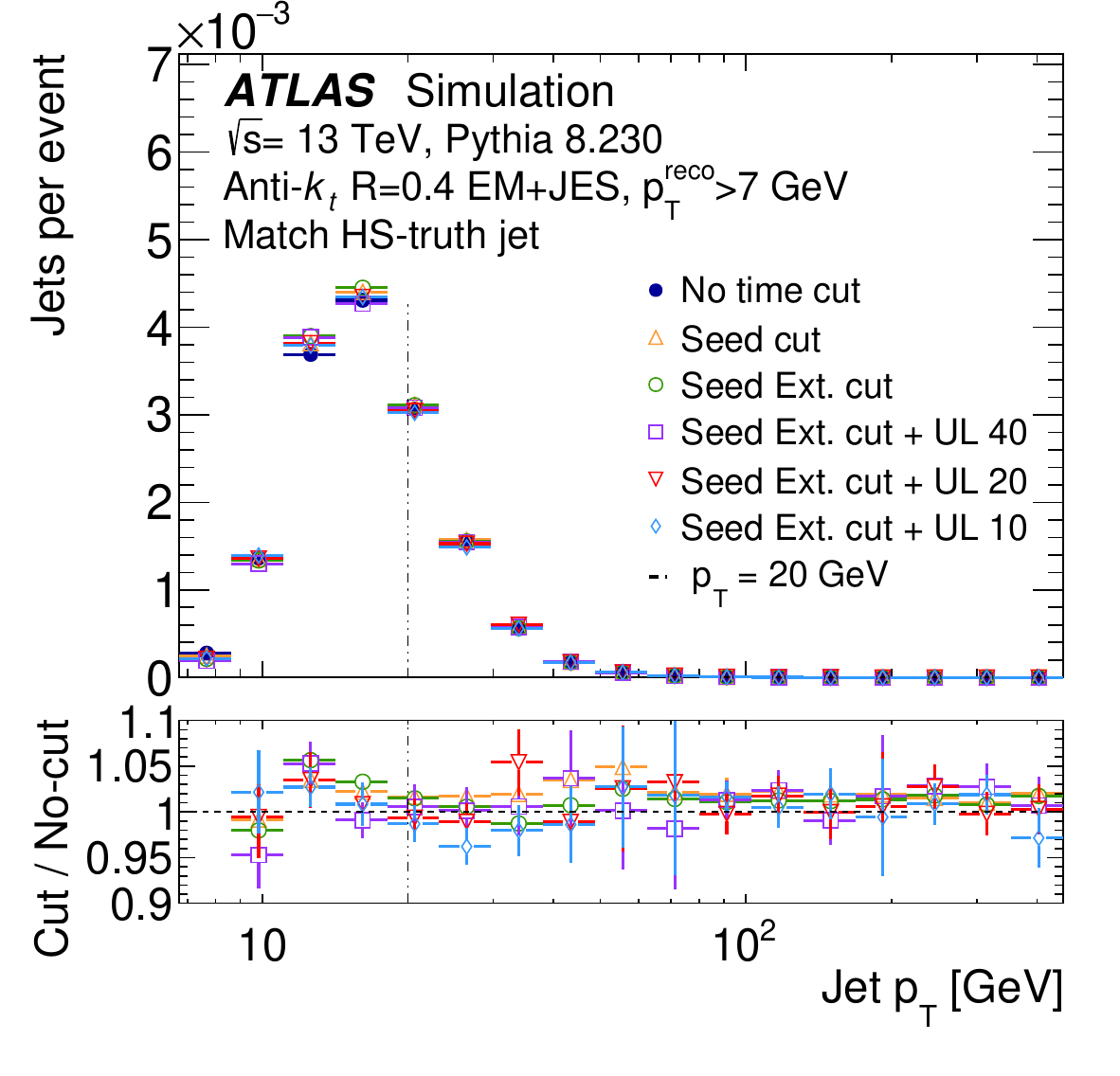}}
\subfigure[]{\label{subfig:JZx_withAllUL_calijetITmPt}
\includegraphics[width=0.49\linewidth]{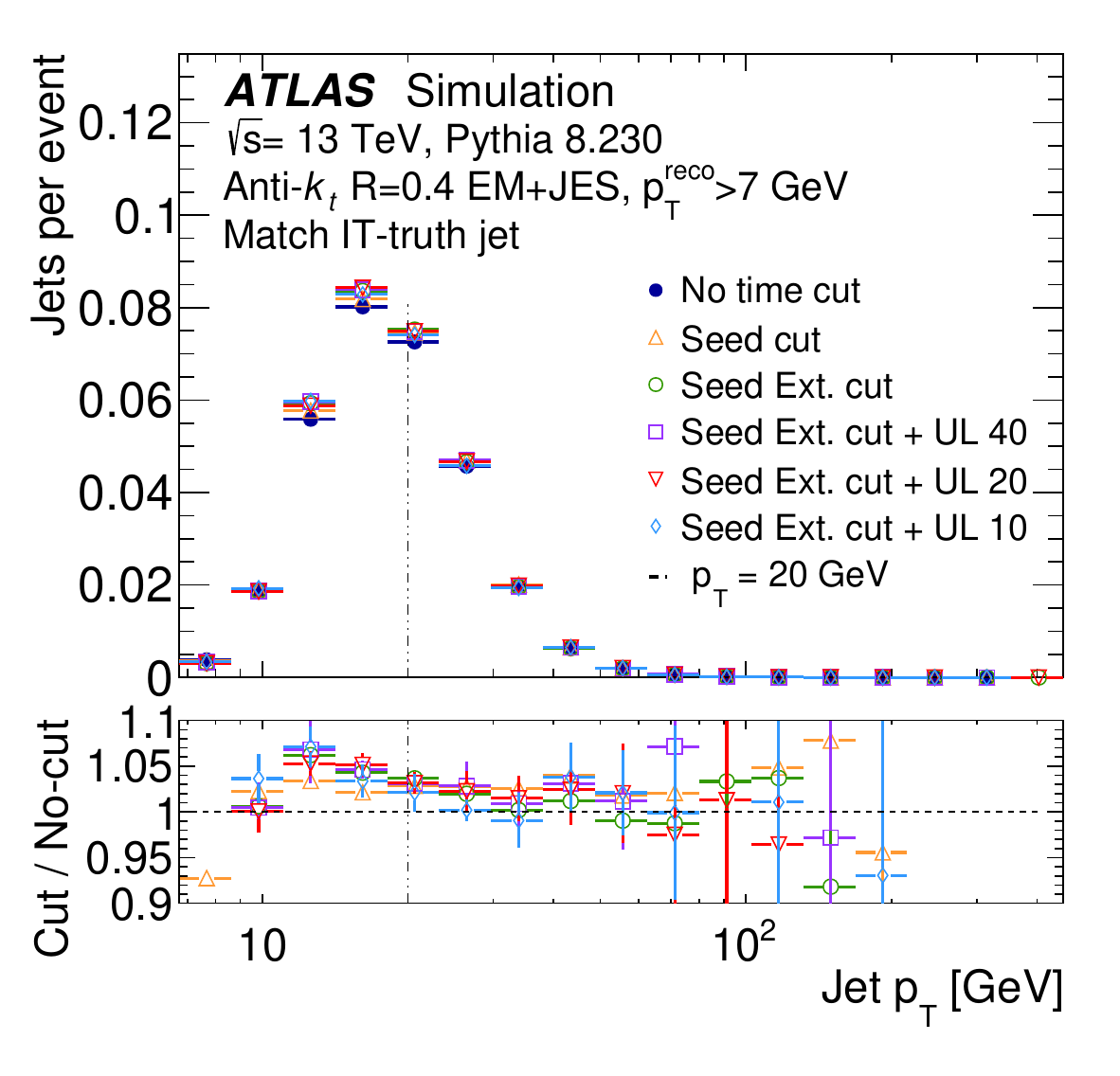}}
\subfigure[]{\label{subfig:JZx_withAllUL_calijetOOTmPt}
\includegraphics[width=0.49\linewidth]{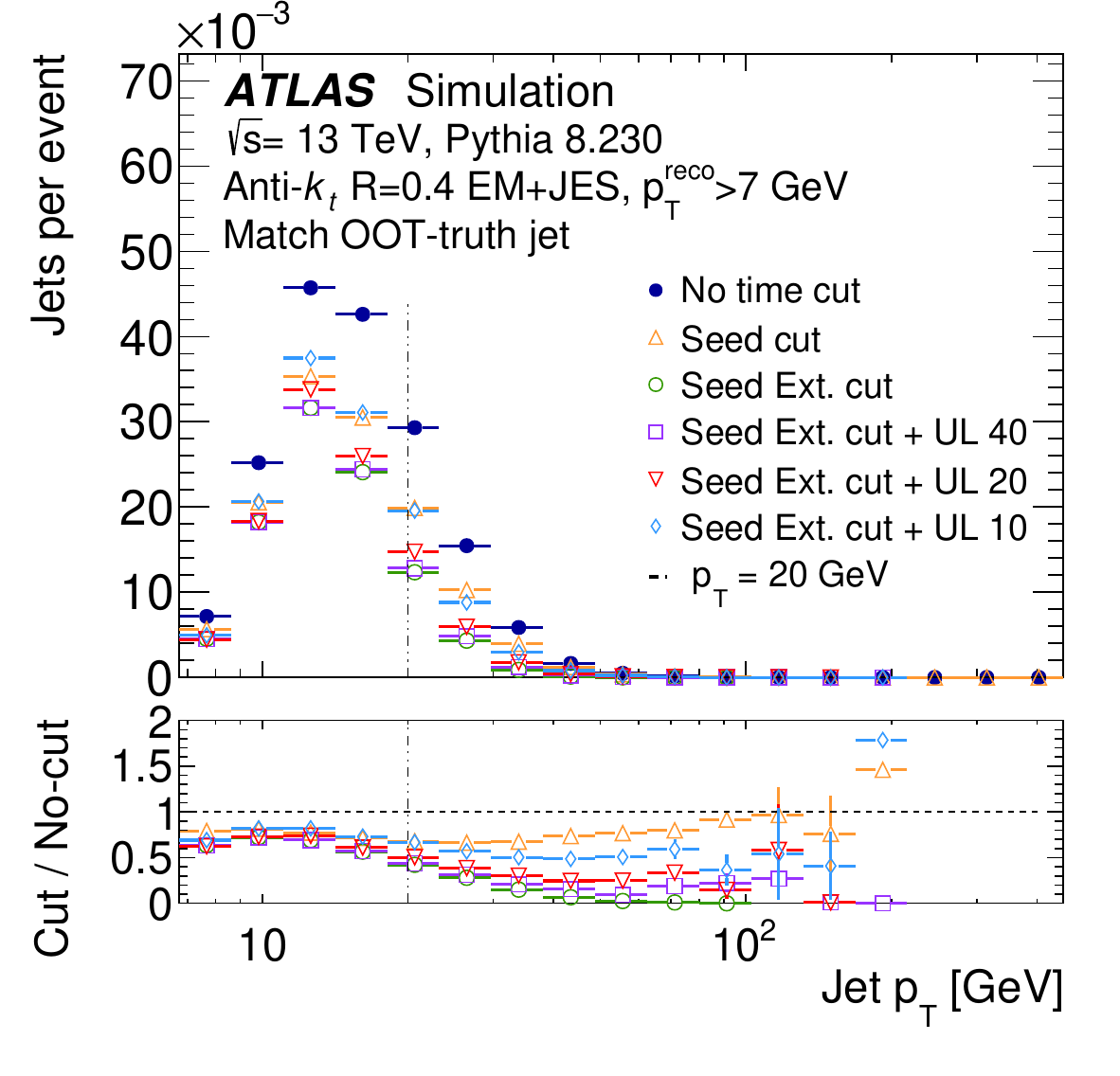}}
\caption{Comparison between jet \pT spectra obtained when the Seed, Seed Extended, or no time cut is used. The Seed Extended cut is also shown in combination with the Upper Limit for $X_{\textrm{UL}}=40$, $X_{\textrm{UL}}=20$, and $X_{\textrm{UL}}=10$. \subref{subfig:JZx_withAllUL_calijetHSmPt}: jets matching HS-truth jets. \subref{subfig:JZx_withAllUL_calijetITmPt}: jets matching IT-truth jets. \subref{subfig:JZx_withAllUL_calijetOOTmPt}: jets matching OOT-truth jets. Uncertainties in the cut~/~no-cut ratios are obtained by splitting the available sample into subsamples and recomputing the cut~/~no-cut ratio for each subsample. The standard deviation of the distribution of the ratio is used to estimate the ratio's uncertainty. The vertical dashed line represents the minimum jet \pT of 20~\GeV\ normally required in ATLAS physics analyses.}
\label{fig:JZx_calijetPt_UL}
\end{figure}
\begin{figure}
\centering
\subfigure[]{\label{subfig:JZx_withAllUL_calijetHSmRapidity}
\includegraphics[width=0.49\linewidth]{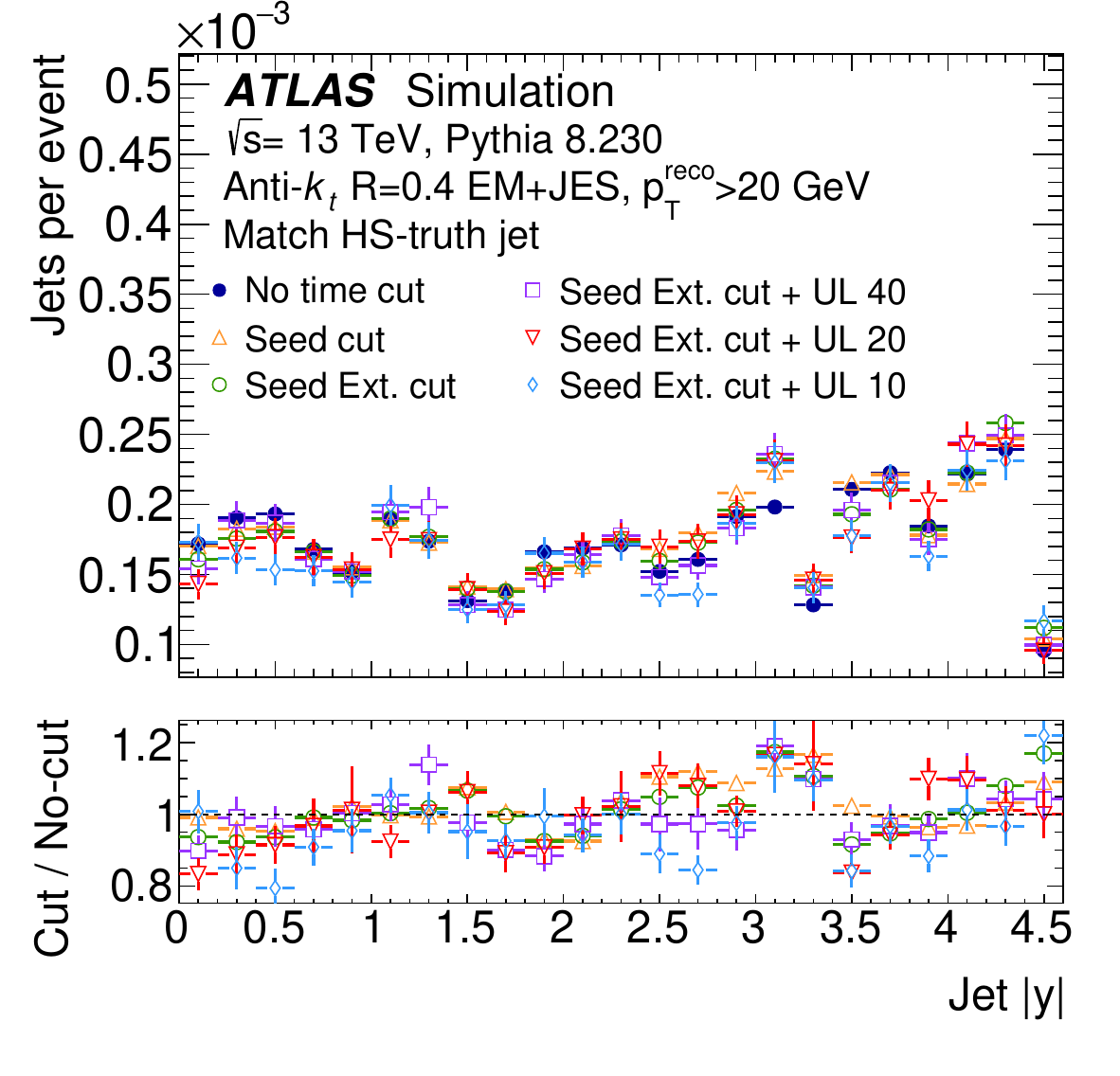}}
\subfigure[]{\label{subfig:JZx_withAllUL_calijetITmRapidity}
\includegraphics[width=0.49\linewidth]{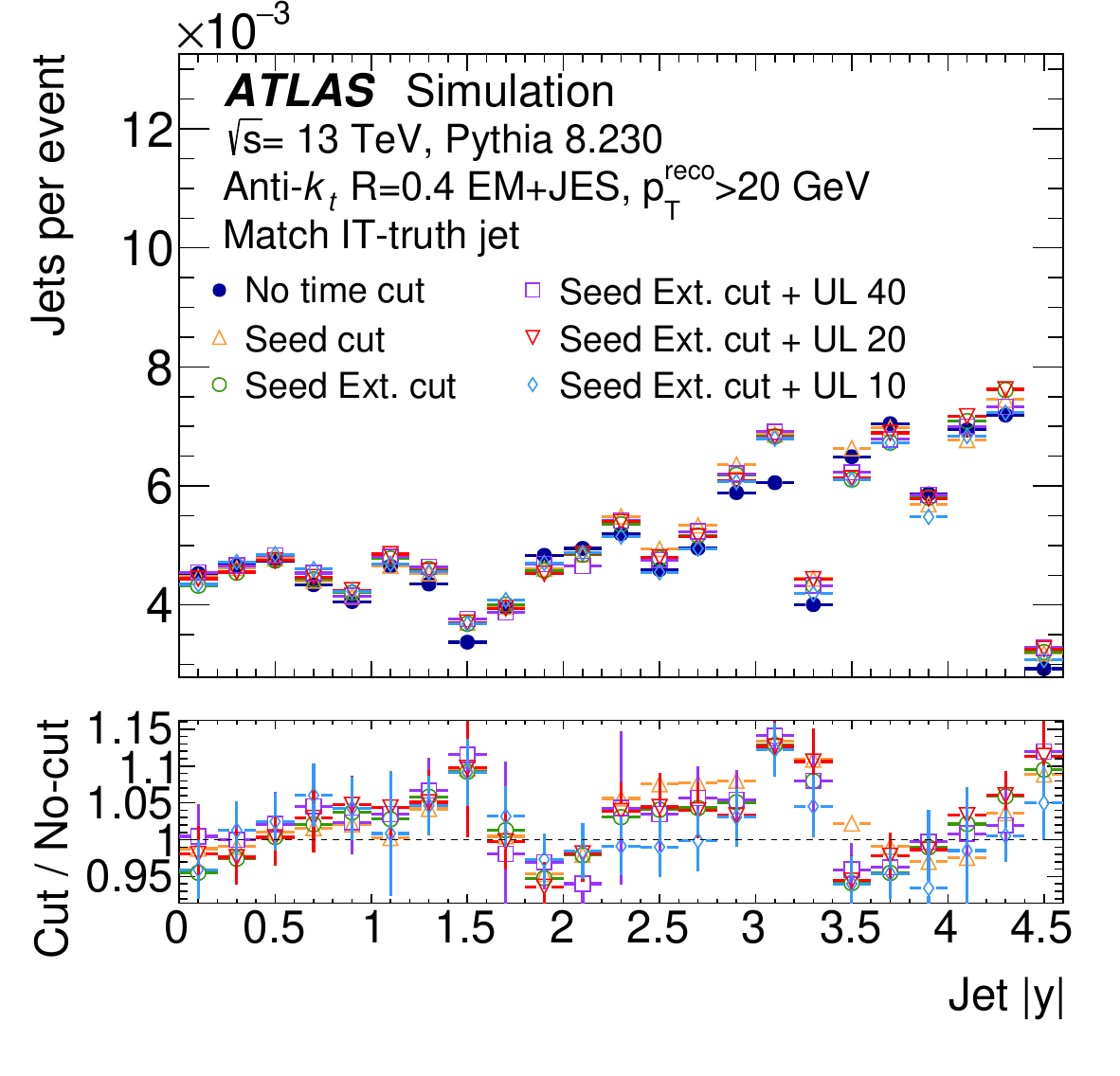}}
\subfigure[]{\label{subfig:JZx_withAllUL_calijetOOTmRapidity}
\includegraphics[width=0.49\linewidth]{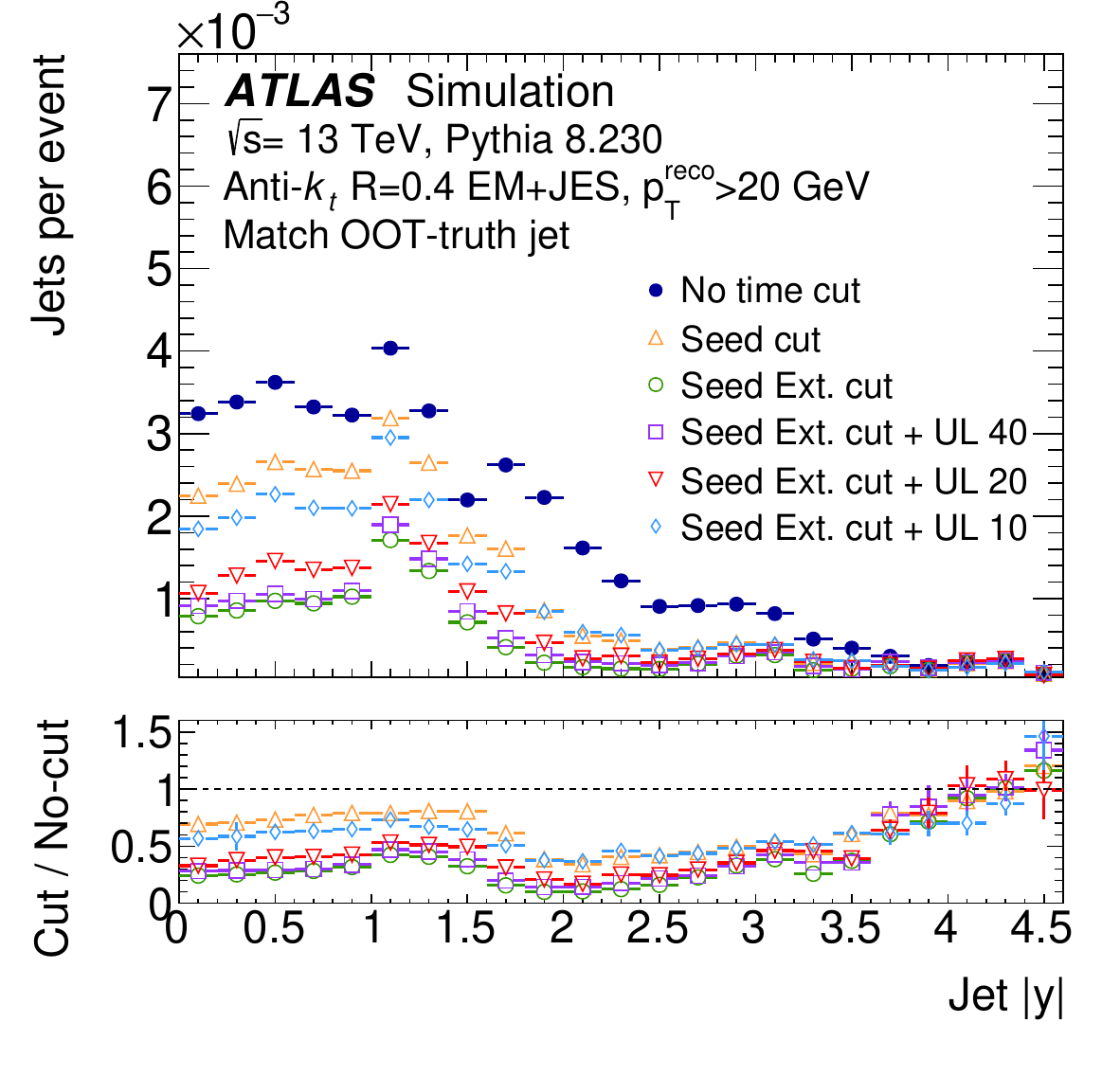}}
\caption{Comparison between jet rapidity spectra obtained when the Seed, Seed Extended, or no time cut is used. The Seed Extended cut is also shown in combination with the Upper Limit for $X_{\textrm{UL}}=40$, $X_{\textrm{UL}}=20$, and$X_{\textrm{UL}}=10$. \subref{subfig:JZx_withAllUL_calijetHSmRapidity}: jets matching HS-truth jets. \subref{subfig:JZx_withAllUL_calijetITmRapidity}: jets matching IT-truth jets. \subref{subfig:JZx_withAllUL_calijetOOTmRapidity}: jets matching OOT-truth jets. Uncertainties in the cut~/~no-cut ratios are obtained by splitting the available sample into subsamples and recomputing the cut~/~no-cut ratio for each subsample. The standard deviation of the distribution of the ratio is used to estimate the ratio's uncertainty.}
\label{fig:JZx_calijetY_UL}
\end{figure}
Figure~\ref{fig:JZx_calijetPt_UL} shows the calibrated transverse momentum distributions for the jet categories described in Section~\ref{sec:jetSelection}.
Both the reconstructed jets matching HS-truth jets and those matching IT-truth jets show only a percent-level impact from the time cut.
It should be pointed out that the time cut can affect the reconstructed jet kinematics, possibly resulting in bin-to-bin migrations in the spectra shown in this subsection. Energy (and \pT) variations can occur in either direction, since the cell energies included in the topo-clusters can be either positive or negative. Out-of-time cells often produce negative energy signals and this explains why the ratio plots in Figures~\ref{subfig:JZx_withAllUL_calijetHSmPt} and~\ref{subfig:JZx_withAllUL_calijetITmPt} have values above unity in some places.
The small \pT-dependent decrease in the fraction of jets matching HS-truth jets was found to be due to random matches.
The fraction of jets that match OOT-truth jets, on the other hand, is drastically reduced by the Seed Extended cut, especially in the high \pT region. The Seed Extended cut reduces the multiplicity of out-of-time jets by ${\sim}60\%$ at $\pT=20~\GeV$, and no out-of-time jets are found above $\sim 45~\GeV$. The Seed cut, on the other hand, reduces the multiplicity of the out-of-time jets by only ${\sim}30\%$ at 20~\GeV\ and only $\sim 20\%$ at 50~\GeV, indicating that the Seed Extended cut rejects out-of-time \pileup more effectively.
Figure~\ref{fig:JZx_calijetPt_UL} also illustrates the effect of the different Upper Limit values. At $\pt \sim 45~\GeV$, the reduction in out-of-time jet multiplicity is ${\sim}90\%$ for an Upper Limit of 40, ${\sim}70\%$ for an Upper Limit of 20, and ${\sim}50\%$ for an Upper Limit of 10, confirming that lower Upper Limit values allow more \pileup in the event.
 
Figure~\ref{fig:JZx_calijetY_UL} shows an analogous comparison for the jet rapidity.
In order to limit the kinematics plots to the phase space of interest in physics analyses, an additional requirement of $\pT>20~\GeV$ is applied to the jets.
Overall, all cuts except the Seed cut and Seed Extended plus Upper Limit 10 cut provide a ${\sim}40\%$ suppression of out-of-time jets within $|y|<1.5$. The effect increases to a 50\%--60\% reduction for $1.5 < |y| < 3.2$ and then decreases until it has negligible impact for $|y|>4$.
From these observations, it can be concluded that the Seed Extended time cut is preferable to the Seed cut because it is more effective in suppressing \pileup.
It can also be seen that the Upper Limit $X_{\textrm{UL}}=10$ allows too much \pileup to remain in the event, so the lowest practical Upper Limit among those tested is $X_{\textrm{UL}}=20$.
 
\subsubsection{Jet resolution}
As discussed in Section~\ref{sec:jetCalib}, the time cut affects the jet calibration because of the different amount of \pileup remaining. To check for its impact on the performance, the jet energy resolution after calibration is compared for the different time cuts.
The jet response is defined as the ratio of the calibrated energy of the reconstructed jet to the energy of the matching truth jet:  $R=E^{\textrm{calibrated}}_j/E^{\textrm{truth}}_j$. Only those jets matched to a HS-truth jet are considered for this study.
The response distribution is computed in bins of truth-jet \pT and rapidity ($|y|$), or truth-jet \pT and the event $\mu$. For each bin, the response distribution is fitted with a Gaussian function. The jet resolution is then computed for each bin, as the ratio of the fitted standard deviation of the jet response $R$ to its fitted mean value ($\sigma_{R}^{\textrm{fit}}/\langle R \rangle ^{\textrm{fit}}$).
 
\begin{figure}
\centering
\subfigure[]{\label{subfig:ResolutionVSy_pt20to30}
\includegraphics[width=0.49\linewidth]{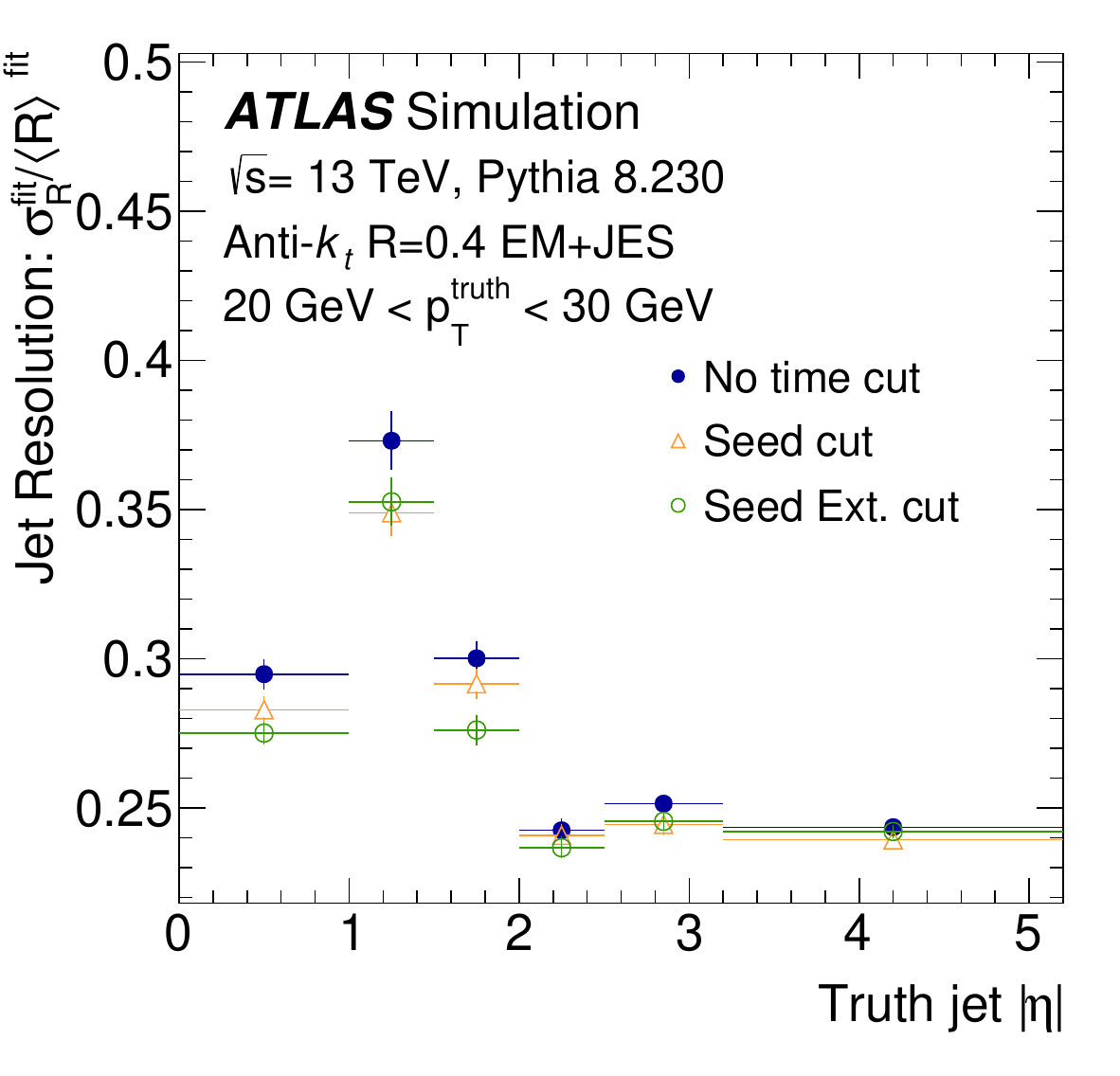}}
\subfigure[]{\label{subfig:ResolutionVSy_pt30to60}
\includegraphics[width=0.49\linewidth]{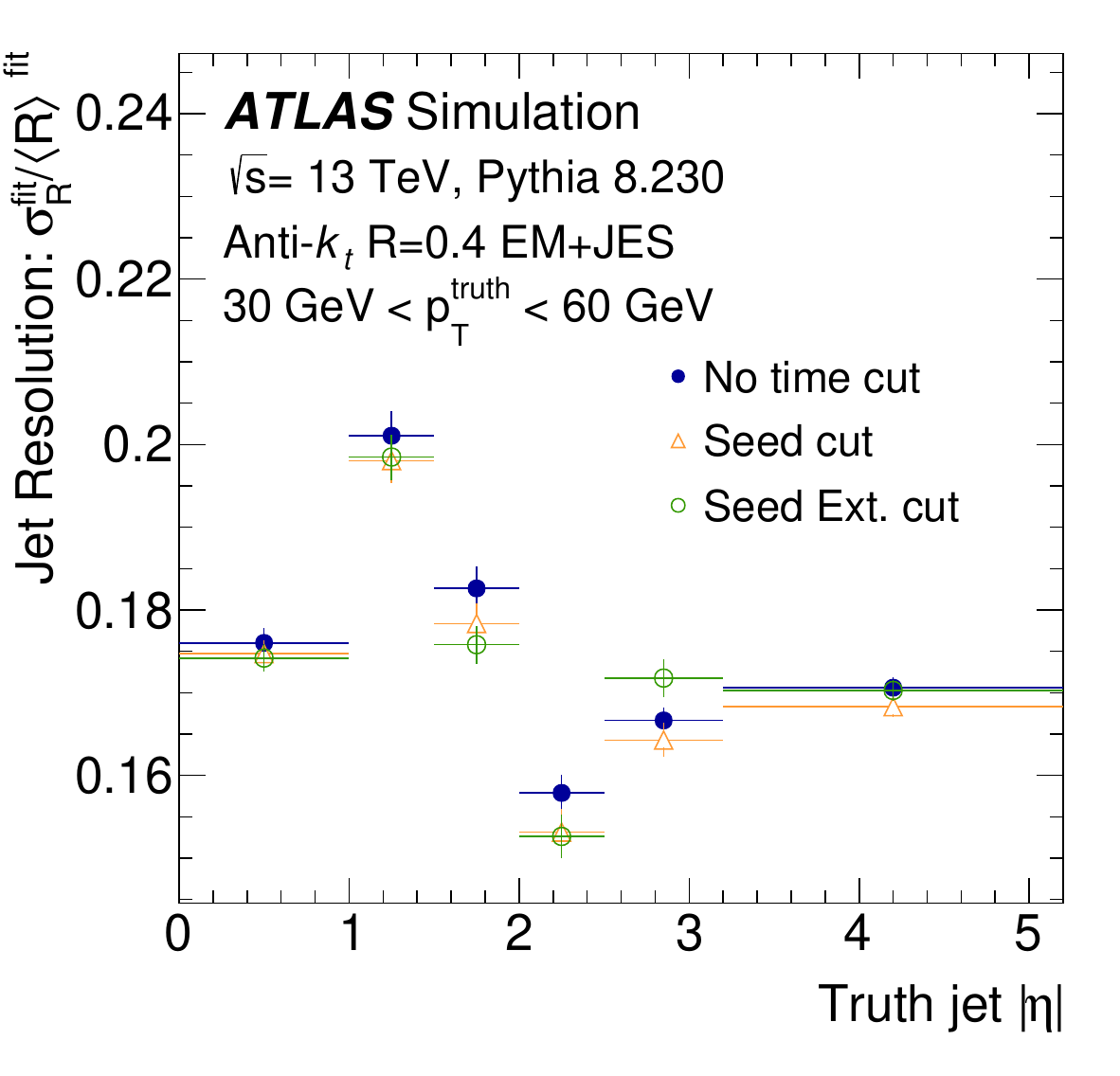}}
\subfigure[]{\label{subfig:ResolutionVSmu_pt20to30}
\includegraphics[width=0.49\linewidth]{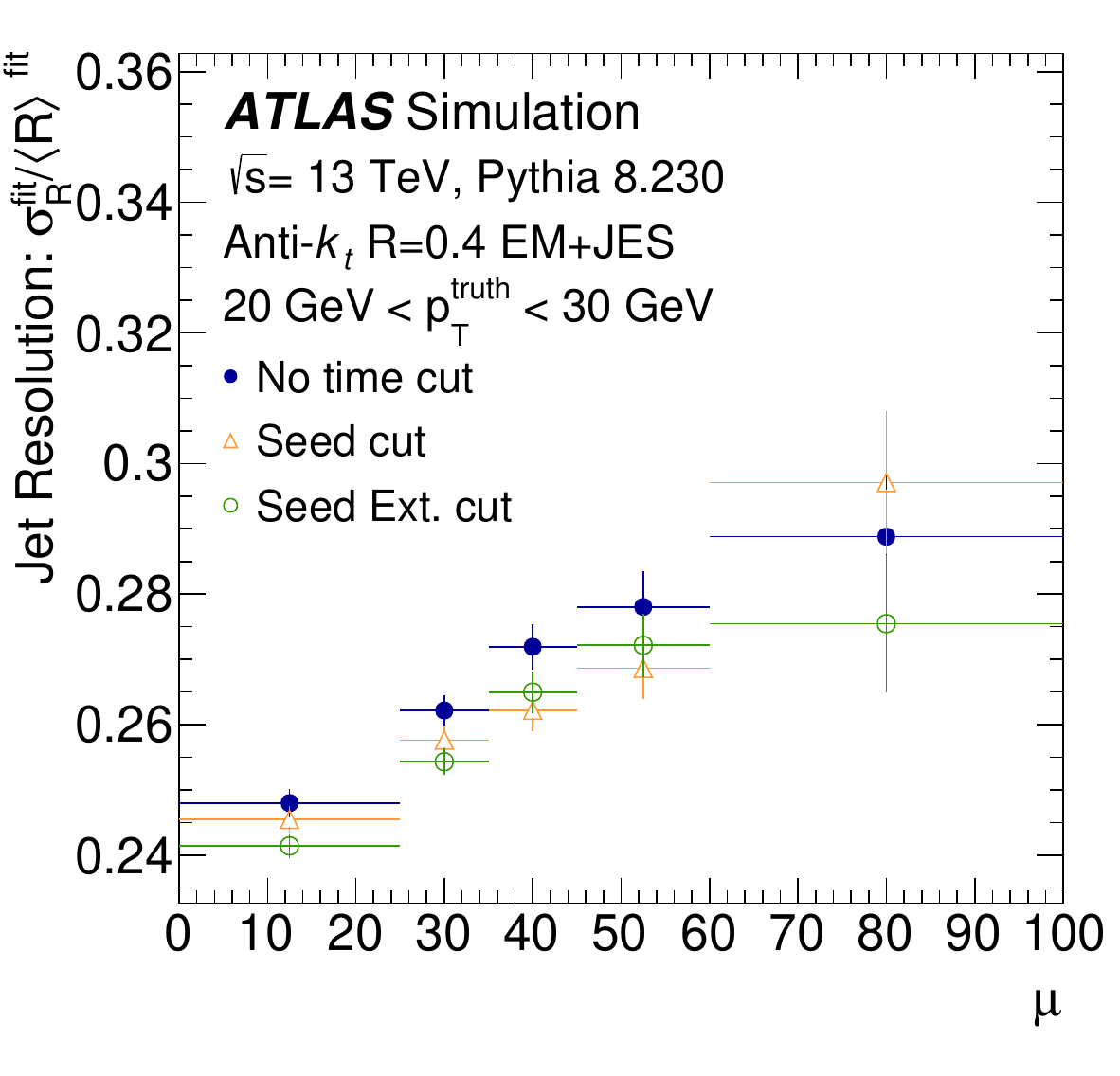}}
\subfigure[]{\label{subfig:ResolutionVSmu_pt30to60}
\includegraphics[width=0.49\linewidth]{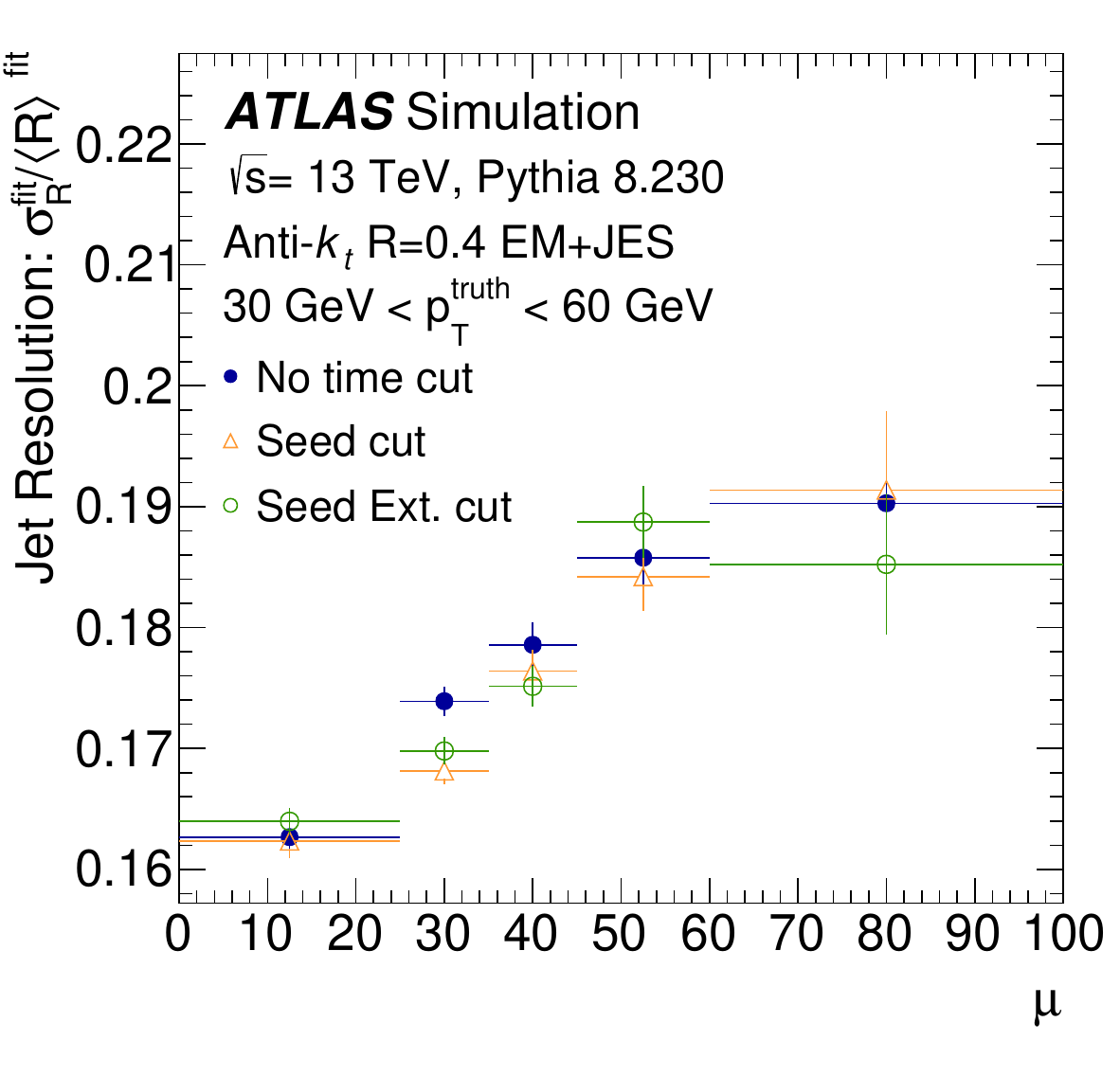}}
\caption{Jet energy resolution after calibration as a function of \subref{subfig:ResolutionVSy_pt20to30}, \subref{subfig:ResolutionVSy_pt30to60} the jet $|\eta|$ and \subref{subfig:ResolutionVSmu_pt20to30}, \subref{subfig:ResolutionVSmu_pt30to60} the number of interactions per bunch-crossing ($\mu$). Results are shown in bins of the jet \pT: \subref{subfig:ResolutionVSy_pt20to30}, \subref{subfig:ResolutionVSmu_pt20to30} $20~\GeV < \pT^{\text{truth}} < 30~\GeV$;  \subref{subfig:ResolutionVSy_pt30to60}, \subref{subfig:ResolutionVSmu_pt30to60} $30~\GeV < \pT^{\text{truth}} < 60~\GeV$. The resolution is defined as the ratio of the standard deviation and mean of a Gaussian function fitted to the response distribution.}
\end{figure}

The estimated resolution is shown for two \pT bins in Figures~\ref{subfig:ResolutionVSy_pt20to30} and~\ref{subfig:ResolutionVSy_pt30to60} as a function of the jet rapidity and in Figures~\ref{subfig:ResolutionVSmu_pt20to30} and~\ref{subfig:ResolutionVSmu_pt30to60} as a function of $\mu$.
Some improvement in the jet resolution is visible, amounting to 5\%--10\% for $20~\GeV \leq \pT \leq 30~\GeV$, but no strong dependence on $\mu$ or $|y|$ is observed. A more accurate estimation of the jet resolution and calibration performance after the time cut is left for future studies, following the implementation of the algorithm in the standard ATLAS reconstruction chain.
 
% End of text imported from the .//MCperformance_pap.tex input file

\FloatBarrier

% The next lines are included from the .//DATAperformance_pap.tex input file
\section{Performance on ATLAS data}
\label{sec:dataperf}
As a further probe of its performance, the time cut was tested on data collected by the ATLAS detector in 2017, corresponding to an integrated luminosity of $\mathcal{L} = 71 \pm 2~\si{\ipb}$ after data quality requirements~\cite{DAPR-2018-01}.
Single-jet triggers were used to collect the data, requiring at least one jet with sufficiently large \pT. Multiple triggers, listed in Table~\ref{tab:triggerset} and implementing different minimum \pT requirements, were studied independently.
Relatively low minimum \pT requirements are considered, as the time cut is known to mostly effect low-energy jets. Triggers with different L1 seeds are also considered: events passing triggers seeded by random L1 triggers are expected to have a larger contamination from out-of-time \pileup than those triggered by single jet at L1. Since all the triggers in Table~\ref{tab:triggerset} are prescaled, the data were reweighted event-by-event by the respective trigger prescale factor.
 
\begin{table}[h!]
\sisetup{group-minimum-digits=4}
\caption{Triggers used for data selection.}
\label{tab:triggerset}
\centering
\begin{tabular}{llcc}
\toprule
Trigger name & L1 trigger seed & Trigger minimum \pT & Offline minimum \pt \\
&	&	{\small ($\geq 1$ jet with $\pT \geq X$)}	&	{\small($\geq 1$ jet with $\pT \geq X$)}\\
\midrule
HLT\_j15 & {\small Random for filled bunches} & $15 \ \GeV$ & $20 \ \GeV$ \\
HLT\_j25 & {\small Random for filled bunches} & $25 \ \GeV$ & $30 \ \GeV$ \\
HLT\_j45 & {\small $\geq 1$ jet with $\pT \geq 15 \ \GeV$} & $45 \ \GeV$ & $50 \ \GeV$ \\
HLT\_j45\_L1RD0\_FILLED & {\small Random for filled bunches} & $45 \ \GeV$ & $50 \ \GeV$ \\
\bottomrule
\end{tabular}
\end{table}
 
\begin{figure}
\centering
\subfigure[]{\label{subfig:HLTj15leadcalijetPtLogPtMin5twoRatios}
\includegraphics[width=0.44\linewidth]{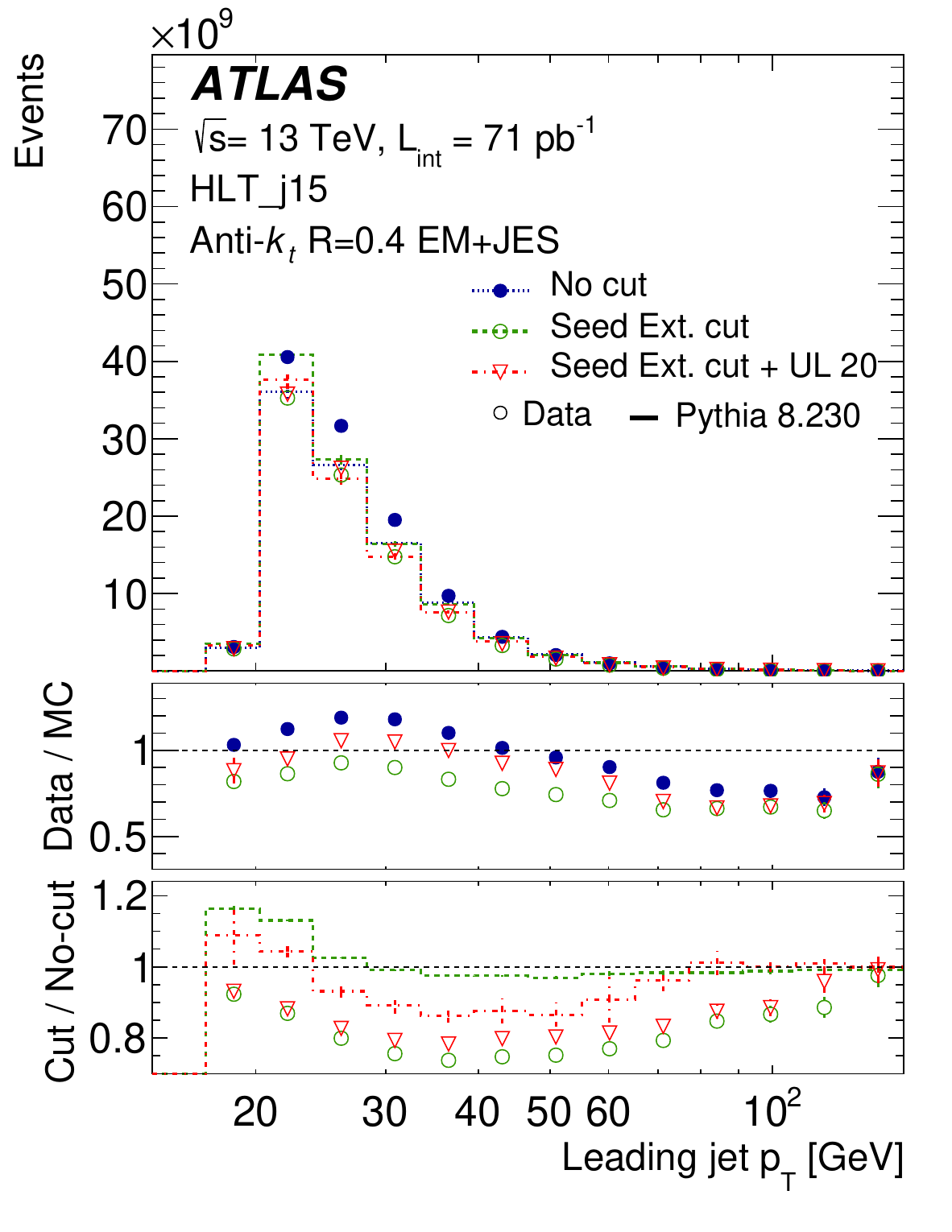}}
\subfigure[]{\label{subfig:HLTj25leadcalijetPtLogPtMin5twoRatios}
\includegraphics[width=0.44\linewidth]{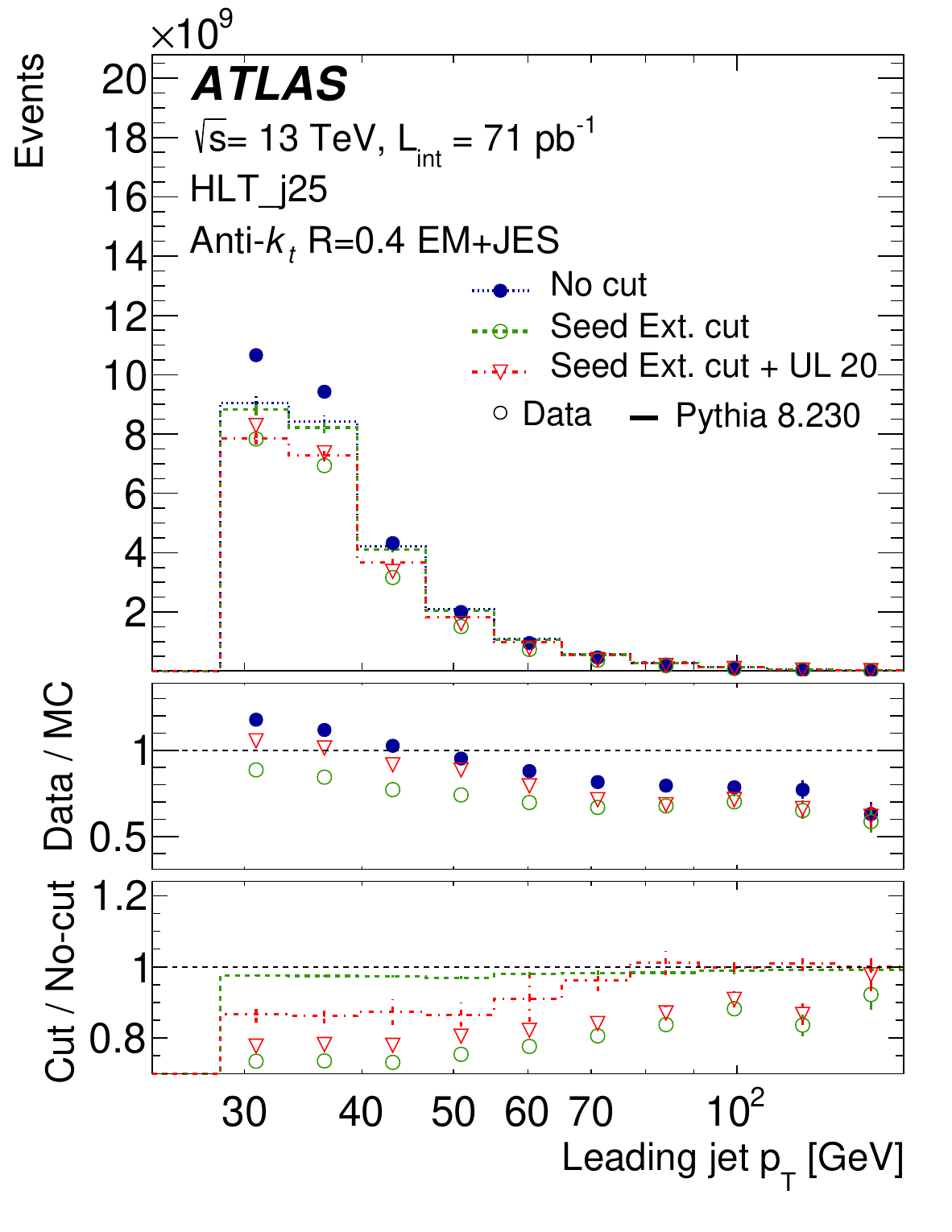}}
\subfigure[]{\label{subfig:HLTj45leadcalijetPtLogPtMin5twoRatios}
\includegraphics[width=0.44\linewidth]{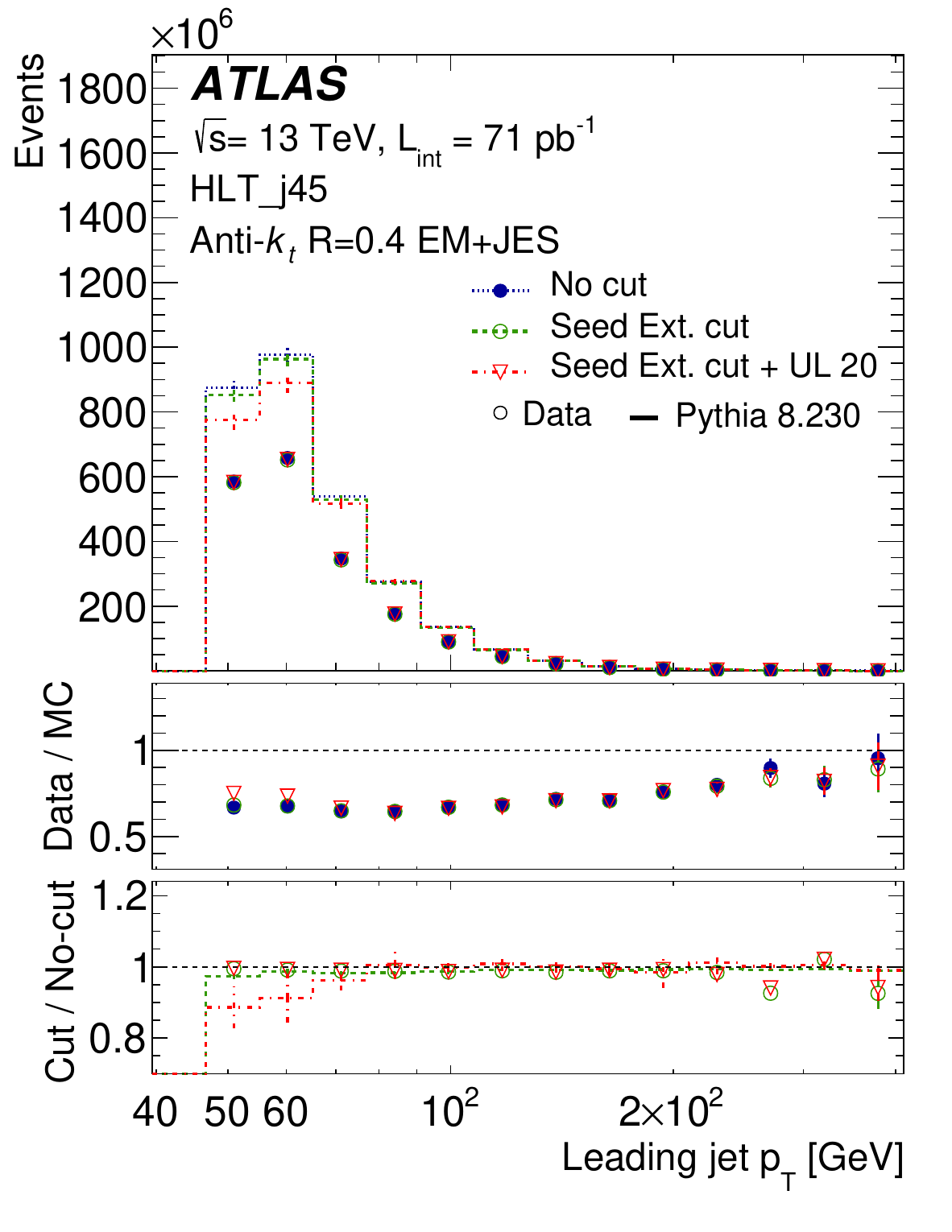}}
\subfigure[]{\label{subfig:HLTj45L1RD0FILLEDleadcalijetPtLogPtMin5twoRatios}
\includegraphics[width=0.44\linewidth]{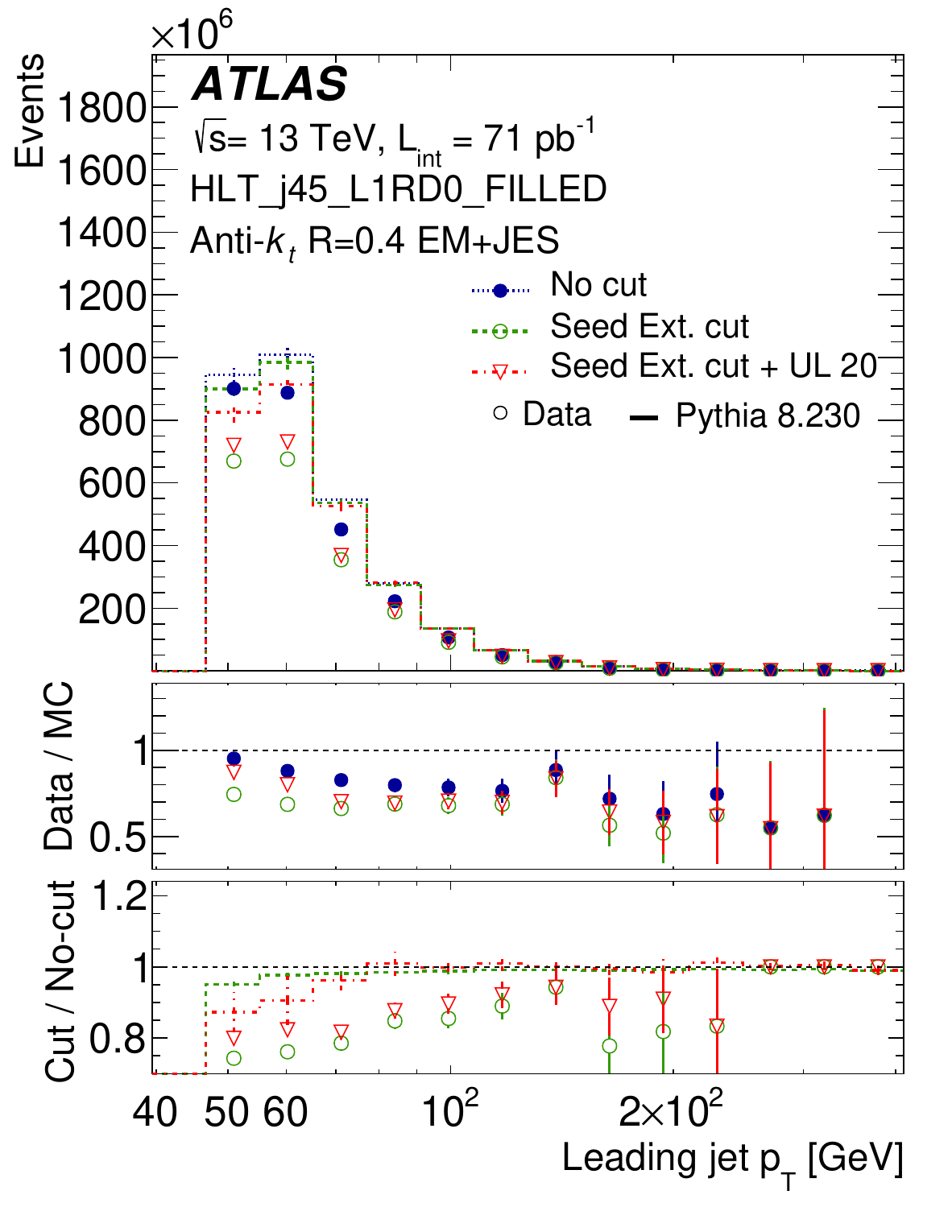}}
\caption{Leading jet \pT spectrum in data and MC selected multi-jet events. Four triggers are compared:  \subref{subfig:HLTj15leadcalijetPtLogPtMin5twoRatios} HLT\_j15, \subref{subfig:HLTj25leadcalijetPtLogPtMin5twoRatios} HLT\_j25,  \subref{subfig:HLTj45leadcalijetPtLogPtMin5twoRatios} HLT\_j45, and \subref{subfig:HLTj45L1RD0FILLEDleadcalijetPtLogPtMin5twoRatios} HLT\_j45\_L1RD0\_FILLED. The error bars convey the statistical and luminosity uncertainties. Uncertainties in the cut / no-cut ratios are obtained by splitting the available sample into subsamples and recomputing the cut / no-cut ratio for each subsample. The standard deviation of the distribution of the ratio is used to estimate the ratio's uncertainty. The luminosity uncertainty does not apply to the cut / no-cut ratio. In the plots, lines represent MC events and markers represent data, different colours and styles represent different cuts.}
\label{fig:dataMC_leadpt}
\end{figure}

\begin{figure}
\centering
\subfigure[]{\label{subfig:HLTj15calijetPtLogPtMin5twoRatios}
\includegraphics[width=0.44\linewidth]{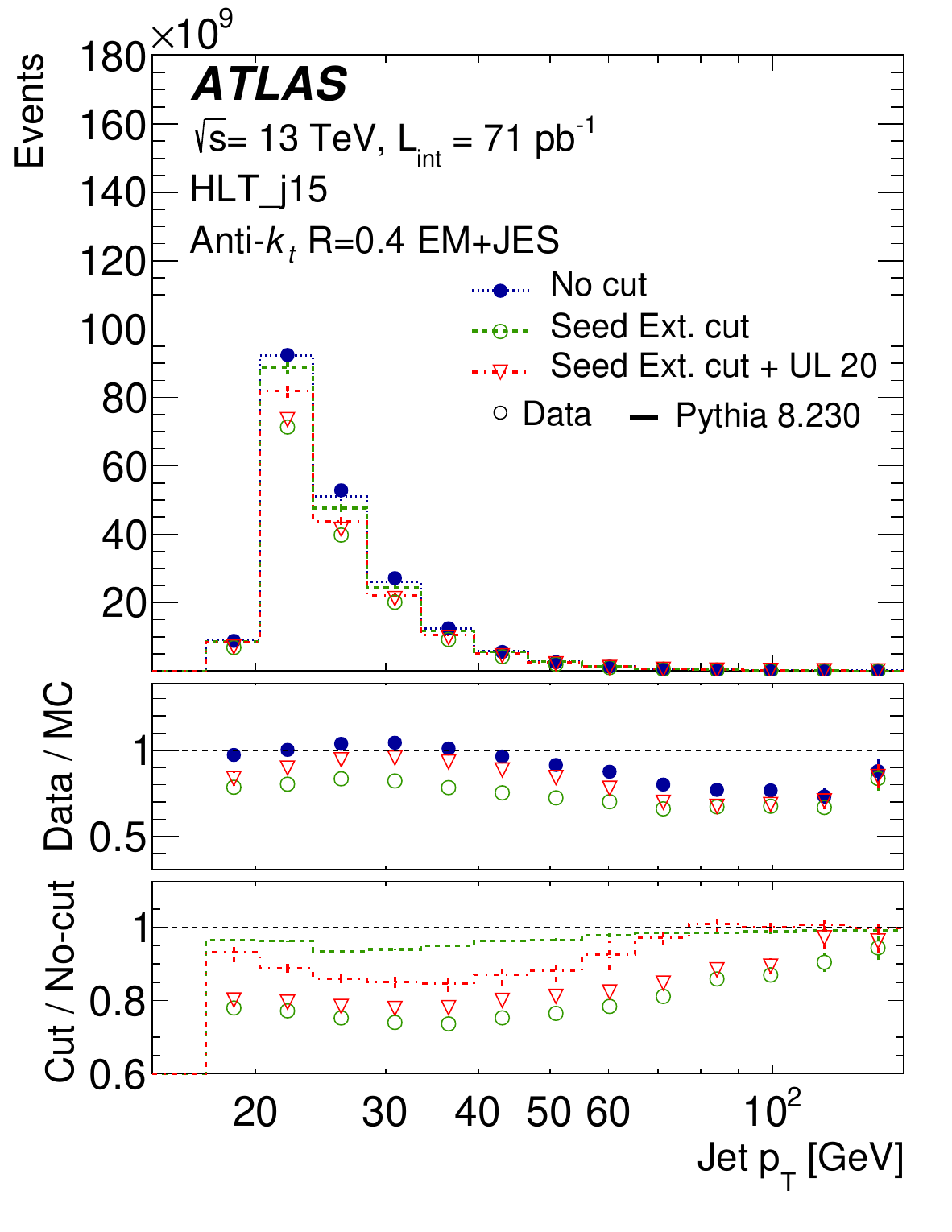}}
\subfigure[]{\label{subfig:HLTj25calijetPtLogPtMin5twoRatios}
\includegraphics[width=0.44\linewidth]{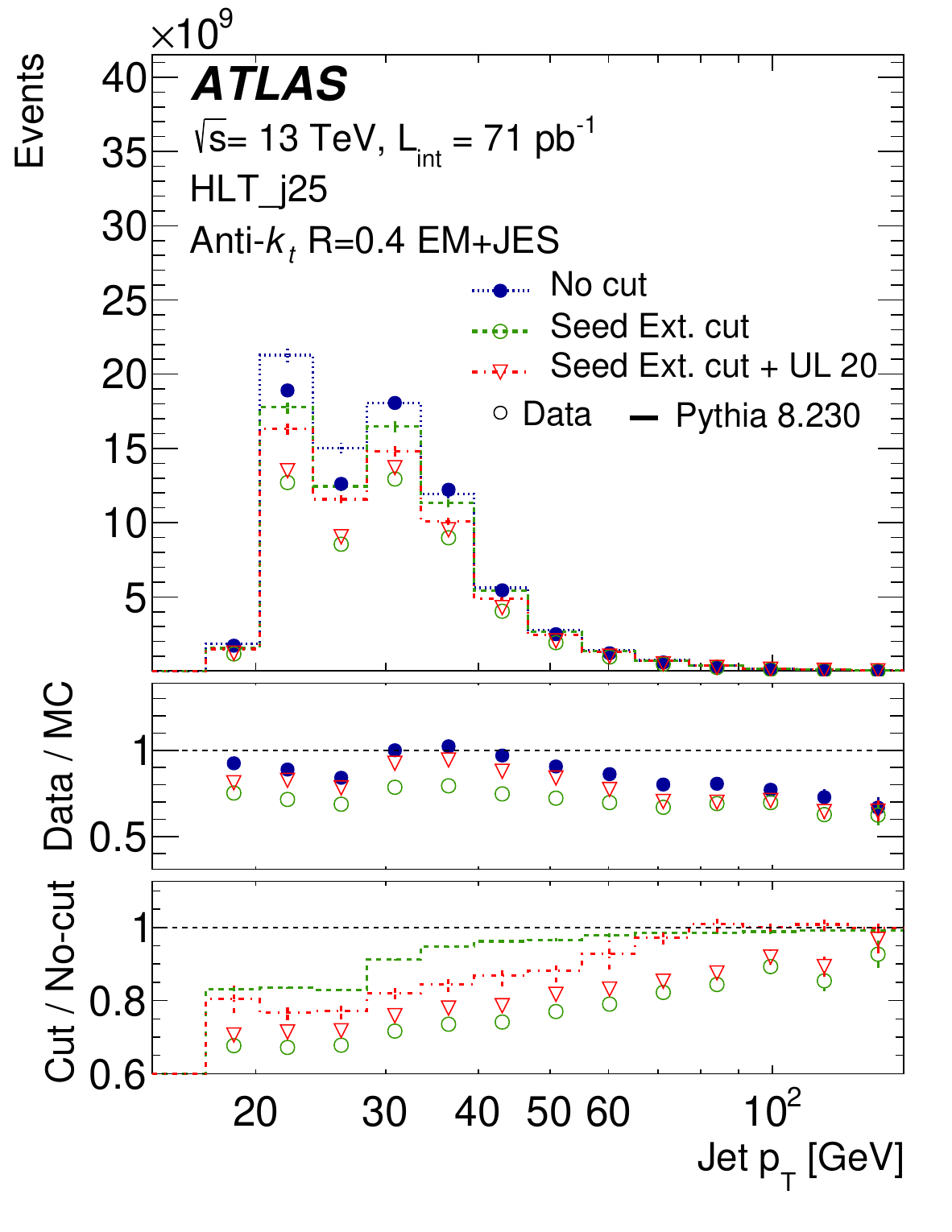}}
\subfigure[]{\label{subfig:HLTj45calijetPtLogPtMin5twoRatios}
\includegraphics[width=0.44\linewidth]{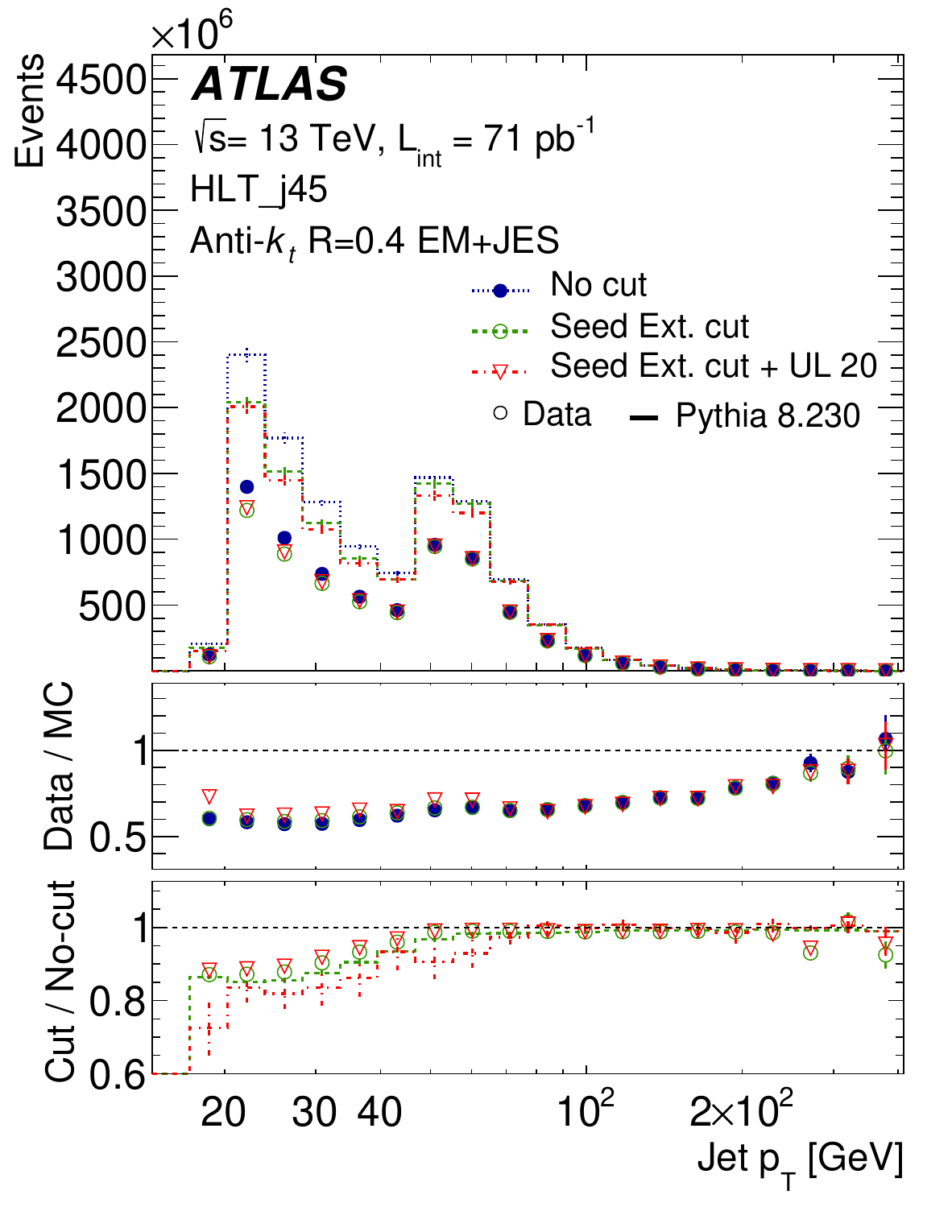}}
\subfigure[]{\label{subfig:HLTj45L1RD0FILLEDcalijetPtLogPtMin5twoRatios}
\includegraphics[width=0.44\linewidth]{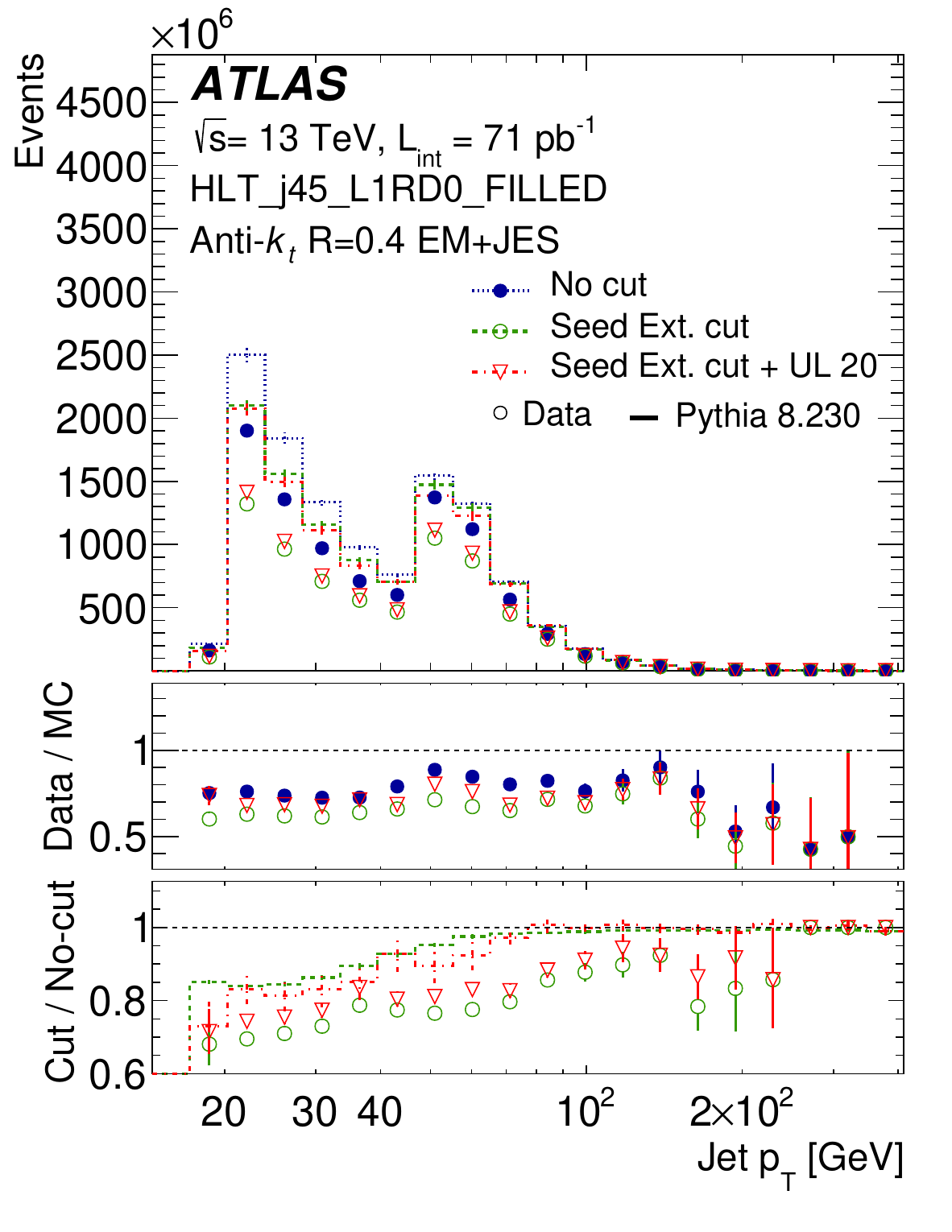}}
\caption{Jet \pT spectrum in data and MC selected multi-jet events. All jets passing the minimum \pT requirement ($\pT > 20~\GeV$) are shown. Four triggers are compared: \subref{subfig:HLTj15calijetPtLogPtMin5twoRatios} HLT\_j15, \subref{subfig:HLTj25calijetPtLogPtMin5twoRatios} HLT\_j25, \subref{subfig:HLTj45calijetPtLogPtMin5twoRatios} HLT\_j45, and  \subref{subfig:HLTj45L1RD0FILLEDcalijetPtLogPtMin5twoRatios} HLT\_j45\_L1RD0\_FILLED. The error bars convey the statistical and luminosity uncertainties. Uncertainties in the cut / no-cut ratios are obtained by splitting the available sample into subsamples and recomputing the cut / no-cut ratio for each subsample. The standard deviation of the distribution of the ratio is used to estimate the ratio's uncertainty. The luminosity uncertainty does not apply to the cut / no-cut ratio. In the plots, lines represent MC events and markers represent data, different colours and styles represent different cuts.}
\label{fig:dataMC_pt}
\end{figure}

\begin{figure}
\centering
\subfigure[]{\label{subfig:HLTj15calijetEtaPtMin5twoRatios}
\includegraphics[width=0.44\linewidth]{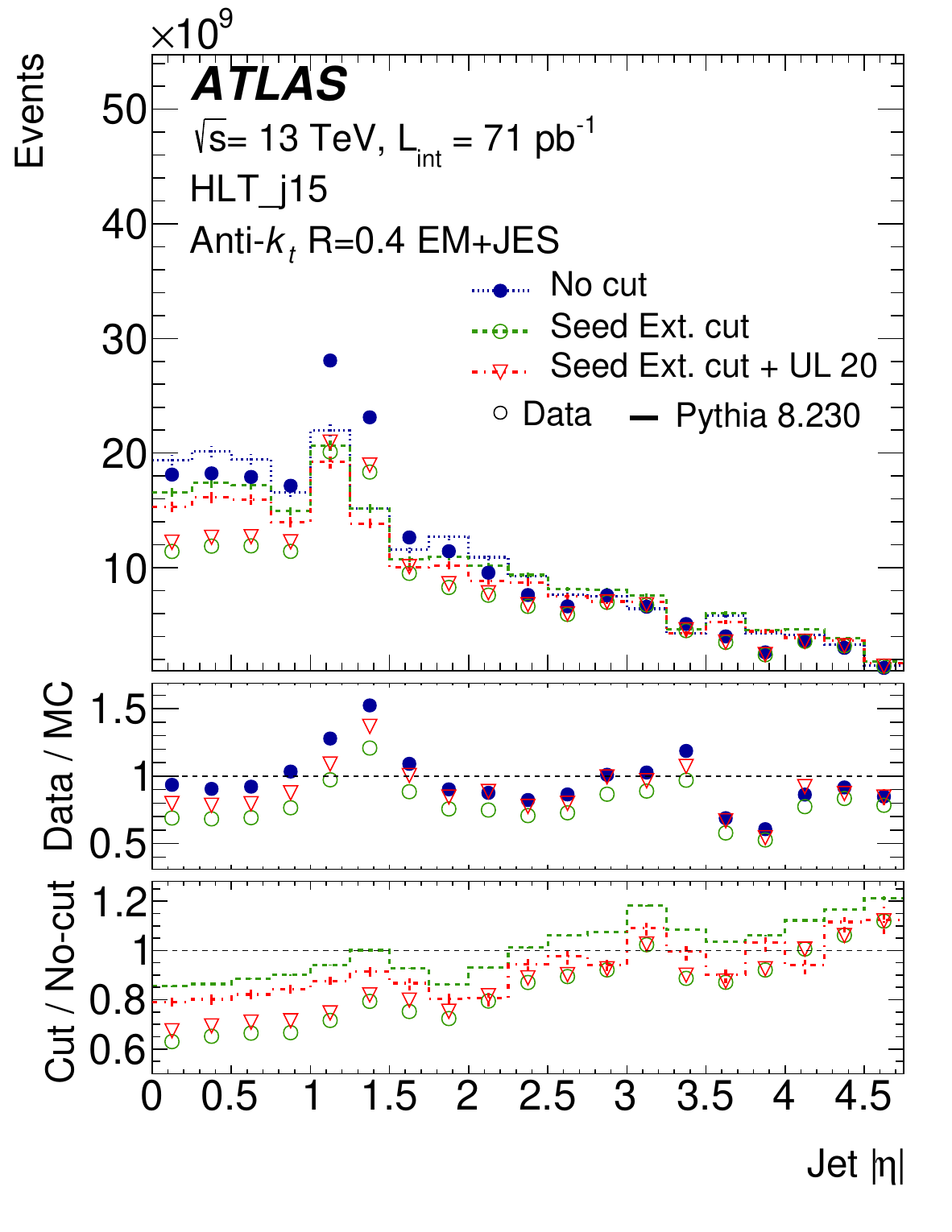}}
\subfigure[]{\label{subfig:HLTj25calijetEtaPtMin5twoRatios}
\includegraphics[width=0.44\linewidth]{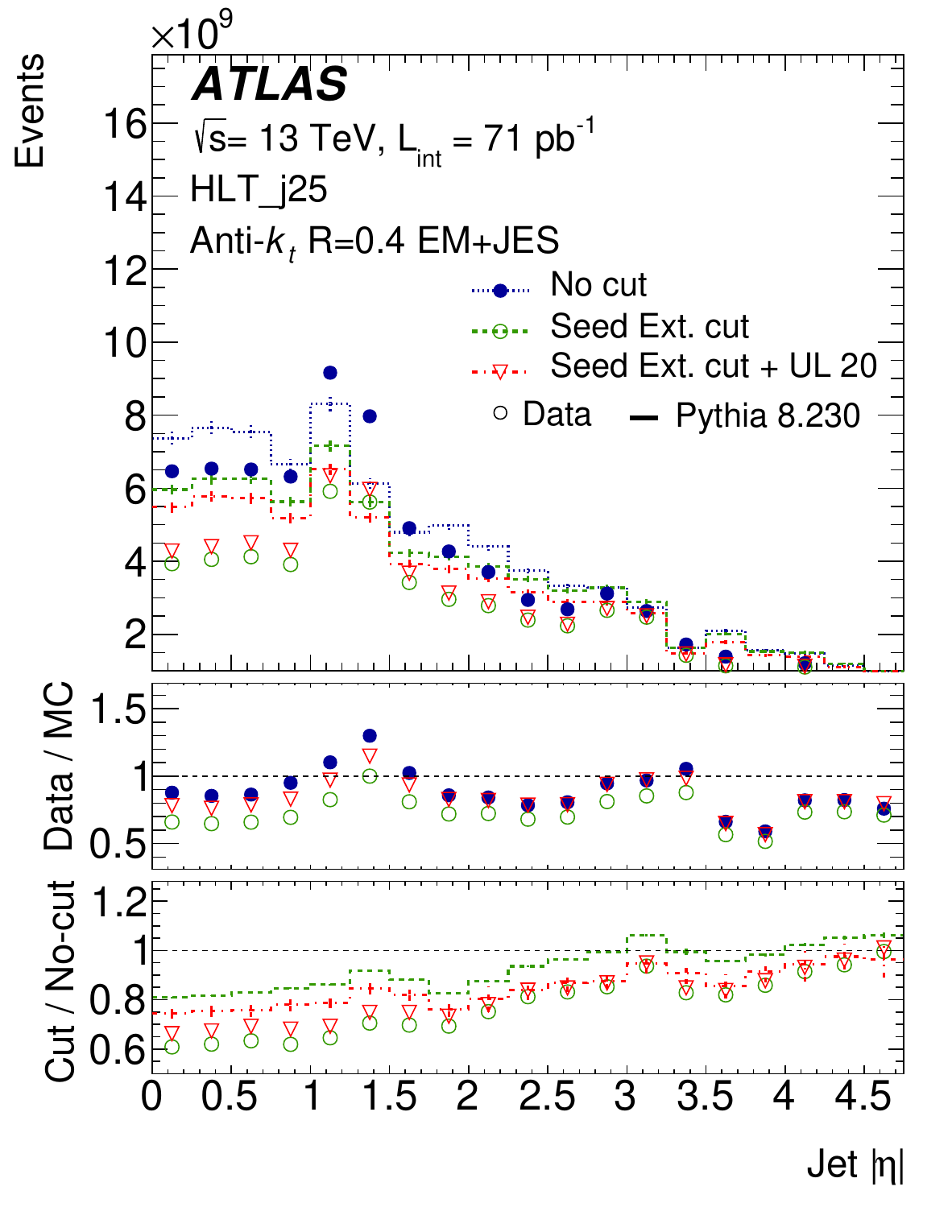}}
\subfigure[]{\label{subfig:HLTj45calijetEtaPtMin5twoRatios}
\includegraphics[width=0.44\linewidth]{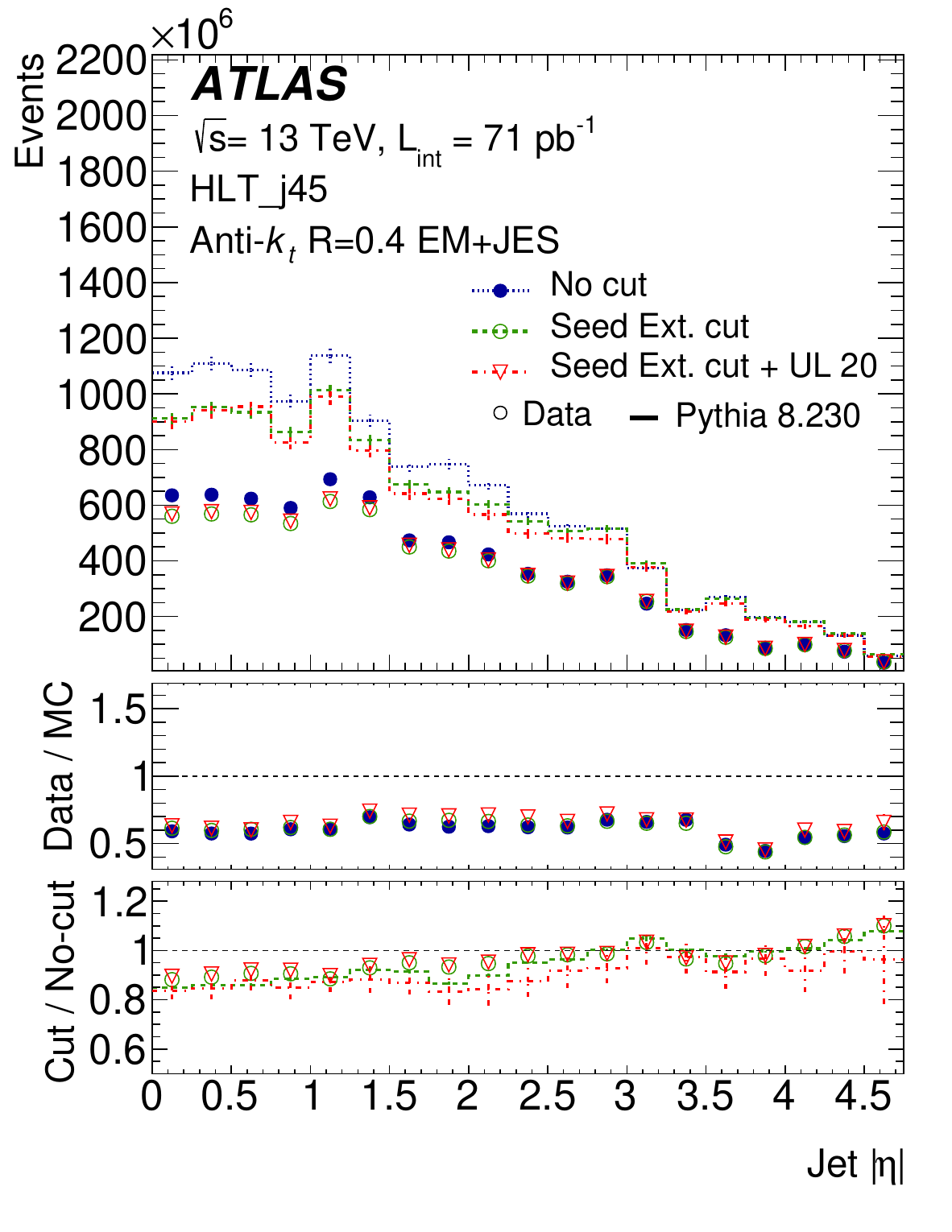}}
\subfigure[]{\label{subfig:HLTj45L1RD0FILLEDcalijetEtaPtMin5twoRatios}
\includegraphics[width=0.44\linewidth]{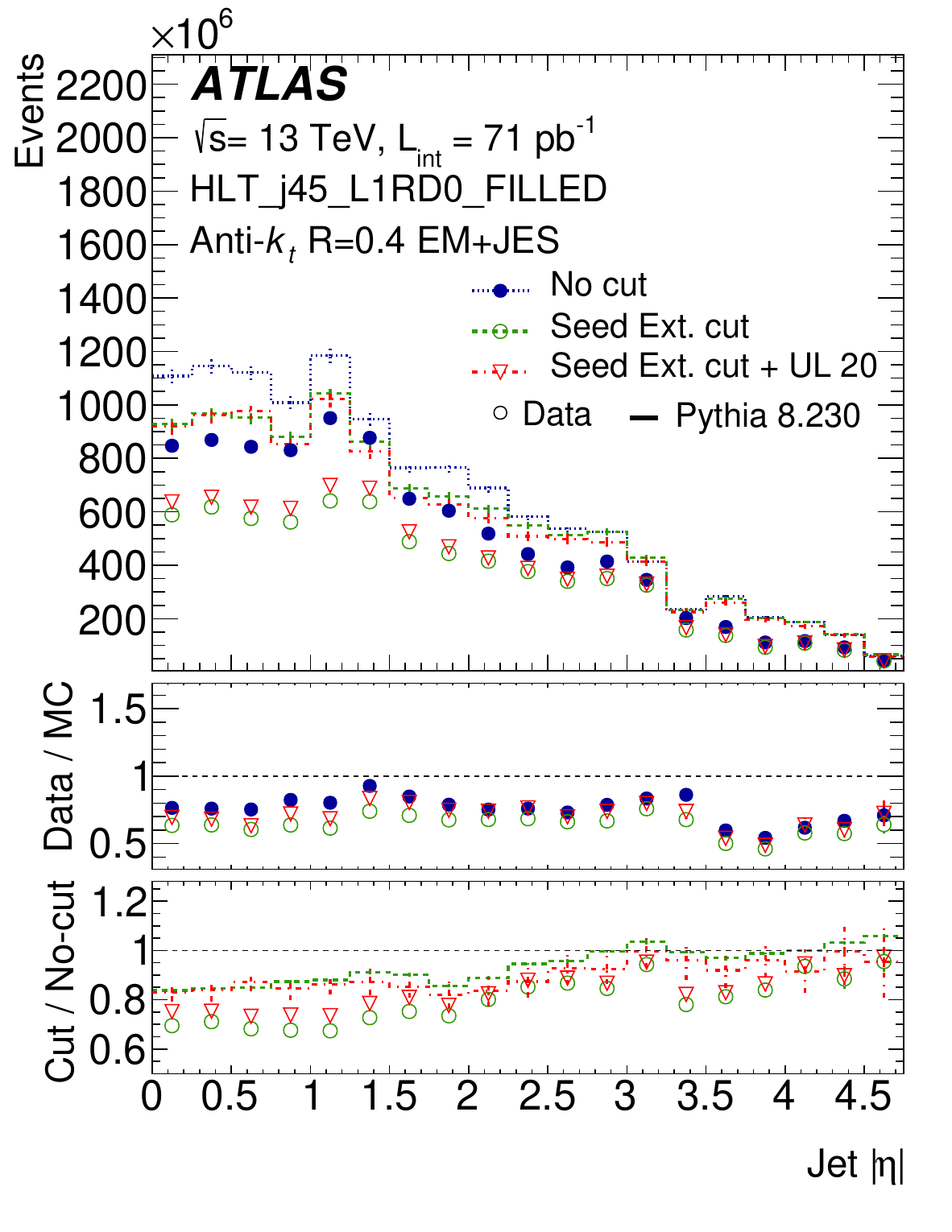}}
\caption{Jet $|\eta|$ spectrum in data and MC selected multi-jet events. All jets passing the minimum \pT requirement ($\pT > 20~\GeV$) are shown. Four triggers are compared: \subref{subfig:HLTj15calijetEtaPtMin5twoRatios} HLT\_j15, \subref{subfig:HLTj25calijetEtaPtMin5twoRatios} HLT\_j25, \subref{subfig:HLTj45calijetEtaPtMin5twoRatios} HLT\_j45, and  \subref{subfig:HLTj45L1RD0FILLEDcalijetEtaPtMin5twoRatios} HLT\_j45\_L1RD0\_FILLED. The error bars convey the statistical and luminosity uncertainties. Uncertainties in the cut / no-cut ratios are obtained by splitting the available sample into subsamples and recomputing the cut / no-cut ratio for each subsample. The standard deviation of the distribution of the ratio is used to estimate the ratio's uncertainty. The luminosity uncertainty does not apply to the cut / no-cut ratio. In the plots, lines represent MC events and markers represent data, different colours and styles represent different cuts.}
\label{fig:dataMC_eta}
\end{figure}

\begin{figure}
\centering
\subfigure[]{\label{subfig:HLTj15clusTPtMin5twoRatios}
\includegraphics[width=0.44\linewidth]{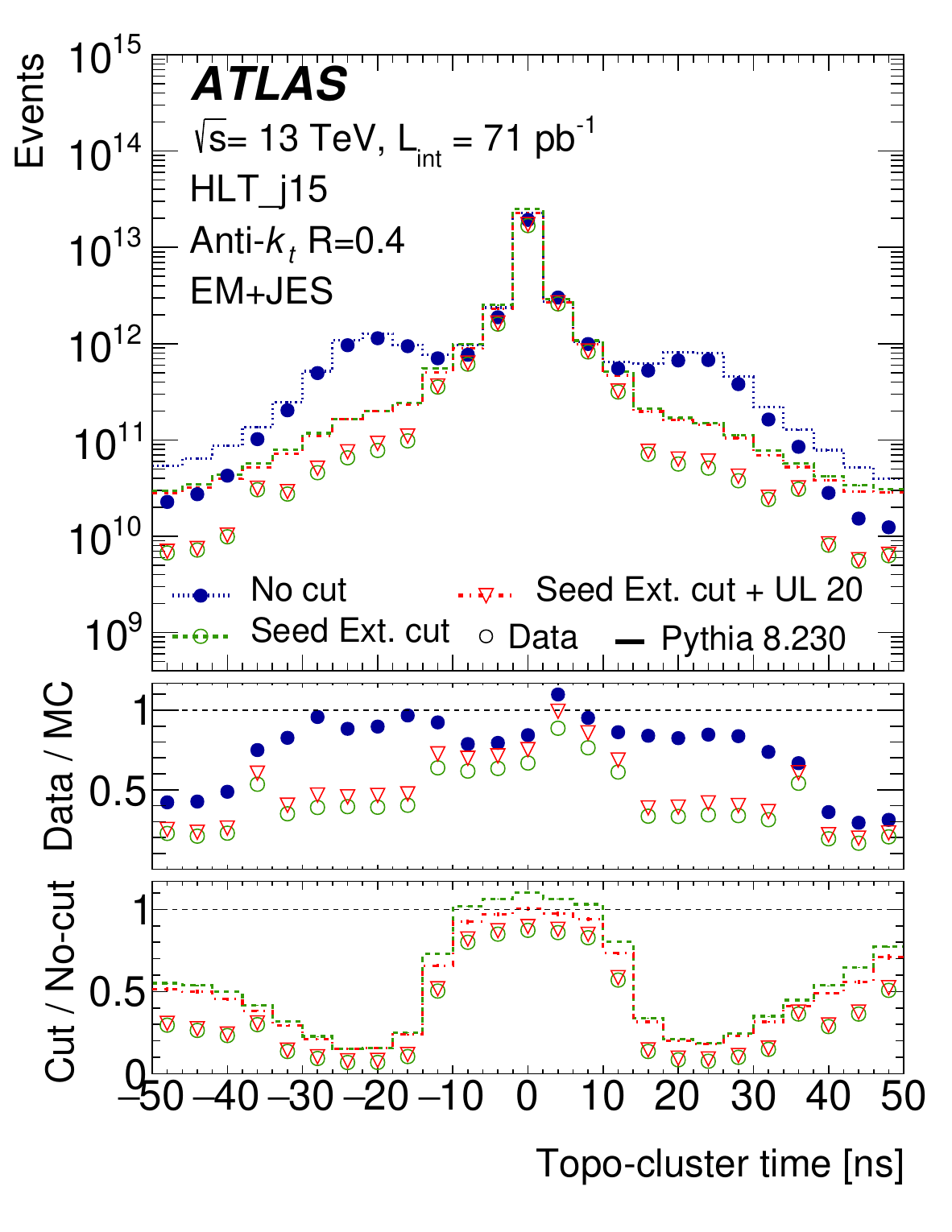}}
\subfigure[]{\label{subfig:HLTj25clusTPtMin5twoRatios}
\includegraphics[width=0.44\linewidth]{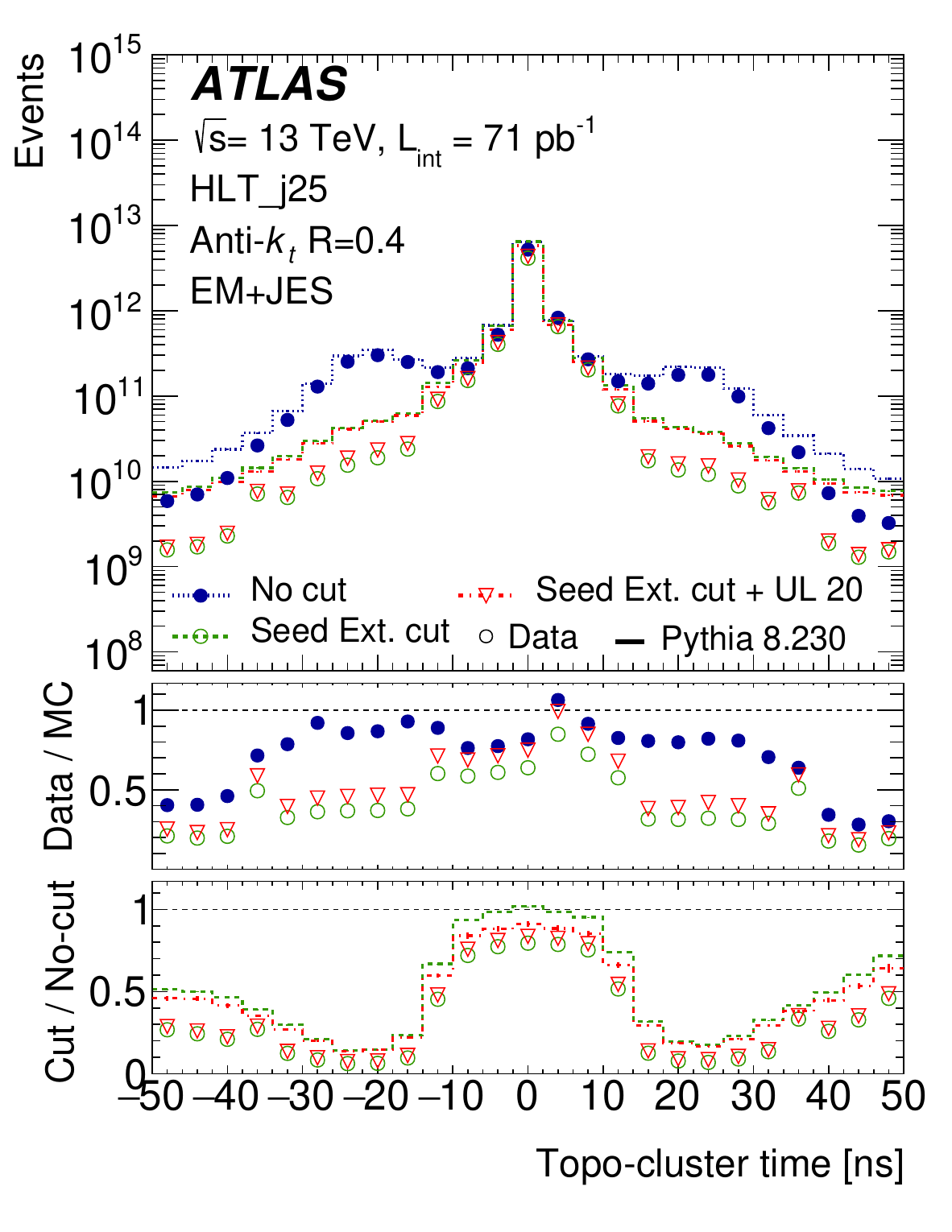}}
\subfigure[]{\label{subfig:HLTj45clusTPtMin5twoRatios}
\includegraphics[width=0.44\linewidth]{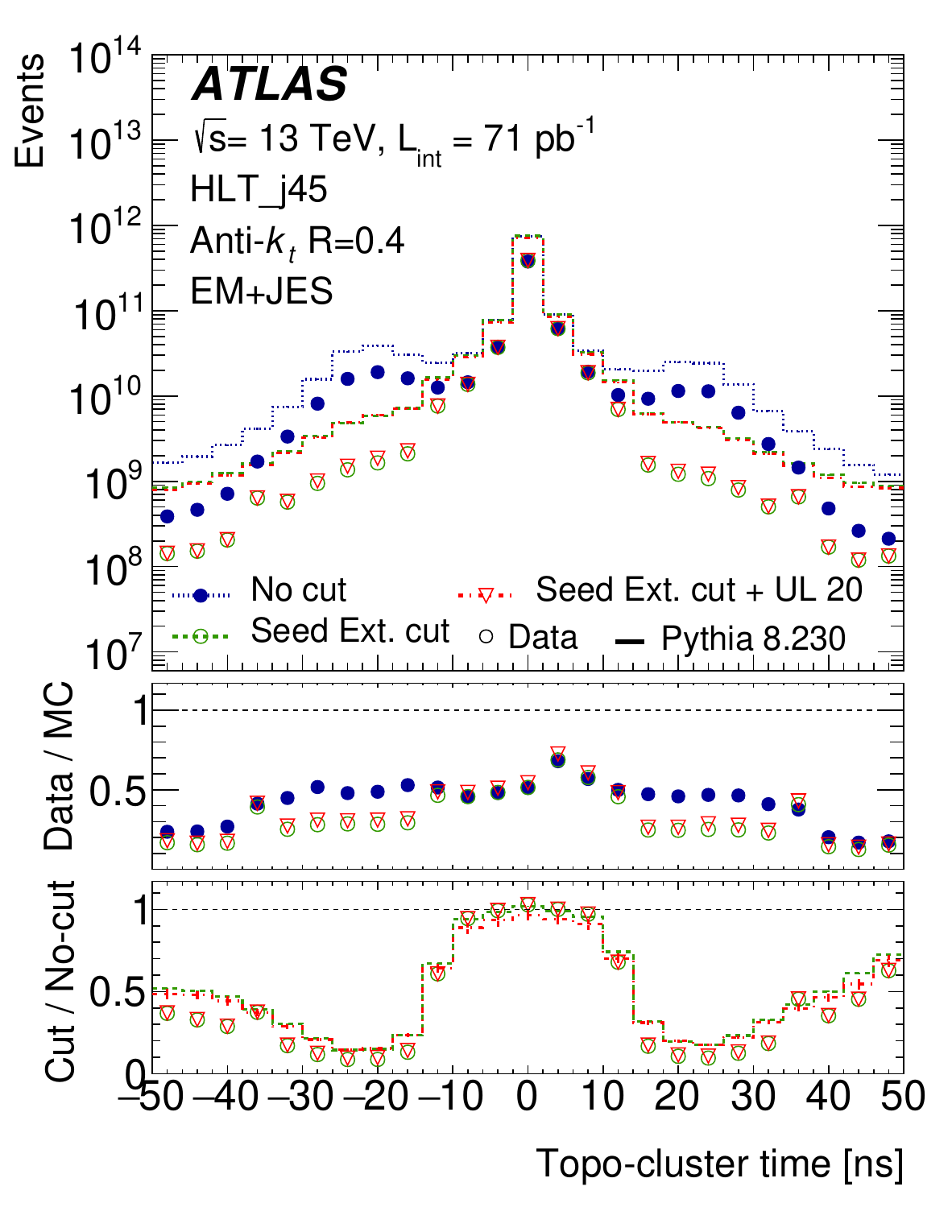}}
\subfigure[]{\label{subfig:HLTj45L1RD0FILLEDclusTPtMin5twoRatios}
\includegraphics[width=0.44\linewidth]{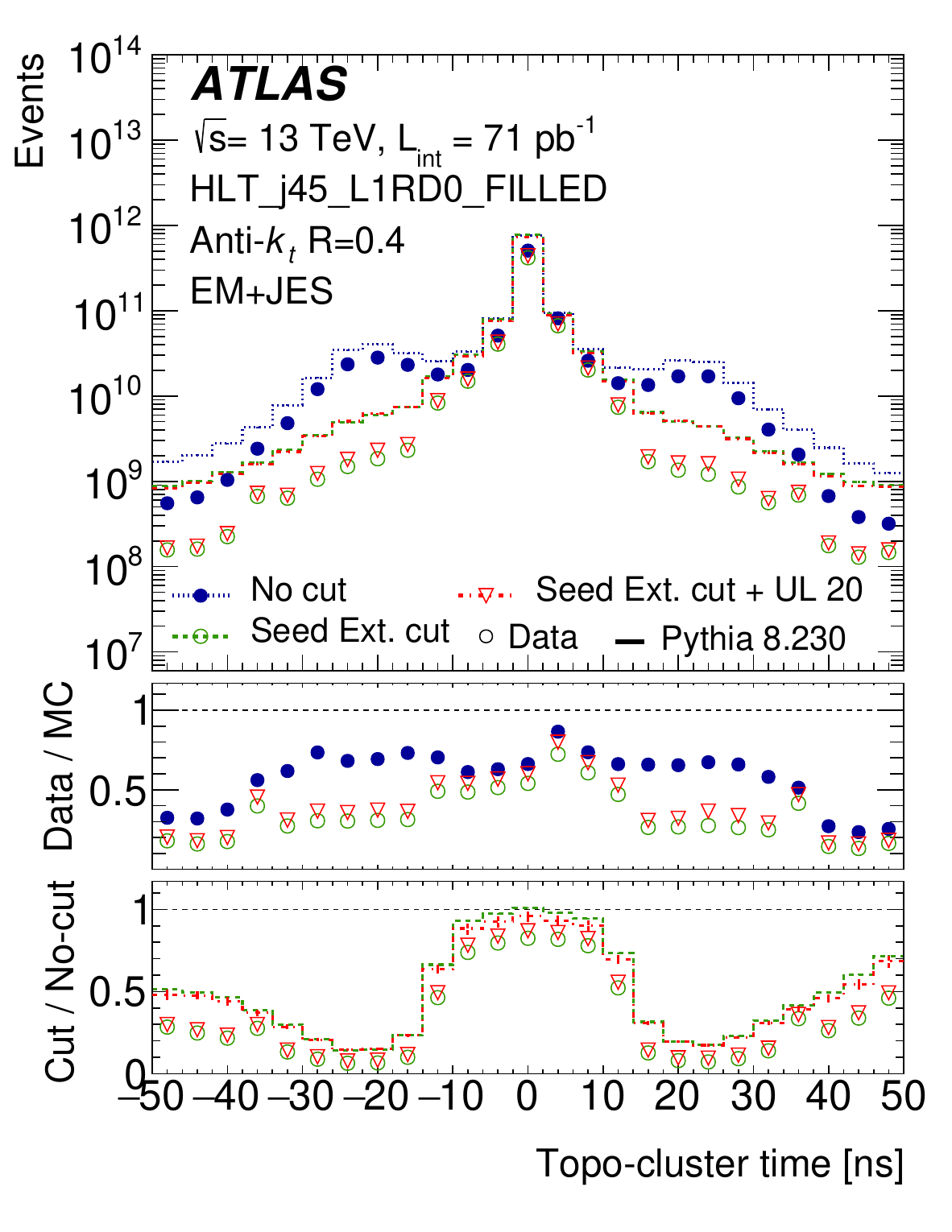}}
\caption{Topo-cluster time spectrum in data and MC selected multi-jet events. Four triggers are compared: \subref{subfig:HLTj15clusTPtMin5twoRatios} HLT\_j15, \subref{subfig:HLTj25clusTPtMin5twoRatios} HLT\_j25, \subref{subfig:HLTj45clusTPtMin5twoRatios} HLT\_j45, and \subref{subfig:HLTj45L1RD0FILLEDclusTPtMin5twoRatios} HLT\_j45\_L1RD0\_FILLED. The error bars convey the statistical and luminosity uncertainties. Uncertainties in the cut / no-cut ratios are obtained by splitting the available sample into subsamples and recomputing the cut / no-cut ratio for each subsample. The standard deviation of the distribution of the ratio is used to estimate the ratio's uncertainty. The luminosity uncertainty does not apply to the cut / no-cut ratio. In the plots, lines represent MC events and markers represent data, different colours and styles represent different cuts.}
\label{fig:dataMC_clusT}
\end{figure}

\begin{figure}
\centering
\subfigure[]{\label{subfig:HLTj15clusEtaPtMin5twoRatios}
\includegraphics[width=0.44\linewidth]{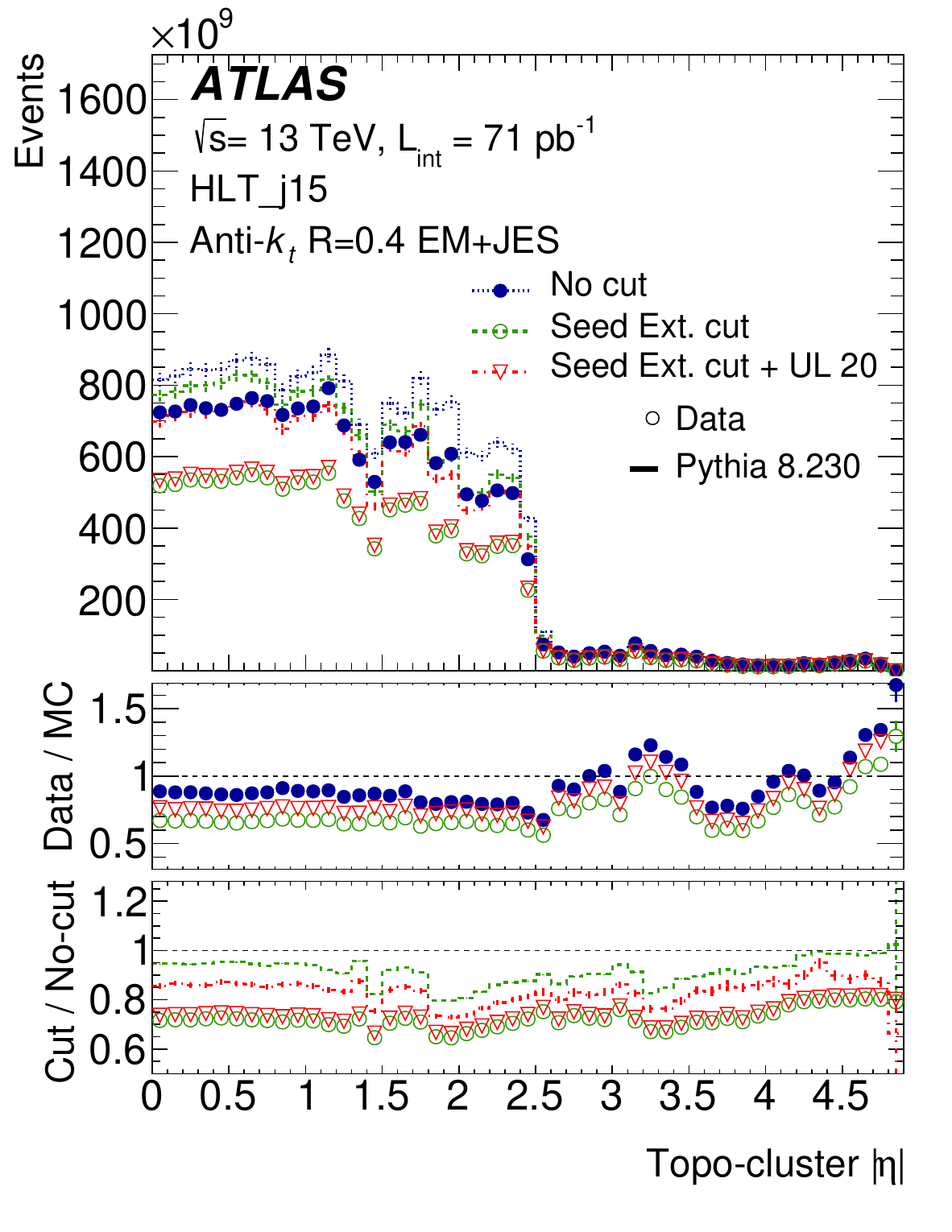}}
\subfigure[]{\label{subfig:HLTj25clusEtaPtMin5twoRatios}
\includegraphics[width=0.44\linewidth]{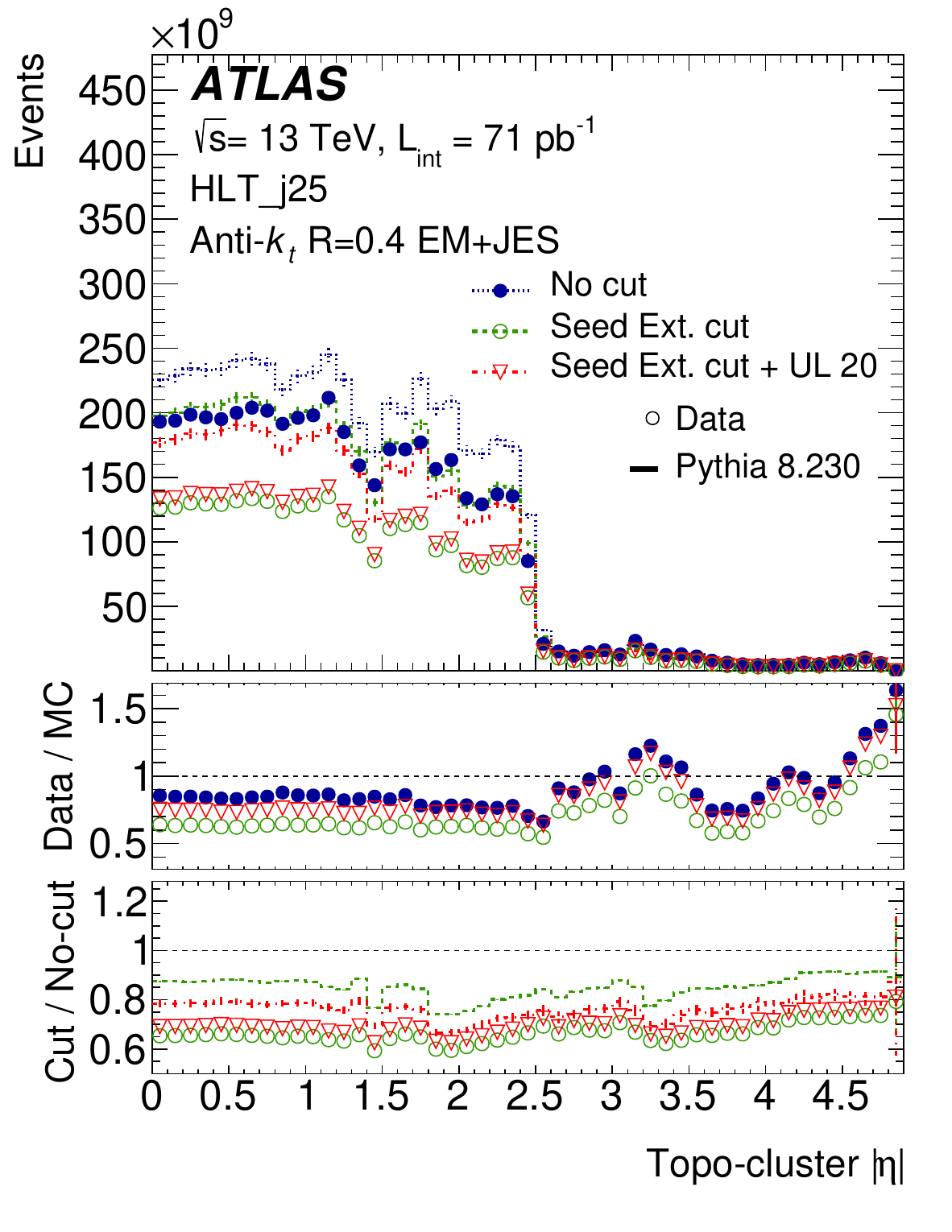}}
\subfigure[]{\label{subfig:HLTj45clusEtaPtMin5twoRatios}
\includegraphics[width=0.44\linewidth]{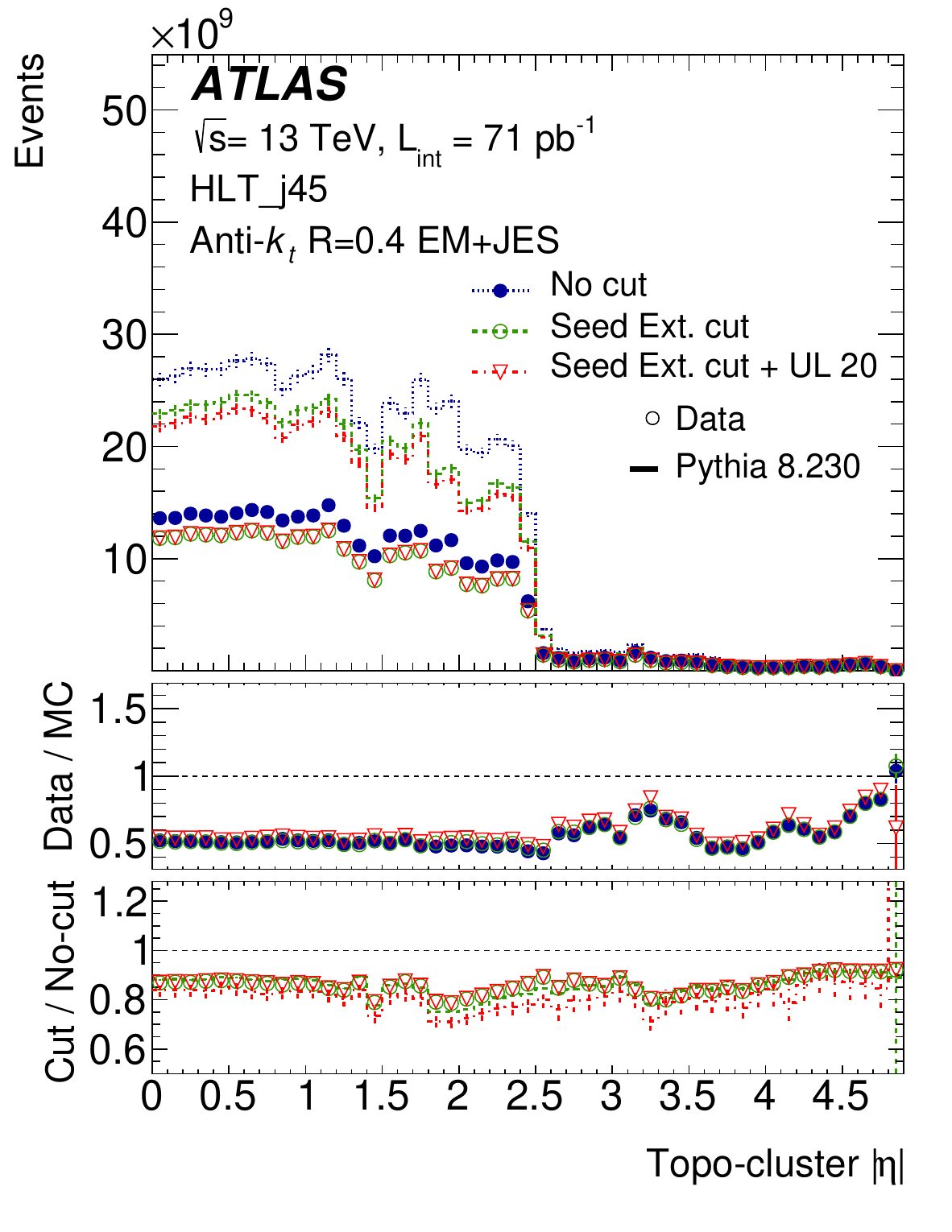}}
\subfigure[]{\label{subfig:HLTj45L1RD0FILLEDclusEtaPtMin5twoRatios}
\includegraphics[width=0.44\linewidth]{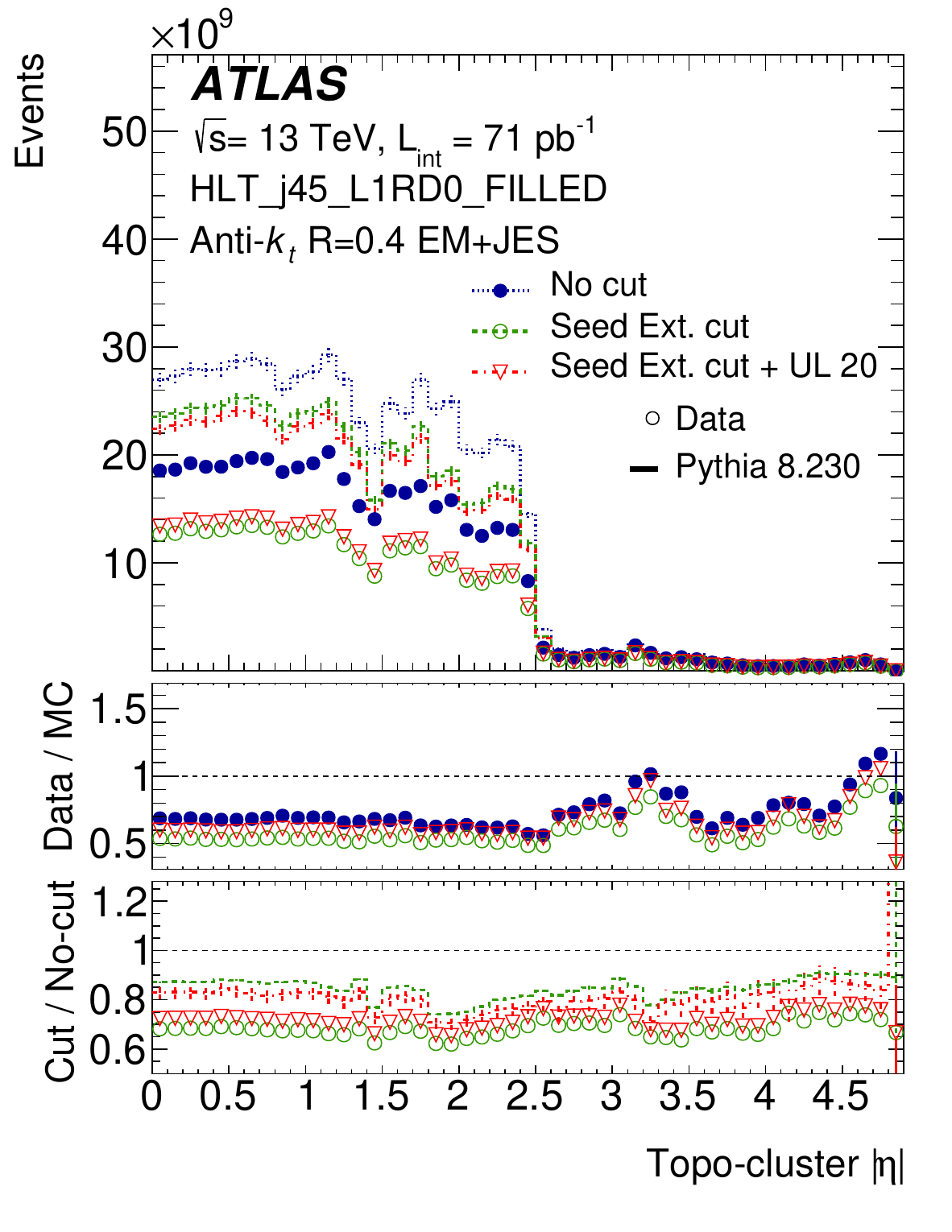}}
\caption{Topo-cluster $|\eta|$ spectrum in data and MC selected multi-jet events. Four triggers are compared: \subref{subfig:HLTj15clusEtaPtMin5twoRatios} HLT\_j15, \subref{subfig:HLTj25clusEtaPtMin5twoRatios} HLT\_j25, \subref{subfig:HLTj45clusEtaPtMin5twoRatios} HLT\_j45, and \subref{subfig:HLTj45L1RD0FILLEDclusEtaPtMin5twoRatios} HLT\_j45\_L1RD0\_FILLED. The error bars convey the statistical and luminosity uncertainties. Uncertainties in the cut / no-cut ratios are obtained by splitting the available sample into subsamples and recomputing the cut / no-cut ratio for each subsample. The standard deviation of the distribution of the ratio is used to estimate the ratio's uncertainty. The luminosity uncertainty does not apply to the cut / no-cut ratio. In the plots, lines represent MC events and markers represent data, different colours and styles represent different cuts.}
\label{fig:dataMC_clusEta}
\end{figure}

\begin{figure}
\centering
\subfigure[]{\label{subfig:HLTj15clusPtLogPtMin5twoRatios}
\includegraphics[width=0.44\linewidth]{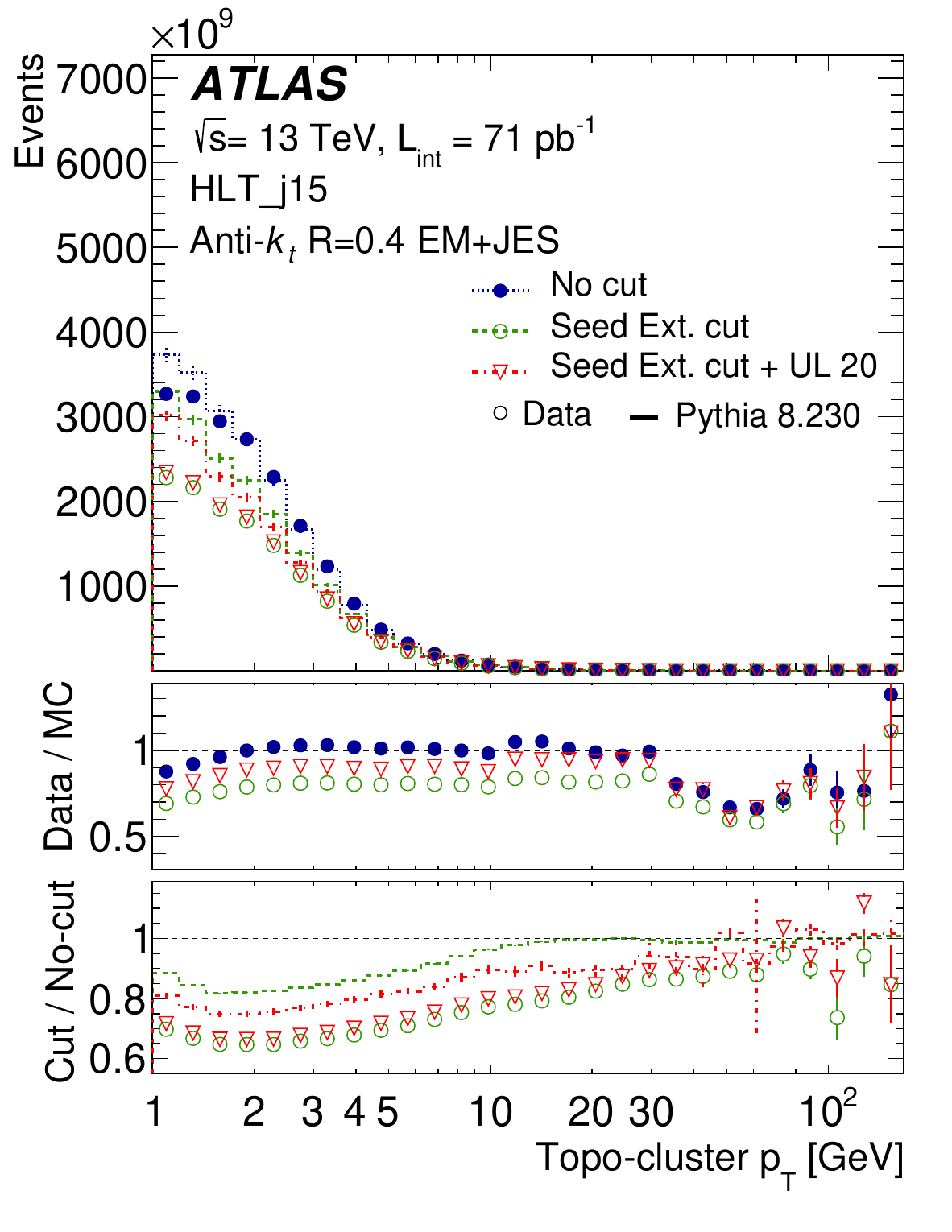}}
\subfigure[]{\label{subfig:HLTj25clusPtLogPtMin5twoRatios}
\includegraphics[width=0.44\linewidth]{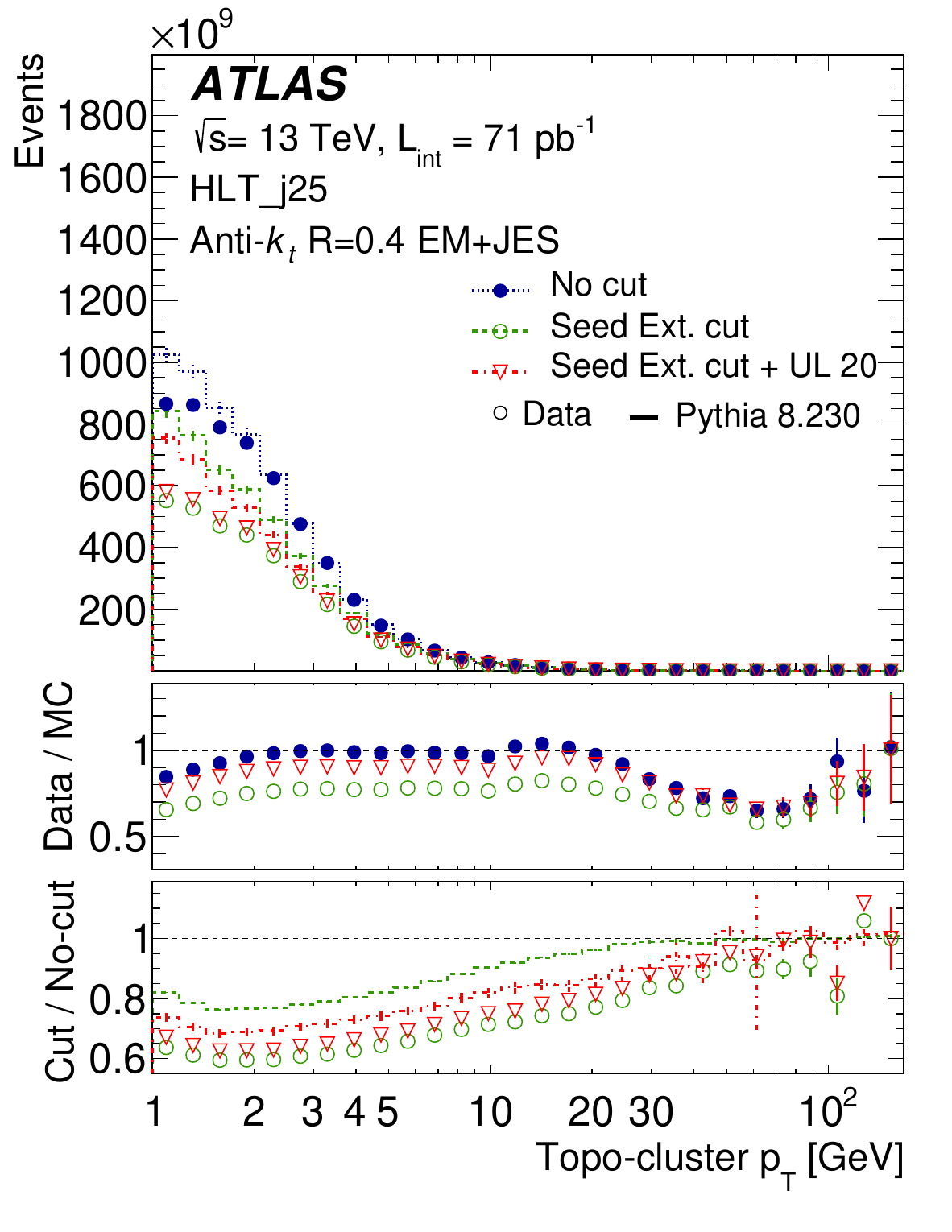}}
\subfigure[]{\label{subfig:HLTj45clusPtLogPtMin5twoRatios}
\includegraphics[width=0.44\linewidth]{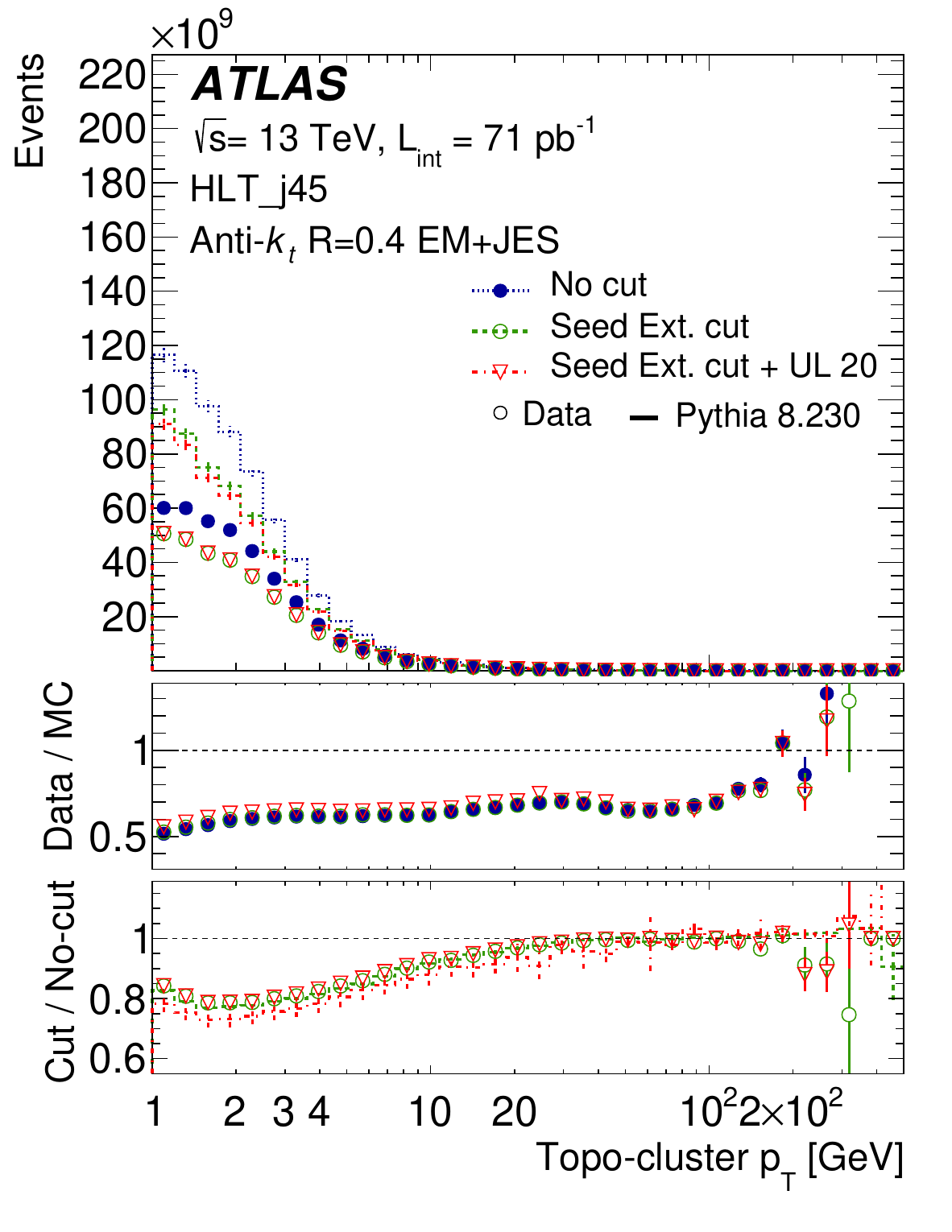}}
\subfigure[]{\label{subfig:HLTj45L1RD0FILLEDclusPtLogPtMin5twoRatios}
\includegraphics[width=0.44\linewidth]{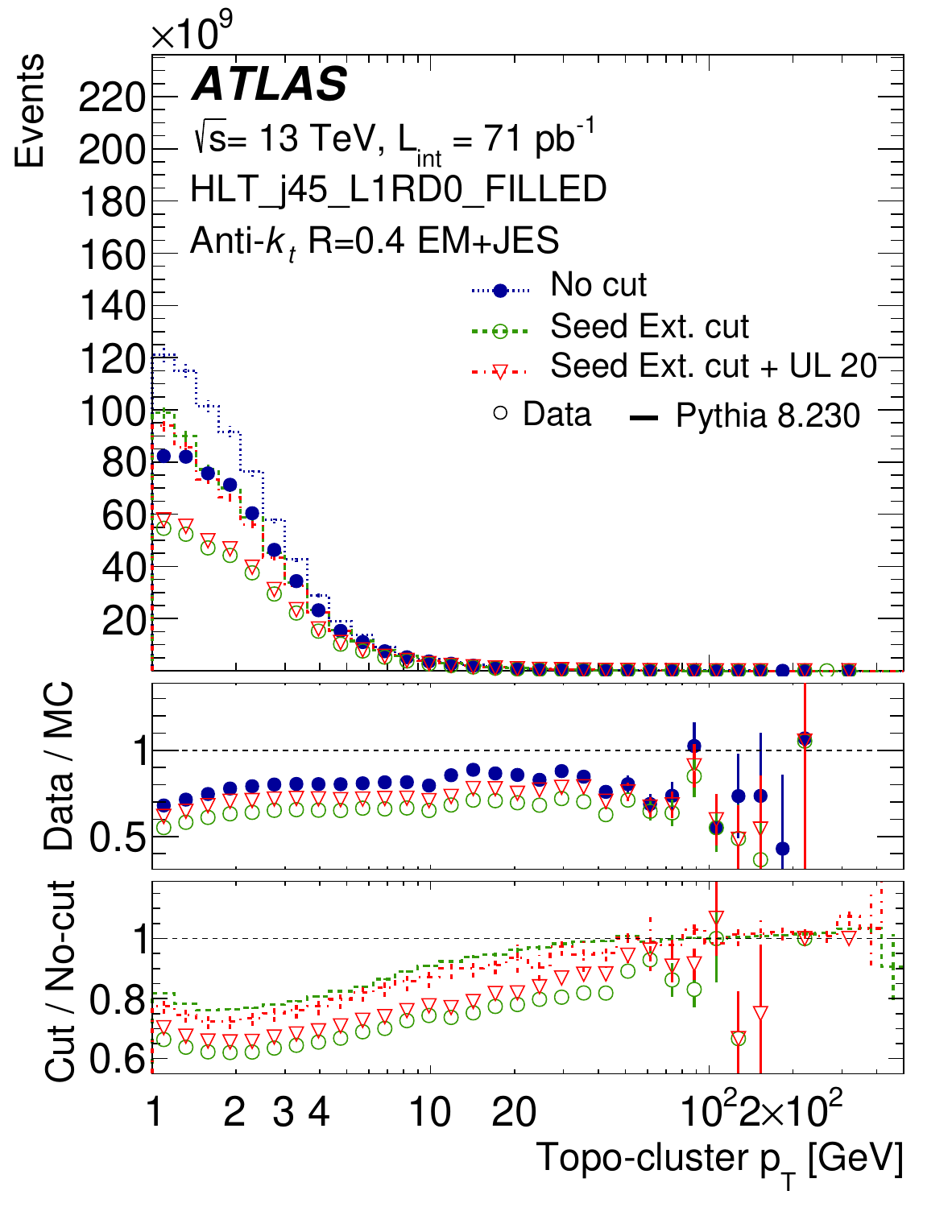}}
\caption{Topo-cluster \pT spectrum in data and MC selected multi-jet events. Four triggers are compared: \subref{subfig:HLTj15clusPtLogPtMin5twoRatios} HLT\_j15, \subref{subfig:HLTj25clusPtLogPtMin5twoRatios} HLT\_j25, \subref{subfig:HLTj45clusPtLogPtMin5twoRatios} HLT\_j45, and \subref{subfig:HLTj45L1RD0FILLEDclusPtLogPtMin5twoRatios} HLT\_j45\_L1RD0\_FILLED. The error bars convey the statistical and luminosity uncertainties. Uncertainties in the cut / no-cut ratios are obtained by splitting the available sample into subsamples and re-omputing the cut / no-cut ratio for each subsample. The standard deviation of the distribution of the ratio is used to estimate the ratio's uncertainty. The luminosity uncertainty does not apply to the cut / no-cut ratio. In the plots, lines represent MC events and markers represent data, different colours and styles represent different cuts.}
\label{fig:dataMC_clusPt}
\end{figure}

Data and simulated events are required to contain at least one primary vertex; the one with the highest sum of squared track \pT is considered in this section. In addition, at least one jet must satisfy a minimum \pT requirement, as listed in Table~\ref{tab:triggerset}. This requirement is needed to restrict selected events to a phase space where the trigger efficiency is within ${\sim}20\%$ of its plateau value. A fully efficient trigger is not necessary for this study, as the time cut is not applied at trigger level and a smaller trigger efficiency would not affect the comparison between the cut and no-cut scenarios. In order to properly compare data and MC events, jets are calibrated using the full calibration chain, including the \textit{in situ} correction.\footnote{As in Section~\ref{sec:mcperf}, the residual \pileup correction is calculated after applying the time cut.} All jets are required to have $\pT > 20~\GeV$.
The Seed Extended cut and Seed Extended plus $X_{\textrm{UL}}=20$ cut are compared with no-cut spectra for data and MC events. Figures~\ref{fig:dataMC_leadpt}, \ref{fig:dataMC_pt} and~\ref{fig:dataMC_eta} show different jet kinematic properties, while topo-cluster properties are shown in Figures~\ref{fig:dataMC_clusT}, \ref{fig:dataMC_clusEta} and~\ref{fig:dataMC_clusPt}. Each histogram shows the distributions for events selected by a different trigger.
Uncertainties shown for MC distributions are a combination of the MC statistical uncertainty and the luminosity uncertainty. The latter is based on the 2.4\% luminosity uncertainty in 2017~\cite{DAPR-2021-01}. Uncertainties shown for data are statistical only.
 
The data and MC distributions are not in perfect agreement, and exhibit relatively uniform differences of about 30\%--40\%. This is to be expected as the MC sample used is a dijet simulation at leading order: even though the hadronisation and jet characteristics are well described by  \PYTHIA[8.230], the overall MC normalisation is likely to be too low due to the missing higher orders. Moreover, low-\pT events are dominated by soft QCD radiation, which is known to be difficult to model.
A complete jet cross-section measurement would have to use an improved simulation and consider additional correction factors. Since this study only seeks to verify that the time cut's behaviour in data reflects what is observed in MC simulation, this level of agreement is considered acceptable.
 
The effect of the time cut depends on the trigger. Events passing HLT\_j45, which has both a higher \pT threshold and a non-random L1 seed, are not affected at all by the time cut.
The numbers of events passing the other three triggers exhibit a consistent reduction of ${\sim}20\%$ for values of the leading-jet \pT between $30$ and $50~\GeV$ (Figure~\ref{fig:dataMC_leadpt}). Effects of similar size can be seen throughout the tested kinematic spectra.
 
Except for the HLT\_j45 trigger, the time cut tends to have a more pronounced effect on data than on MC events, the difference being of $\mathcal{O}(10\%)$ in most of the phase space. This is most likely due to the simulation not accounting for events consisting exclusively of \pileup. Such events are expected to be present in non-negligible numbers for the other three triggers, which have relatively low \pT thresholds and are seeded by random events.
In addition, the MC simulation does not reproduce bunch-to-bunch luminosity fluctuations, which have a more significant effect on the reconstruction of out-of-time signals~\cite{PERF-2014-07}.
 
% End of text imported from the .//DATAperformance_pap.tex input file

\FloatBarrier

% The next lines are included from the .//ineffCheck_pap.tex input file
\subsection{Checks for time cut inefficiencies}
The time cut relies on the accurate measurement of signal timing provided by the calorimeters. The precision of the calorimeter timing measurement is guaranteed by periodic realignment and constant monitoring, as discussed in Section~\ref{sec:calotiming}.
As a further safety check, the robustness of the cut against a local time miscalibration was tested. The events selected for this check were those in which one calorimeter channel had been flagged as producing fake out-of-time signals due to cross-talk. The effect of the time cut on particles hitting the affected region was then compared with the effect of the time cut on particles hitting unaffected regions of the same subdetector.
 
\begin{figure}
\centering
\subfigure[]{\label{subfig:2DLArEnone}
\includegraphics[width=0.4\linewidth]{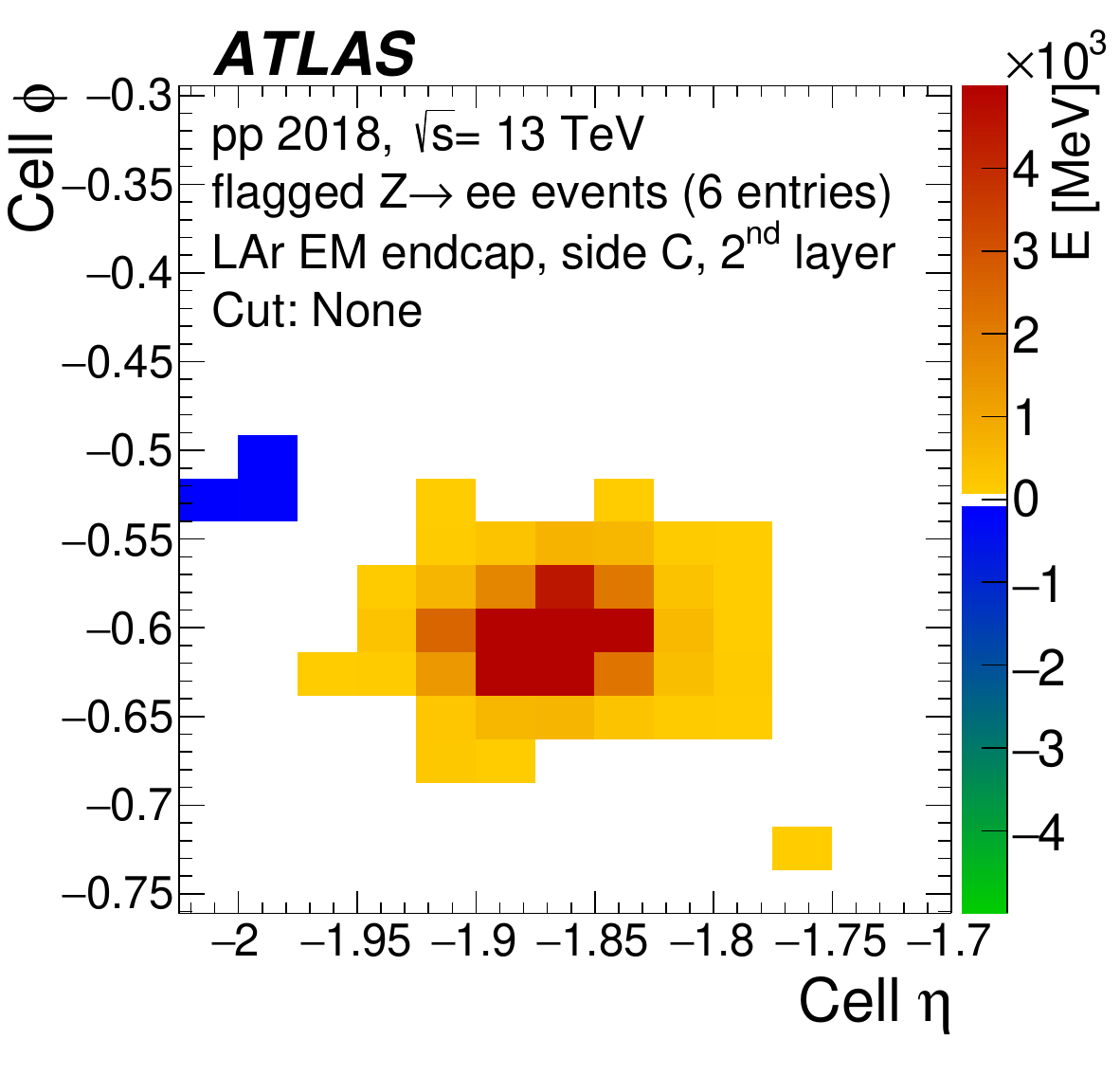}}
\subfigure[]{\label{subfig:2DLArEseedext}
\includegraphics[width=0.4\linewidth]{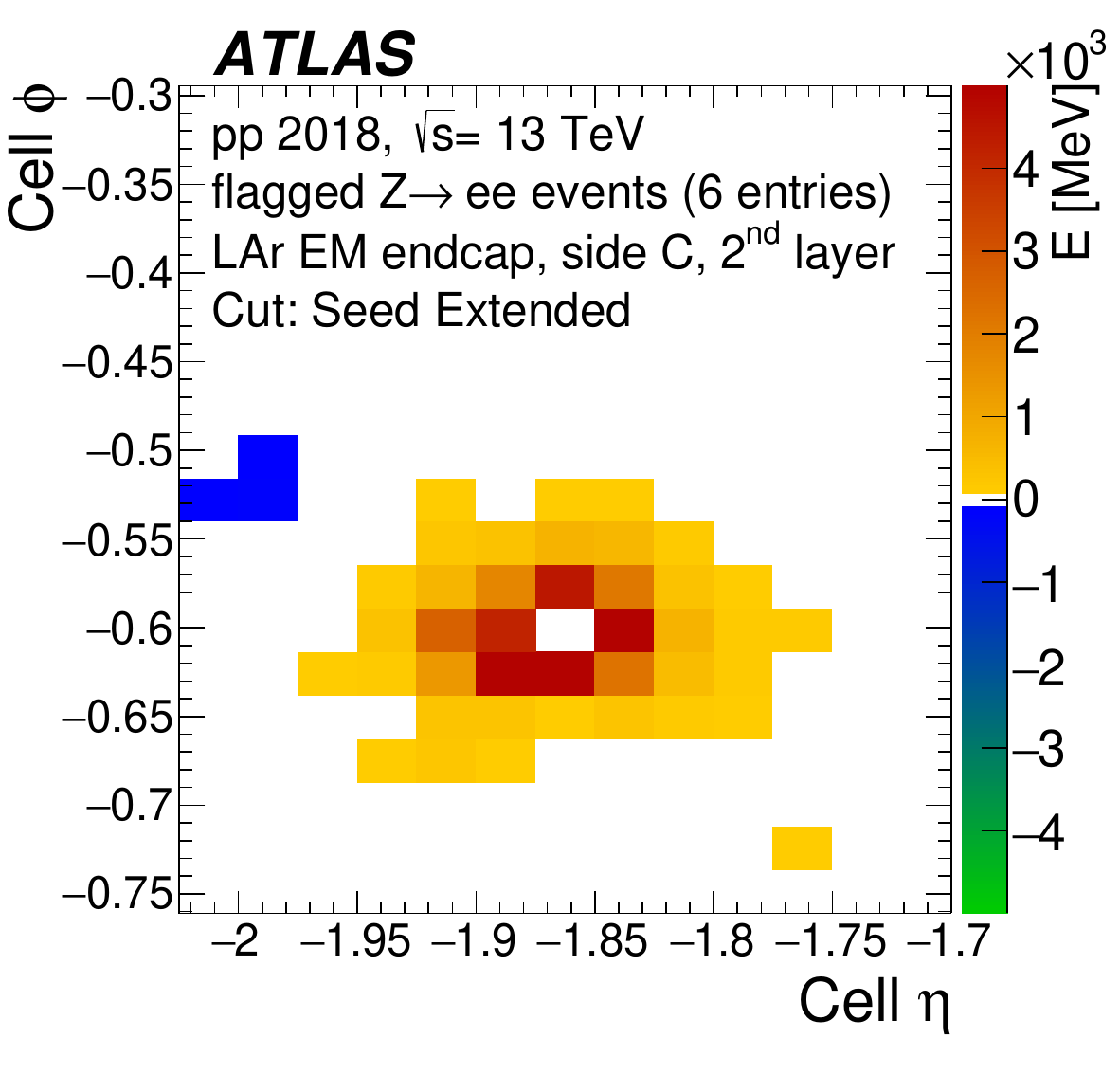}}
\subfigure[]{\label{subfig:2DLArEseedextul20}
\includegraphics[width=0.4\linewidth]{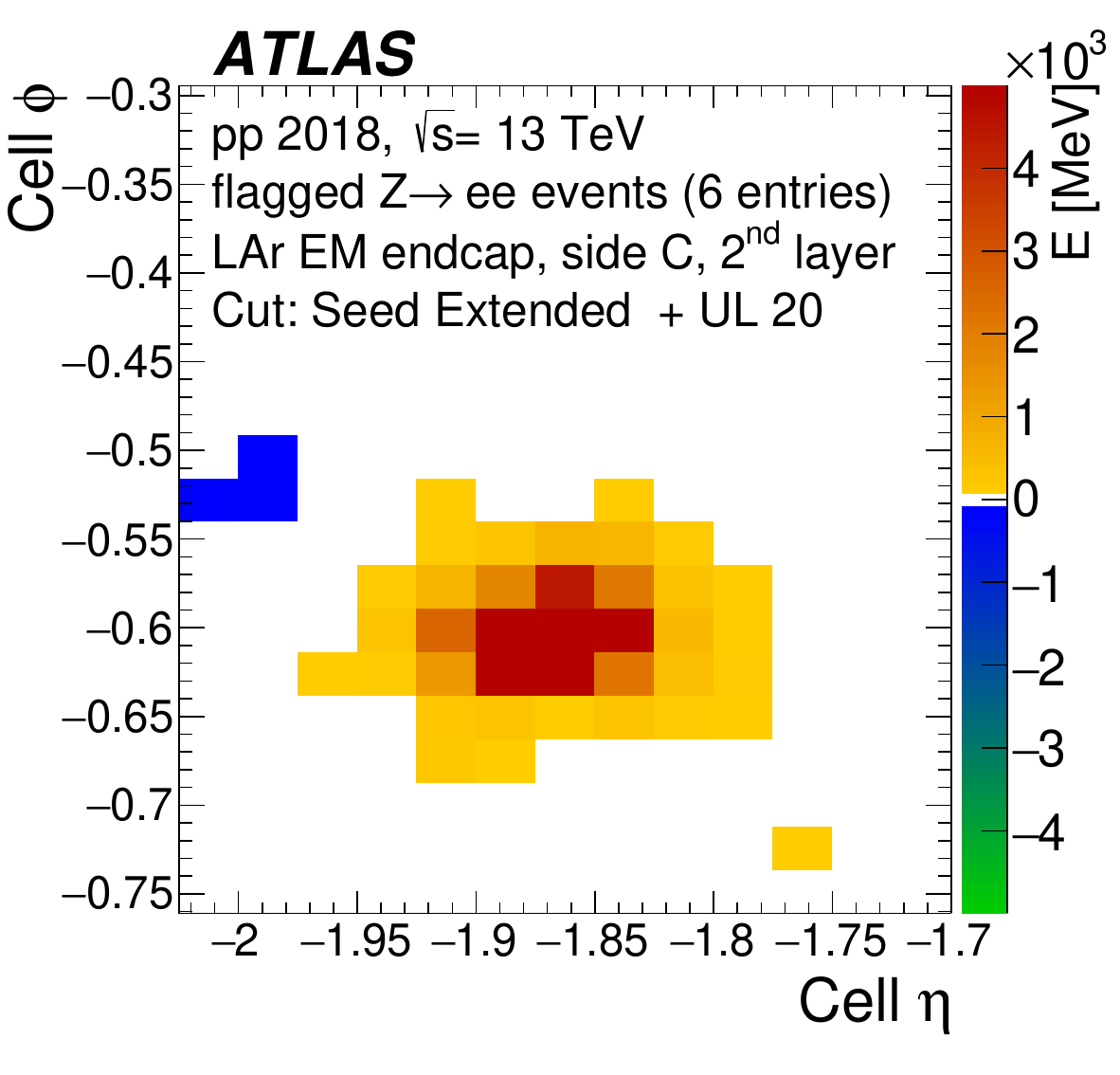}}
\caption{The $\eta \times \phi$ distribution of cells belonging to topo-clusters in the second layer of the LAr EM endcap calorimeter, side C. The colour scale represents the recorded energy. Six events, in which $Z\rightarrow ee$ electrons hit the affected area, are averaged. The three panels show results with \subref{subfig:2DLArEnone} no time cut, \subref{subfig:2DLArEseedext} the Seed Extended cut and \subref{subfig:2DLArEseedextul20} the Seed Extended plus $X_{\textrm{UL}}=20$ cut.}
\label{fig:2DLArE}
\end{figure}

The affected channel was located in the second layer of the LAr EM endcap calorimeter, on the C-side (along the negative $z$-axis). As an example, Figure~\ref{fig:2DLArE} shows a combination of six events in which an electron from a $Z\rightarrow ee$ decay hit the affected area. Events were reconstructed with no cut, the Seed Extended cut and the Seed Extended plus $X_{\textrm{UL}}=20$ cut.
These events were identified first by applying a standard $Z\rightarrow ee$ event selection. Events are required to pass photon triggers analogous to those used in diphoton resonance searches~\cite{SUSY-2019-14, SUSY-2020-28} and to contain at least two electrons with $\pT > 10$~\GeV\ and $|\eta| < 2.47$ (excluding the transition region $1.37 < |\eta| < 1.52$ between the LAr EM barrel and endcap calorimeters), satisfying the Medium identification selection and FCLoose isolation criteria~\cite{EGAM-2018-01}. The reconstructed track matched to each electron candidate must be consistent with having originated from the primary vertex: its longitudinal impact parameter $z_0$ and transverse impact parameter $d_0$ must satisfy $|z_0 \cdot \sin{\theta}| < 0.5$~mm and $|d_0|/\sigma_{d_0} < 5$~mm respectively.
Furthermore, the dielectron pair is required to have an invariant mass $m_{ee}$ satisfying $68~\GeV < m_{ee} < 108~\GeV$ and a pseudorapidity separation $|\Delta\eta(e_1, e_2)| > 0.1$. Electrons hitting the affected channel were then flagged by requiring that the most energetic cell in the second layer of the EM calorimeter has high enough energy ($E_{\text{cell}}^{\text{max}}>5~\GeV$) and its measured time satisfies $|t_{\text{cell}}^{\text{max}}|>12.5~\textrm{ns}$.
Figure~\ref{fig:2DLArE} shows that the Seed Extended time cut does reject the energy from one spot in the centre of the topo-cluster. Despite this, the overall topo-cluster is not lost. Moreover, the addition of the Upper Limit restores most of the lost energy, so that no cell is entirely missing from the electron topo-cluster.
 
To better quantify the effect, the average $\eta \times \phi$ location of flagged $Z\rightarrow ee$ electrons was used to define a $3 \times 3$-cell \textit{Test Region} containing the affected channel in the C-side LAr EM endcap's second layer. The $\eta \times \phi$ ranges corresponding to this area are listed in Table~\ref{tab:regions}.
In order to compare the time cut's effect in the Test Region with the normal time-cut behaviour, three \textit{Control Regions} were also defined by inverting the sign of either $\eta$, $\phi$ or both in the Test Region's definition. The Control Regions'  $\eta \times \phi$ boundaries are also listed in Table~\ref{tab:regions}.
\begin{table}[htbp]
\caption{Definition of Test and Control Regions.}
\label{tab:regions}
\begin{center}
\begin{tabular}{lcc}
\toprule
& $\eta$ range & $\phi$ range \\
\midrule
Test Region & $[-1.9,-1.825]$ & $[-0.638,-0.565]$ \\
Control Region 1 & $[+1.9,+1.825]$ & $[-0.638,-0.565]$ \\
Control Region 2 & $[-1.9,-1.825]$ & $[+0.638,+0.565]$ \\
Control Region 3 & $[+1.9,+1.825]$ & $[+0.638,+0.565]$ \\
\bottomrule
\end{tabular}
\end{center} \end{table}
 
Events from one Luminosity Block of 2018 data, during which the Test Region was known to be affected by cross-talk, were used to test the time cut's behaviour. Rather than restricting the set of events to those passing a $Z\rightarrow ee$ selection, a more generic event selection is applied: for each region, events are considered if at least one cell of that region has a recorded transverse momentum of $\pT > 10~\GeV$.
 
Four variables of interest were considered. The cell occupancy $N^\textrm{cells}$ and the cells' total energy $E^\textrm{cells}$ are respectively defined as the number and total energy of cells in a given region that are part of a topo-cluster: they quantify whether single cells are lost due to the miscalibrated channel. The topo-cluster occupancy $N^\textrm{clus}$ and topo-cluster total energy $E^\textrm{clus}$ are the number and total energy of reconstructed topo-clusters that share at least one cell with a given region: they provide a way to evaluate whether entire topo-clusters are lost or their energy is significantly impacted by the time cut.
In order to better quantify the effect of the time cut, the fraction $\mathcal{F}_{\text{unchanged}}$ of events for which a given variable in a given region is unchanged by the cut is computed. For each event, $N^{\textrm{cells}}$ ($N^{\textrm{clus}}$) is considered unchanged if the number of cells in the region (topo-clusters overlapping with the region) remains the same when applying the time cut. The total energies $E^{\textrm{cells}}$ and $E^{\textrm{clus}}$ are considered unchanged if the difference between the total energies computed with and without the time cut ($E^{\textrm{cells}(\textrm{clus})}_{\text{cut}}$ and $E^{\textrm{cells}(\textrm{clus})}_{\text{no cut}}$ respectively) satisfies:
\begin{equation*}
|E^{\textrm{cells}(\textrm{clus})}_{\text{cut}} - E^{\textrm{cells}(\textrm{clus})}_{\text{no cut}}| < \sigma(E^{\textrm{cells}(\textrm{clus})}_{\text{no cut}}),
\end{equation*}
where $\sigma(E^{\textrm{cells}(\textrm{clus})}_{\text{no cut}})$ is the expected calorimeter resolution computed for either the cells' total energy ($E^{\textrm{cells}}$) or total cluster energy ($E^{\textrm{clus}}$) without any cut. The resolution is taken from Ref.~\cite{PERF-2007-01} to be:
\begin{equation*}
\frac{\sigma(E^{\textrm{cells}(\textrm{clus})}_{\text{no cut}})}{E^{\textrm{cells}(\textrm{clus})}_{\text{no cut}}} = \frac{10\%}{\sqrt{|E^{\textrm{cells}(\textrm{clus})}_{\text{no cut}}|}} \oplus 0.7\%.
\end{equation*}
 
\begin{figure}
\centering
\subfigure[]{\label{subfig:cellOccupancySE}
\includegraphics[width=0.49\linewidth]{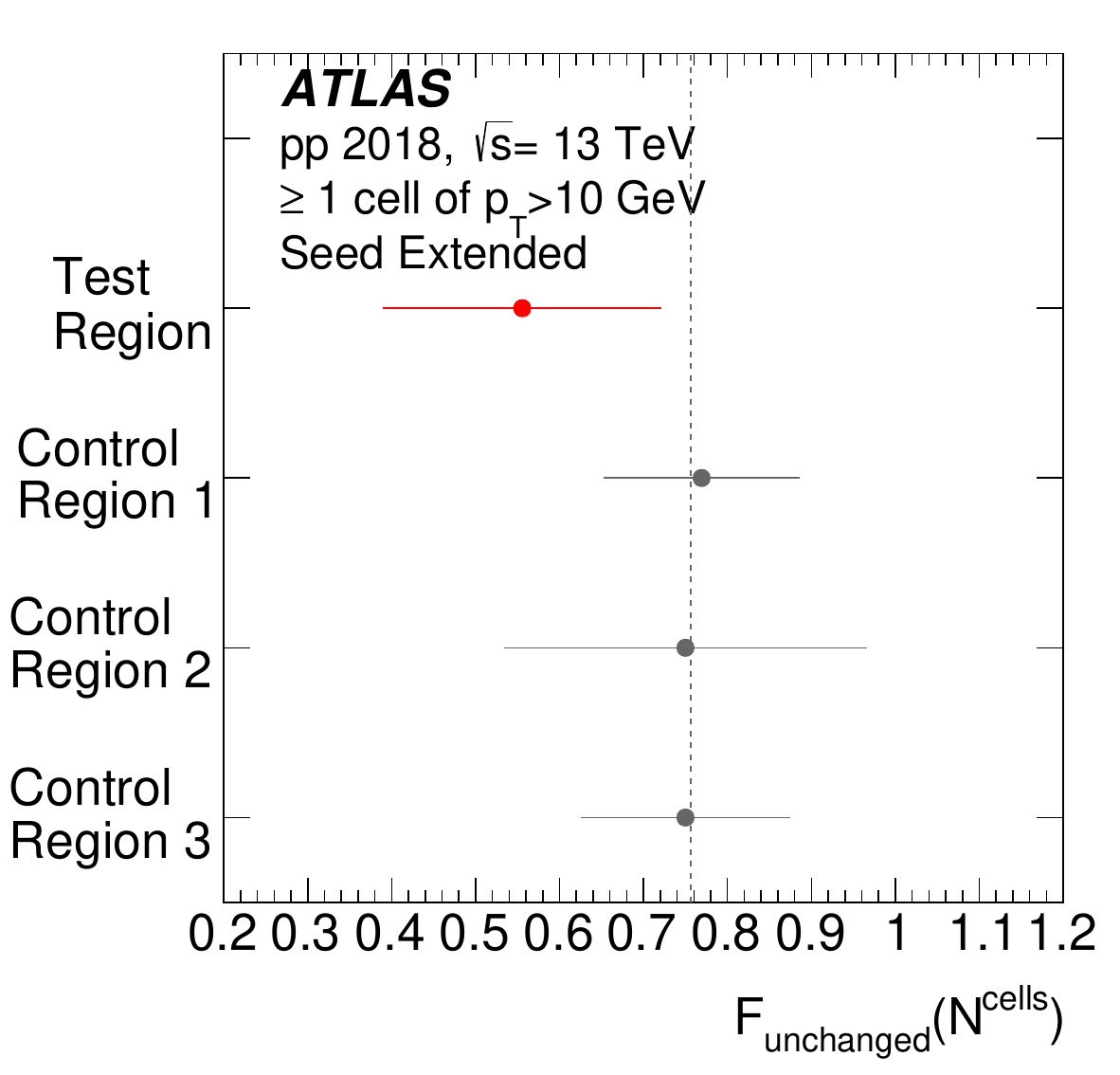}}
\subfigure[]{\label{subfig:cellOccupancySEUL}
\includegraphics[width=0.49\linewidth]{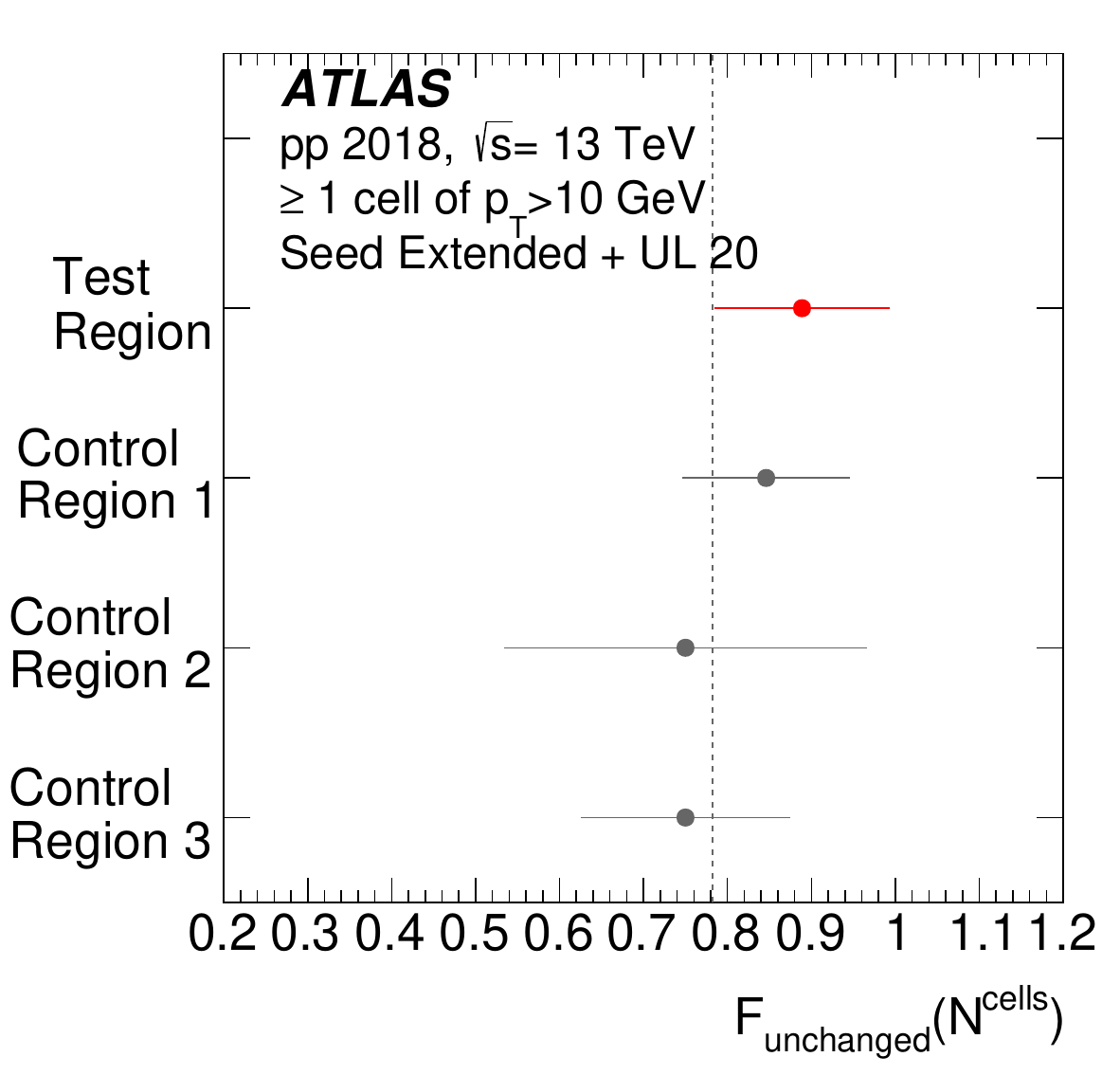}}
\subfigure[]{\label{subfig:cellTotESE}
\includegraphics[width=0.49\linewidth]{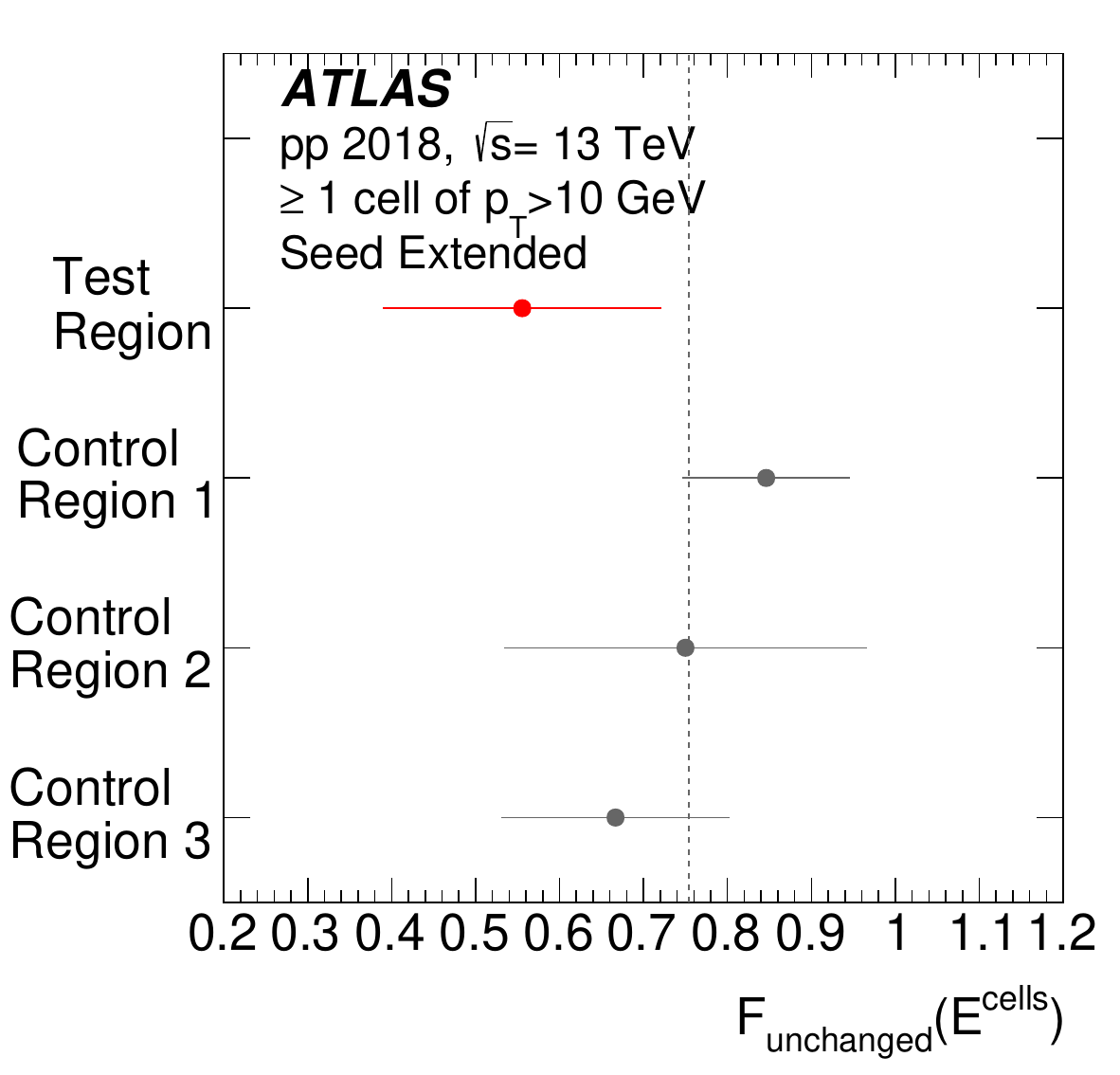}}
\subfigure[]{\label{subfig:cellTotESEUL}
\includegraphics[width=0.49\linewidth]{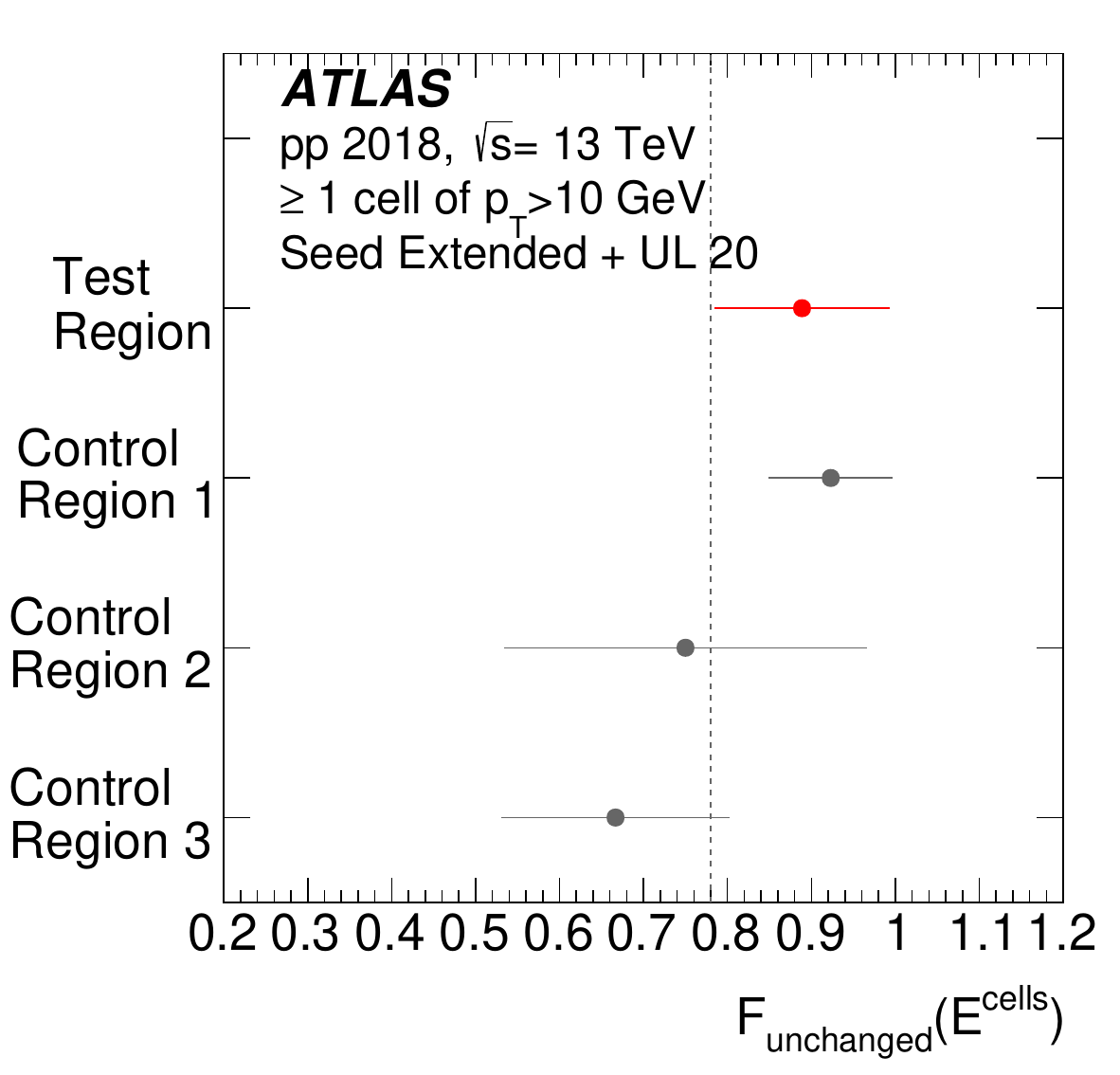}}
\caption{Comparison between the Test Region and three Control Regions for \subref{subfig:cellOccupancySE} the fraction of events in which the cell occupancy $N^\textrm{cells}$ is left unchanged by the Seed Extended cut, \subref{subfig:cellOccupancySEUL} the fraction of events in which $N^\textrm{cells}$ is left unchanged by the Seed Extended plus $X_{\textrm{UL}}=20$ cut, \subref{subfig:cellTotESE} the fraction of events in which the cells' total energy $E^\textrm{cells}$ is left unchanged by the Seed Extended cut, and~\subref{subfig:cellTotESEUL} the fraction of events in which $E^\textrm{cells}$ is left unchanged by the Seed Extended plus $X_{\textrm{UL}}=20$ cut. The cell occupancy (cells' total energy) in a given region is defined as the number (energy sum) of cells in the region that are included in a topo-cluster. Uncertainties on the event fractions are computed as the standard deviation of a binomial distribution. The dashed line represents the average of the three Control Regions.}
\label{fig:REGocc_cells_pt100}
\end{figure}

\begin{figure}
\centering
\subfigure[]{\label{subfig:clusOccupancy_acell_SE}
\includegraphics[width=0.49\linewidth]{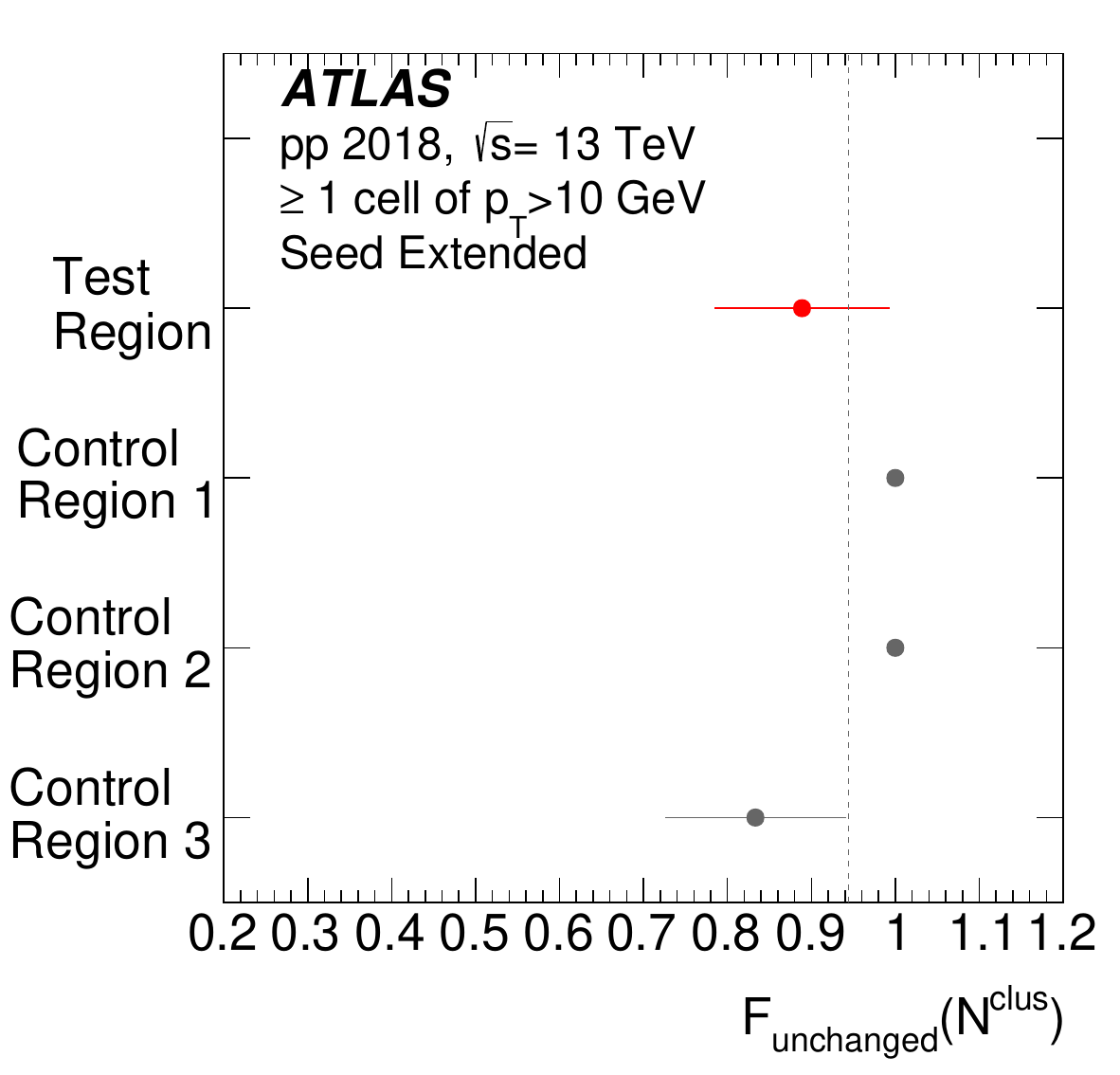}}
\subfigure[]{\label{subfig:clusOccupancy_acell_SEUL}
\includegraphics[width=0.49\linewidth]{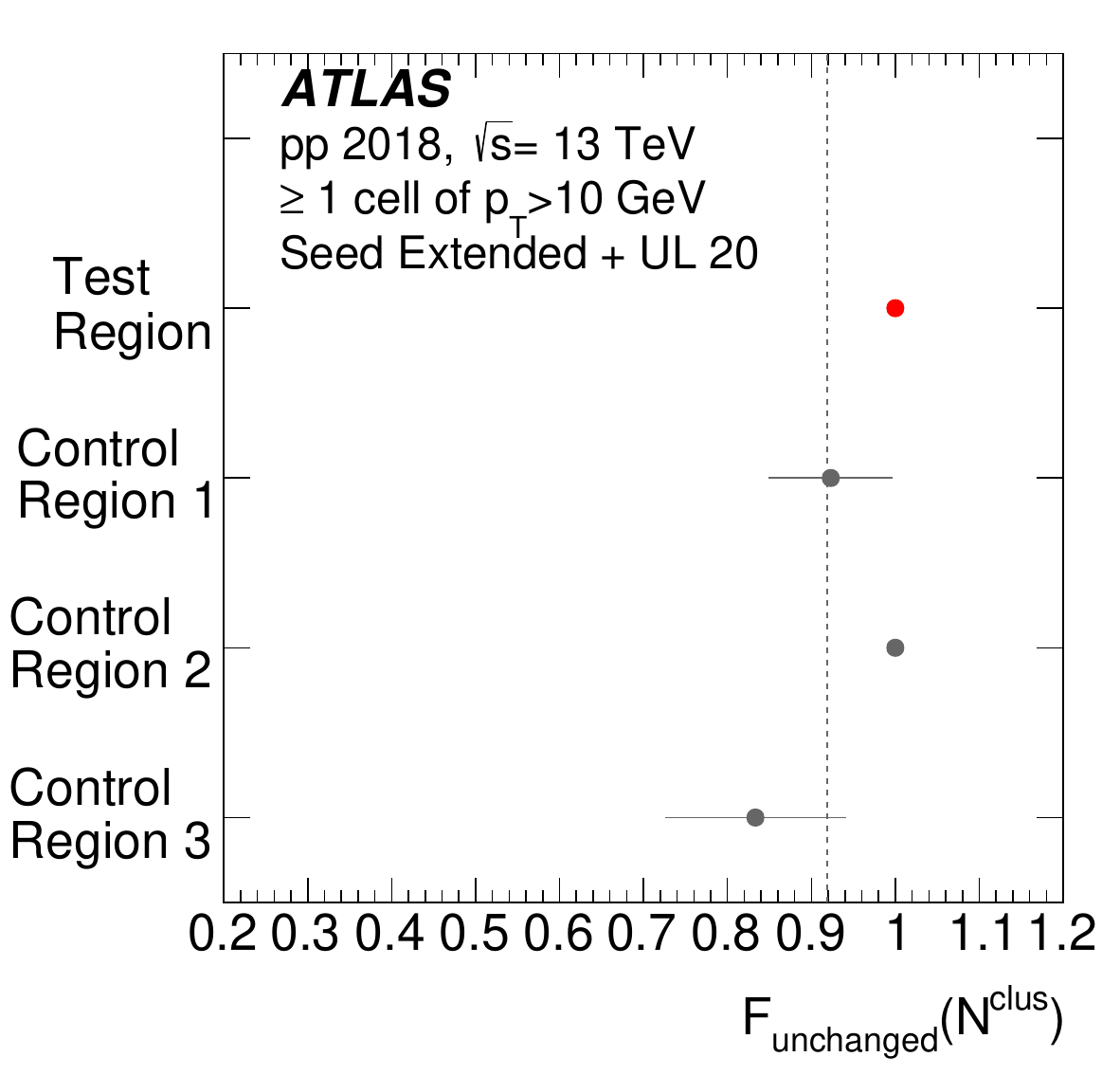}}
\subfigure[]{\label{subfig:clusTotE_acell_SE}
\includegraphics[width=0.49\linewidth]{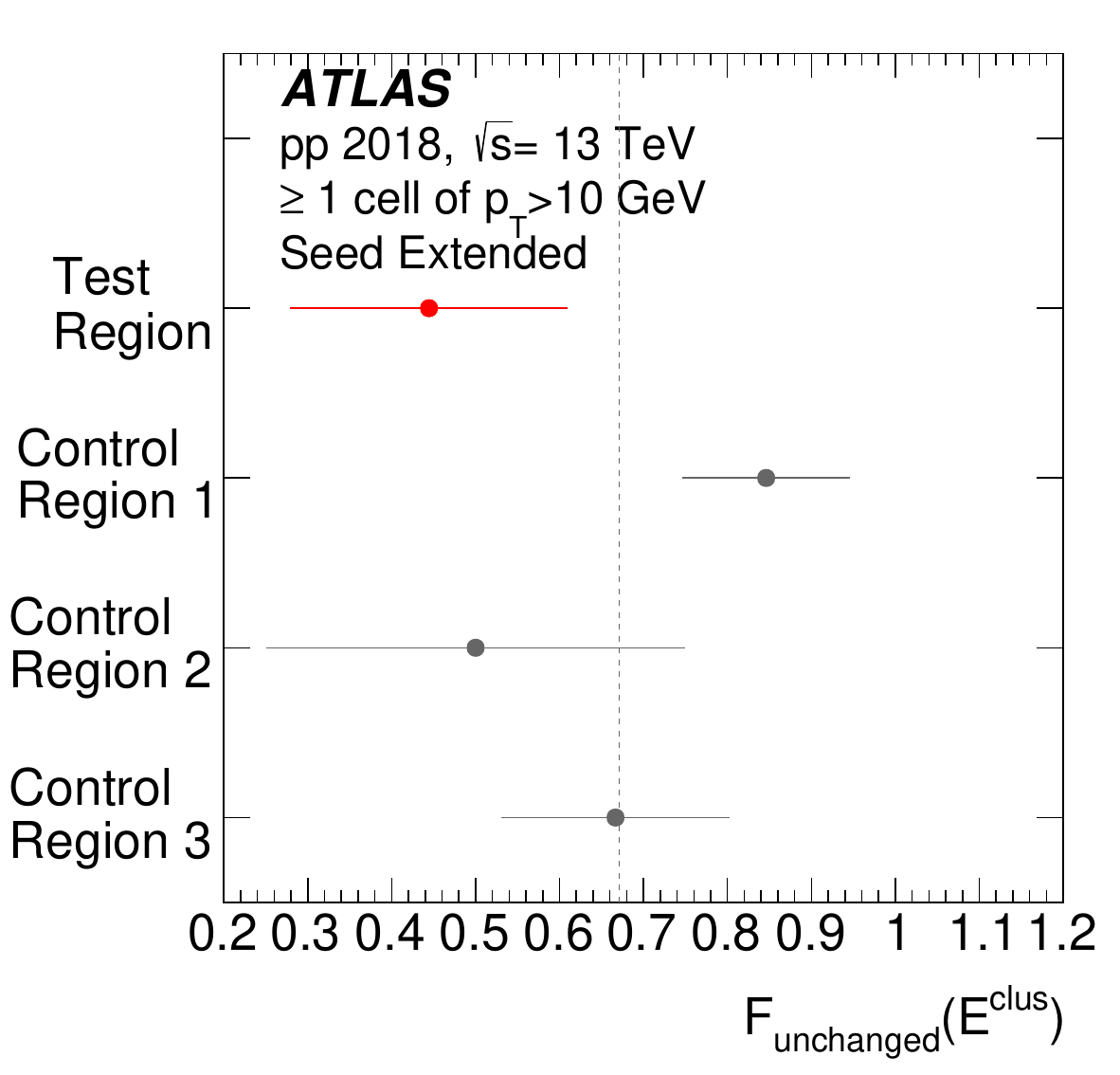}}
\subfigure[]{\label{subfig:clusTotE_acell_SEUL}
\includegraphics[width=0.49\linewidth]{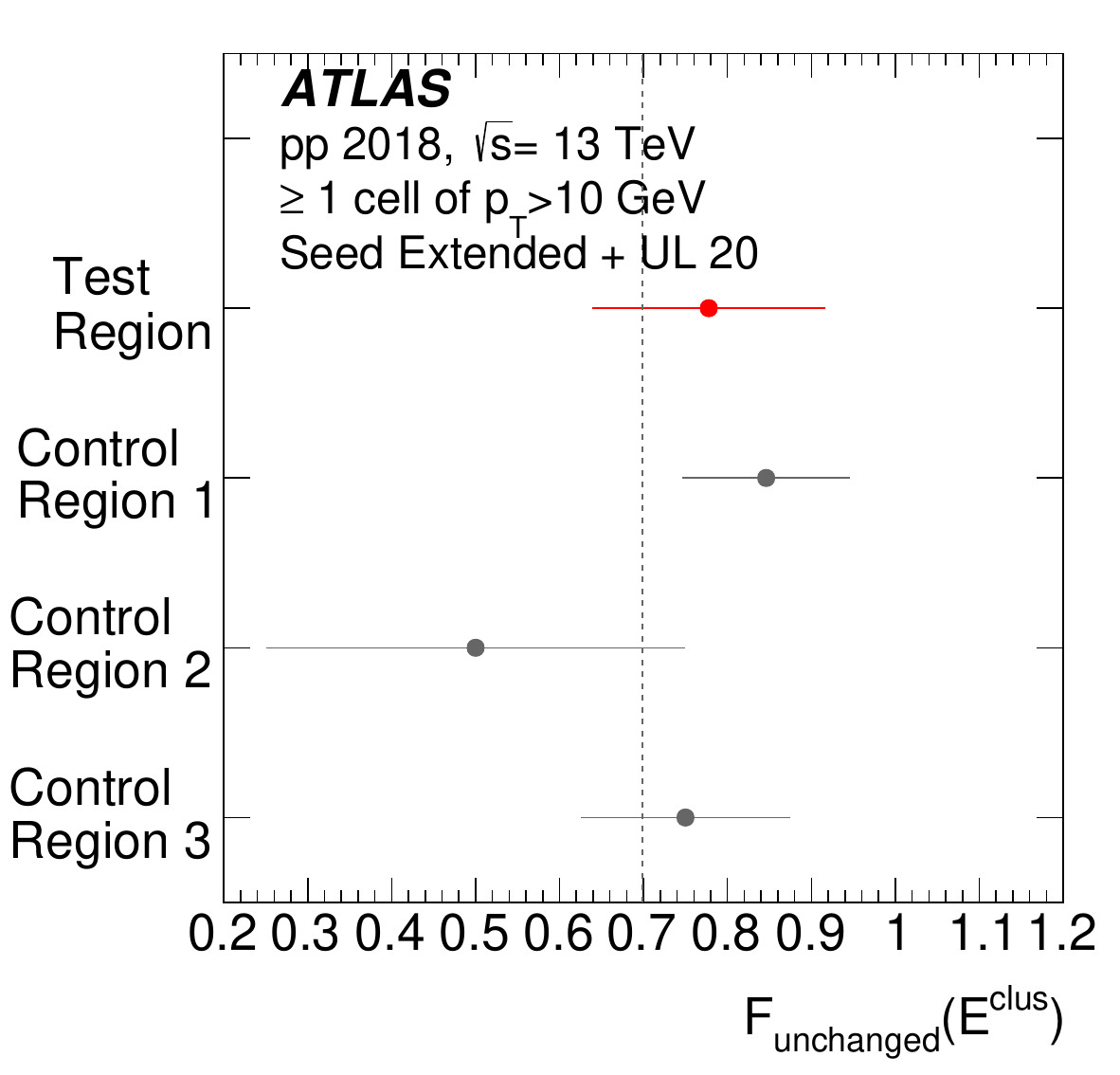}}
\caption{
Comparison between the Test Region and three Control Regions for \subref{subfig:clusOccupancy_acell_SE} the fraction of events in which the topo-cluster occupancy $N^\textrm{clus}$ is left unchanged by the Seed Extended cut, \subref{subfig:clusOccupancy_acell_SEUL} the fraction of events in which $N^\textrm{clus}$ is left unchanged by the Seed Extended plus $X_{\textrm{UL}}=20$ cut, \subref{subfig:clusTotE_acell_SE} the fraction of events in which the topo-cluster total energy $E^\textrm{clus}$ is left unchanged by the Seed Extended cut, and \subref{subfig:clusTotE_acell_SEUL} the fraction of events in which $E^\textrm{clus}$ is left unchanged by the Seed Extended plus $X_{\textrm{UL}}=20$ cut. The topo-cluster occupancy (total energy) in a given region is defined as the number (energy sum) of topo-clusters that contain at least one cell belonging to the region. Uncertainties on the event fractions are computed as the standard deviation of a binomial distribution. The dashed line represents the average of the three Control Regions.}
\label{fig:REGocc_clus_pt100}
\end{figure}

Figures~\ref{fig:REGocc_cells_pt100} and~\ref{fig:REGocc_clus_pt100} show $\mathcal{F}_{\text{unchanged}}$ for the four variables of interest. Figure~\ref{fig:REGocc_cells_pt100} is indicative of the time cut's impact on the cell content for clusters in the Test Region compared to that for clusters in the Control Regions.
The fraction of events with unchanged $N^\textrm{cells}$ is shown in Figures~\ref{subfig:cellOccupancySE} and~\ref{subfig:cellOccupancySEUL}: for the Seed Extended cut, $\mathcal{F}_{\text{unchanged}}$  in the Test Region is smaller than the average of the control values by slightly more than $1\sigma$. Even though the phenomenon has a small significance due to the limited sample size, the effect of the time cut can be seen to disappear when the same events are processed with the Seed Extended plus $X_{\textrm{UL}}=20$ cut, confirming that cell losses are cured by the Upper Limit.
The same behaviour is visible for $E^\textrm{cells}$ (Figures~\ref{subfig:cellTotESE} and~\ref{subfig:cellTotESEUL}), indicating that the loss of a cell, when it occurs, causes a significant energy variation, while no decrease in $E^\textrm{cells}$ can be seen in the case of the Seed Extended plus $X_{\textrm{UL}}=20$ cut.
Figure~\ref{fig:REGocc_clus_pt100} shows the fraction of events in which the topo-cluster occupancy and total energy are unchanged, thus investigating the effect of the time cut on topo-clusters as a whole, when they overlap the Test or Control Regions.
The topo-cluster occupancy (Figures~\ref{subfig:clusOccupancy_acell_SE} and~\ref{subfig:clusOccupancy_acell_SEUL}) remains consistent with the control values even for the Seed Extended cut, confirming that the signal topo-cluster is not entirely lost. The topo-cluster total energy $E^\textrm{clus}$ is impacted by the Seed Extended cut (Figure~\ref{subfig:clusTotE_acell_SE}), reflecting the loss of a cell in a way similar to that causing losses in the cells' total energy, while also being restored to its original value by applying the Upper Limit (Figure~\ref{subfig:clusTotE_acell_SEUL}).
 
% End of text imported from the .//ineffCheck_pap.tex input file

\FloatBarrier

% The next lines are included from the .//eventSize_pap.tex input file
\subsection{Impact on the ATLAS event size}
Calorimeter topo-clusters are used as input to the reconstruction of particles other than jets, most notably electrons, photons and $\tau$-leptons. The time cut will then have \pileup-suppressing effects on these particles as well. A complete discussion of the effects of the time cut on electrons, photons and $\tau$-leptons is beyond the scope of this \mypaper.
However, the widespread usage of topo-clusters in the ATLAS event reconstruction implies that the time cut can have a large beneficial effect on the overall consumption of computing resources by ATLAS data and MC samples.
 
The change in the ATLAS event size due to the time cut was evaluated for the introduction of the Seed Extended plus $X_{\textrm{UL}}=20$ time cut into the default ATLAS reconstruction at the beginning of Run~3, before the start of 2023 data-taking. The test is performed by reconstructing one data sample and one MC sample ($t\overline{t}$ production in the fully hadronic decay) before and after the introduction of the time cut and later comparing the average event sizes on disk. In both samples the detector conditions are those of Run~3.
Results are shown in Tables~\ref{tab:esizedata} and~\ref{tab:esizemc}. The principal particle collections are shown, together with their relative size change. As expected, the largest effects are observed for the particle-flow object and topo-cluster collections, which are reduced in size by ${\sim}16\%$ to ${\sim}18\%$, while smaller differences are present for electrons/photons and $\tau$-leptons\footnote{A very small difference in the \textquoteleft Tracking\textquoteright \ category is to be expected since specific information related to electron tracks is grouped in this category.}. The overall event size is reduced by about $6\%$ in data and $7\%$ in MC simulation.
 
\begin{table}[htb]
\sisetup{group-minimum-digits=4}
\caption{Collection size changes for data events on disk with the introduction of the time cut. Changes of $\mathcal{O}(100~\textrm{b})$ are not significant for most categories since they can arise from changes in the alignment of the (otherwise unchanged) objects and the subsequent compression when writing to disk. Larger size changes are reported relative to the original size on disk. The \enquote{Cells in topo-clusters} line is limited to clusters matched to tracks, electrons and muons.}
\label{tab:esizedata}
\centering
\begin{tabular}{llllc}
\toprule
Category & Fraction  & Size on disk  & Size on disk  & Size  \\
&  (before cut) & before cut &  after cut & change \\
\midrule
Total & 1.000 & 288.7 kb & 271.1 kb & $-6.1\%$ \\
\midrule
Trigger & 0.302 & \phantom{0}87.11 kb & \phantom{0}87.01 kb & - \\
Tracking & 0.218 & \phantom{0}62.95 kb & \phantom{0}62.84 kb & - \\
Topo-clusters & 0.150 & \phantom{0}43.21 kb & \phantom{0}35.95 kb & $-17\%$\phantom{0.0} \\
Particle-flow \cite{PERF-2015-09, JETM-2018-05} objects & 0.103 & \phantom{0}29.75 kb & \phantom{0}24.50 kb & $-18\%$\phantom{0.0} \\
$\tau$-leptons & 0.078 & \phantom{0}22.42 kb & \phantom{0}20.58 kb & $-8.2\%$ \\
Electrons / photons & 0.071 & \phantom{0}20.37 kb & \phantom{0}18.79 kb & $-7.8\%$ \\
Cells in topo-clusters & 0.057 & \phantom{0}16.60 kb & \phantom{0}15.13 kb & $-8.8\%$ \\
Muons & 0.018 & \phantom{00}5.293 kb & \phantom{00}5.295 kb & - \\
ATLAS forward proton detector \cite{ATLAS-TDR-24, ATL-PHYS-PUB-2017-012} & 0.003 & \phantom{00}0.810 kb & \phantom{00}0.810 kb & - \\
Metadata & 0.001 & \phantom{00}0.154 kb & \phantom{00}0.154 kb & - \\
\bottomrule
\end{tabular}
\end{table}

\begin{table}[htbp]
\sisetup{group-minimum-digits=4}
\caption{Collection size changes for MC events on disk with the introduction of the time cut. Changes of $\mathcal{O}(100~\textrm{b})$ are not significant for most categories since they can arise from changes in the alignment of the (otherwise unchanged) objects and the subsequent compression when writing to disk. The \enquote{Cells in topo-clusters} line is limited to clusters matched to tracks, electrons and muons.}
\label{tab:esizemc}
\centering
\begin{tabular}{llllc}
\toprule
Category & Fraction  & Disk Size & Disk Size  & Size  \\
&  (before cut) &  before cut &  after cut & change \\
\midrule
Total & \phantom{$<~\:\!$}1.000 & 428.8 kb & 398.3 kb & $-7.1\%$ \\
\midrule
Tracking & \phantom{$<~\:\!$}0.222 & \phantom{0}94.98 kb & \phantom{0}94.51 kb & $-0.5\%$ \\
Topo-clusters & \phantom{$<~\:\!$}0.160 & \phantom{0}68.79 kb & \phantom{0}57.34 kb & $-17\%$\phantom{0.0} \\
Particle-flow \cite{PERF-2015-09, JETM-2018-05} objects & \phantom{$<~\:\!$}0.114 & \phantom{0}49.00 kb & \phantom{0}39.94 kb & $-18\%$\phantom{0.0} \\
$\tau$-leptons & \phantom{$<~\:\!$}0.108 & \phantom{0}46.38 kb & \phantom{0}43.24 kb & $-6.8\%$ \\
Trigger & \phantom{$<~\:\!$}0.106 & \phantom{0}45.26 kb & \phantom{0}45.25 kb & - \\
Generator-level event & \phantom{$<~\:\!$}0.103 & \phantom{0}44.28 kb & \phantom{0}44.29 kb & - \\
Electrons / photons & \phantom{$<~\:\!$}0.086 & \phantom{0}36.94 kb & \phantom{0}33.74 kb & $-8.6\%$ \\
Cells in topo-clusters & \phantom{$<~\:\!$}0.082 & \phantom{0}34.98 kb & \phantom{0}31.80 kb & $-9.1\%$ \\
Muons & \phantom{$<~\:\!$}0.016 & \phantom{00}6.927 kb & \phantom{00}6.931 kb & - \\
Metadata & \phantom{$<~\:\!$}0.003 & \phantom{00}1.201 kb & \phantom{00}1.202 kb & - \\
ATLAS forward proton detector \cite{ATLAS-TDR-24, ATL-PHYS-PUB-2017-012} & $<0.001$ & \phantom{00}0.041 kb & \phantom{00}0.041 kb & - \\
\bottomrule
\end{tabular}
\end{table}

% End of text imported from the .//eventSize_pap.tex input file

\FloatBarrier
 
\section{Conclusion}
\label{sec:conclusion}
 
This \mypaper\ presents the design and evaluation of a cell-level time criterion used in the ATLAS topo-cluster reconstruction algorithm. This criterion (the \enquote{time cut}) removes cells compatible with out-of-time \pileup signals. A second requirement, the Upper Limit, is imposed to prevent cells with a large amount of energy from being removed, including those expected from long-lived particles.
The impact of the time cut on the reconstruction of hadronic signals was studied at both cluster level and jet level, using a combination of MC samples and ATLAS data recorded in 2017 and 2018.
 
The time cut was found to significantly reduce the contribution from out-of-time \pileup while not hindering signal reconstruction.
Studies of MC events indicate that the multiplicity of out-of-time \pileup jets is reduced by ${\sim}50\%$ at $\pT\sim20~\GeV$ and by ${\sim}80\%$ above $\pT \sim 50~\GeV$, across the rapidity region $|y|\lesssim 3.5$.
Studies of data events found that the time cut has negligible effect on events accepted by a single-jet trigger, requiring one jet with $\pT > 45~\GeV$ and seeded by a single jet at L1. On the other hand, the time cut does affect events enriched in out-of-time \pileup, typically those accepted by a random L1 trigger. In this case, the effect of the time cut is largest for $\pT \lesssim 30$~\GeV\ and it is found to have a stronger effect on data events than on MC events, typical differences being of $\mathcal{O}(10\%)$. The effect is believed to be caused by the absence of \pileup-only events in the MC simulation.
Additional tests for signal inefficiency in data were carried out and show that the time cut would not cause a loss of clusters in the case of a calorimeter channel miscalibration. The addition of the Upper Limit is also found to prevent the loss of individual cells within clusters in such a case.
 
The time cut was also found to impact the jet calibration, a repetition of the \pileup correction step being necessary. The time cut provides percent-level improvement in the jet resolution at $\pt \lesssim 30~\GeV$. More extensive investigation is needed to evaluate in detail the effect of the time cut on the jet energy resolution and jet energy scale uncertainty, e.g.\ the \pileup-related JES uncertainty components. The full calibration chain will be re-evaluated after the introduction of the time cut into the ATLAS reconstruction software.
 
The Seed Extended plus $X_{\textrm{UL}}=20$ cut was found to be the best option among those studied. This cut was adopted as the default for offline topo-cluster reconstruction in ATLAS Run~3 data. The introduction of the Seed Extended plus $X_{\textrm{UL}}=20$ cut into the ATLAS offline reconstruction is also found to reduce the ATLAS event size on disk by $6\%$ in data and by $7\%$ in MC simulation.
Further studies are necessary before the cut can be introduced at trigger level.
 
\section*{Acknowledgements}

% The next lines are included from the .//acknowledgements/Acknowledgements.tex input file

We thank CERN for the very successful operation of the LHC, as well as the
support staff from our institutions without whom ATLAS could not be
operated efficiently.
 
We acknowledge the support of
ANPCyT, Argentina;
YerPhI, Armenia;
ARC, Australia;
BMWFW and FWF, Austria;
ANAS, Azerbaijan;
CNPq and FAPESP, Brazil;
NSERC, NRC and CFI, Canada;
CERN;
ANID, Chile;
CAS, MOST and NSFC, China;
Minciencias, Colombia;
MEYS CR, Czech Republic;
DNRF and DNSRC, Denmark;
IN2P3-CNRS and CEA-DRF/IRFU, France;
SRNSFG, Georgia;
BMBF, HGF and MPG, Germany;
GSRI, Greece;
RGC and Hong Kong SAR, China;
ISF and Benoziyo Center, Israel;
INFN, Italy;
MEXT and JSPS, Japan;
CNRST, Morocco;
NWO, Netherlands;
RCN, Norway;
MEiN, Poland;
FCT, Portugal;
MNE/IFA, Romania;
MESTD, Serbia;
MSSR, Slovakia;
ARRS and MIZ\v{S}, Slovenia;
DSI/NRF, South Africa;
MICINN, Spain;
SRC and Wallenberg Foundation, Sweden;
SERI, SNSF and Cantons of Bern and Geneva, Switzerland;
MOST, Taipei;
TENMAK, T\"urkiye;
STFC, United Kingdom;
DOE and NSF, United States of America.
In addition, individual groups and members have received support from
BCKDF, CANARIE, CRC and DRAC, Canada;
PRIMUS 21/SCI/017 and UNCE SCI/013, Czech Republic;
COST, ERC, ERDF, Horizon 2020, ICSC-NextGenerationEU and Marie Sk{\l}odowska-Curie Actions, European Union;
Investissements d'Avenir Labex, Investissements d'Avenir Idex and ANR, France;
DFG and AvH Foundation, Germany;
Herakleitos, Thales and Aristeia programmes co-financed by EU-ESF and the Greek NSRF, Greece;
BSF-NSF and MINERVA, Israel;
Norwegian Financial Mechanism 2014-2021, Norway;
NCN and NAWA, Poland;
La Caixa Banking Foundation, CERCA Programme Generalitat de Catalunya and PROMETEO and GenT Programmes Generalitat Valenciana, Spain;
G\"{o}ran Gustafssons Stiftelse, Sweden;
The Royal Society and Leverhulme Trust, United Kingdom.
 
The crucial computing support from all WLCG partners is acknowledged gratefully, in particular from CERN, the ATLAS Tier-1 facilities at TRIUMF/SFU (Canada), NDGF (Denmark, Norway, Sweden), CC-IN2P3 (France), KIT/GridKA (Germany), INFN-CNAF (Italy), NL-T1 (Netherlands), PIC (Spain), RAL (UK) and BNL (USA), the Tier-2 facilities worldwide and large non-WLCG resource providers. Major contributors of computing resources are listed in Ref.~\cite{ATL-SOFT-PUB-2023-001}.

% End of text imported from the .//acknowledgements/Acknowledgements.tex input file

\printbibliography
 
\clearpage
% ATLAS Collaboration author list
% Reference date of JETM-2023-01 is 2023-07-07
% Author list last updated on date 23-JAN-24
% Data extracted on 23-Jan-2024 for paper reference JETM-2023-01
% at 3:20pm
 
\begin{flushleft}
\hypersetup{urlcolor=black}
{\Large The ATLAS Collaboration}

\bigskip

\AtlasOrcid[0000-0002-6665-4934]{G.~Aad}$^\textrm{\scriptsize 102}$,
\AtlasOrcid[0000-0002-5888-2734]{B.~Abbott}$^\textrm{\scriptsize 120}$,
\AtlasOrcid[0000-0002-1002-1652]{K.~Abeling}$^\textrm{\scriptsize 55}$,
\AtlasOrcid[0000-0001-5763-2760]{N.J.~Abicht}$^\textrm{\scriptsize 49}$,
\AtlasOrcid[0000-0002-8496-9294]{S.H.~Abidi}$^\textrm{\scriptsize 29}$,
\AtlasOrcid[0000-0002-9987-2292]{A.~Aboulhorma}$^\textrm{\scriptsize 35e}$,
\AtlasOrcid[0000-0001-5329-6640]{H.~Abramowicz}$^\textrm{\scriptsize 151}$,
\AtlasOrcid[0000-0002-1599-2896]{H.~Abreu}$^\textrm{\scriptsize 150}$,
\AtlasOrcid[0000-0003-0403-3697]{Y.~Abulaiti}$^\textrm{\scriptsize 117}$,
\AtlasOrcid[0000-0002-8588-9157]{B.S.~Acharya}$^\textrm{\scriptsize 69a,69b,m}$
% ...
\bigskip
\\

% ...
$^{29}$Physics Department, Brookhaven National Laboratory, Upton NY; United States of America.\\
% ...
$^{35}$$^{(a)}$Facult\'e des Sciences Ain Chock, R\'eseau Universitaire de Physique des Hautes Energies - Universit\'e Hassan II, Casablanca;$^{(b)}$Facult\'{e} des Sciences, Universit\'{e} Ibn-Tofail, K\'{e}nitra;$^{(c)}$Facult\'e des Sciences Semlalia, Universit\'e Cadi Ayyad, LPHEA-Marrakech;$^{(d)}$LPMR, Facult\'e des Sciences, Universit\'e Mohamed Premier, Oujda;$^{(e)}$Facult\'e des sciences, Universit\'e Mohammed V, Rabat;$^{(f)}$Institute of Applied Physics, Mohammed VI Polytechnic University, Ben Guerir; Morocco.\\
% ...
$^{49}$Fakult\"{a}t Physik , Technische Universit{\"a}t Dortmund, Dortmund; Germany.\\
% ...
$^{55}$II. Physikalisches Institut, Georg-August-Universit\"{a}t G\"ottingen, G\"ottingen; Germany.\\
% ...
$^{69}$$^{(a)}$INFN Gruppo Collegato di Udine, Sezione di Trieste, Udine;$^{(b)}$ICTP, Trieste;$^{(c)}$Dipartimento Politecnico di Ingegneria e Architettura, Universit\`a di Udine, Udine; Italy.\\
% ...
$^{102}$CPPM, Aix-Marseille Universit\'e, CNRS/IN2P3, Marseille; France.\\
% ...
$^{117}$Department of Physics, New York University, New York NY; United States of America.\\
% ...
$^{120}$Homer L. Dodge Department of Physics and Astronomy, University of Oklahoma, Norman OK; United States of America.\\
% ...
$^{150}$Department of Physics, Technion, Israel Institute of Technology, Haifa; Israel.\\
$^{151}$Raymond and Beverly Sackler School of Physics and Astronomy, Tel Aviv University, Tel Aviv; Israel.\\
% ...
$^{a}$ Also Affiliated with an institute covered by a cooperation agreement with CERN.\\
$^{b}$ Also at An-Najah National University, Nablus; Palestine.\\
$^{c}$ Also at Borough of Manhattan Community College, City University of New York, New York NY; United States of America.\\
$^{d}$ Also at Center for High Energy Physics, Peking University; China.\\
$^{e}$ Also at Center for Interdisciplinary Research and Innovation (CIRI-AUTH), Thessaloniki; Greece.\\
$^{f}$ Also at Centro Studi e Ricerche Enrico Fermi; Italy.\\
% ...
$^{m}$ Also at Department of Physics, King's College London, London; United Kingdom.\\
% ...

\end{flushleft}

% Created with Glance <Atlas.Glance@cern.ch>

\clearpage

\end{document}